%% file: neurips_2026.tex
\documentclass{article}

\usepackage[preprint]{neurips_2026}
\input{commands}
\usepackage[utf8]{inputenc} 
\usepackage[T1]{fontenc}    
\usepackage[draft]{hyperref} 
\usepackage{url}            
\usepackage{booktabs}       
\usepackage{amsmath}        
\usepackage{amsfonts}       
\usepackage{nicefrac}       
\usepackage{microtype}      
\usepackage{multirow}       
\usepackage[table]{xcolor}  
\usepackage{natbib}
\usepackage{tcolorbox}      
\usepackage{listings}       
\usepackage{float}          
\usepackage{pifont}         
\usepackage{color}          
\usepackage{xspace}         
\usepackage{tikz}           
\usetikzlibrary{arrows.meta,positioning,shapes.geometric,fit,backgrounds,calc,decorations.pathreplacing}
\usepackage{alltt}          
\usepackage{fancyvrb}       
\usepackage{titletoc}       
\newcommand{\caseStudyBarrier}{\clearpage}
\newcommand{\mainFigureBarrier}{\par\medskip}
\definecolor{stageProb}{RGB}{220,220,220}    
\definecolor{stageSpec}{RGB}{255,210,160}    
\definecolor{stageExec}{RGB}{180,220,235}    
\definecolor{stageEval}{RGB}{200,230,200}    
\definecolor{stageProbD}{RGB}{120,120,120}
\definecolor{stageSpecD}{RGB}{200,120,30}
\definecolor{stageExecD}{RGB}{50,120,150}
\definecolor{stageEvalD}{RGB}{60,130,70}

\newtcolorbox{artifactbox}[3]{%
  width=3.3cm, height=4.55cm,
  colback=white, colframe=black!50,
  arc=2pt, boxrule=0.4pt,
  title={#1},
  fonttitle=\footnotesize\bfseries,
  colbacktitle=#2, coltitle=black,
  boxsep=0pt, top=2pt, bottom=2pt, left=3pt, right=3pt,
  toptitle=1pt, bottomtitle=1pt,
  valign=top,
  before skip=0pt, after skip=0pt,
  #3
}

\lstdefinelanguage{Verus}{
  keywords=[1]{fn,let,pub,struct,use,mod,impl,trait,enum,match,if,else,
    for,while,loop,return,true,false,self,Self,super,crate,as,in,mut,
    ref,where,type,const,static,unsafe,extern,move,async,await,dyn,
    spec,proof,exec,requires,ensures,decreases,recommends,
    open,closed,tracked,ghost,trigger,verus,assert,assume,reveal,
    reveal_with_fuel,forall,exists},
  keywords=[2]{bool,i64,i128,u8,u16,u32,u64,usize,isize,char,str,
    String,Vec,Option,Some,None,Result,Ok,Err,Box,
    Seq,Map,Set,In1,Out,ExecIn1,ExecOut,
    exec_spec_unverified,deep_view,int,nat},
  sensitive=true,
  morecomment=[l]{//},
  morecomment=[s]{/*}{*/},
  morestring=[b]",
}

\title{Verus-SpecGym: An Agentic Environment for Evaluating Specification Autoformalization}

\author{%
Anmol Agarwal\\
  CMU \\
  \And
  Natalie Neamtu\\
  CMU \\
  \And
  Pranjal Aggarwal\\
  CMU \\
  \AND
  Seungone Kim\\
  CMU\\
  \And
  Jannis Limperg\thanks{Work performed while at Amazon.}\\
  Amazon\\
  \And
  Cedric Flamant\\
  Amazon\\
  \And
  Kanna Shimizu\\
  Amazon\\
  \AND
  Bryan Parno\\
  CMU\\
  \And
  Sean Welleck\\
  CMU
}

\begin{document}

\maketitle

\begin{abstract}
AI coding agents are increasingly used to write real-world software, but ensuring that their outputs are correct
remains a fundamental challenge. Formal verification offers a promising path: an agent generates code together with a
machine-checked proof, guaranteeing that the code satisfies a formal specification.
However, there is no guarantee that the formal specification itself matches the user's intent.
In this work, we study \textit{specification autoformalization}: whether language-model agents can translate informal programming problems into faithful formal specifications.
We introduce \vsb, a benchmark of \numproblems{} specification-writing tasks derived from Codeforces problems and targeting Verus, a verifier for Rust, and \vsg, an agentic environment in which models interact with Verus, bash, and the filesystem to develop these specifications.
The central challenge is evaluation: expert-written reference specifications are expensive to write, and LLM judges can miss subtle mistakes.
We address this by (a) extending Verus's \texttt{exec\_spec} mechanism so that generated specifications can be executed as Rust code, and (b) testing them against official Codeforces tests and adversarial cases extracted from Codeforces ``hacks'', which are edge cases written by competitors to break incorrect solutions.
On \vsb, the strongest frontier model, \geminithreeonepro, solves 77.8\% of tasks, other frontier models solve 51.1--57.8\%, and open-source models reach only 21.5--25.5\%.
Our analysis of failure modes shows that model-generated specifications can omit important input assumptions, accept incorrect outputs, and reject valid ones.
Separately, we find that LLM-as-a-judge evaluation misses 26\% of the failures our evaluator catches.
Overall, our results suggest that specification autoformalization is within reach for frontier agents but remains brittle even on problems where they can already generate correct code. The code for running the benchmark, links to the tasks, and the dashboard for trajectory logs can be found at \url{https://github.com/formal-verif-is-cool/verus-spec-gym}.
\todoanmol{PDF note p.~1: Check whether this URL needs to be changed or switched.}

\end{abstract}
\input{paper_sections/main_paper/00_introduction_v2}

\input{paper_sections/main_paper/01_verus_gym_v2}

\input{paper_sections/main_paper/01_z_data}

\input{paper_sections/main_paper/02_experiments}

\input{paper_sections/main_paper/03_related_work}

\input{paper_sections/main_paper/04_conclusion}

\input{paper_sections/main_paper/acknowledgements}

\clearpage




\bibliographystyle{plainnat}
\bibliography{bibliography}

\newpage

\appendix
\raggedbottom  

\startcontents[appendices]
\section*{Appendix Table of Contents}
\printcontents[appendices]{}{0}

\input{paper_sections/appendix/appendix_verus_binary_search}

\input{paper_sections/appendix/related_work_table}

\input{paper_sections/appendix/appendix_skeleton}

\input{paper_sections/appendix/appendix_exec_spec}

\input{paper_sections/appendix/additional_data_creation}

\input{paper_sections/appendix/additional_dataset_statistics}
\input{paper_sections/appendix/additional_insights}

\input{paper_sections/appendix/failure_modes}

\input{paper_sections/appendix/appendix_prompt}



\end{document}

%% file: commands.tex
\newcommand{\PreSpecCompleteness}{\colorbox{pink!25}{\textsc{pre\_spec completeness}}}
\newcommand{\PostSpecCompleteness}{\colorbox{yellow!20}{\textsc{post\_spec completeness}}}
\newcommand{\PreSpecSoundness}{\colorbox{green!20}{\textsc{pre\_spec soundness}}}
\newcommand{\PostSpecSoundness}{\colorbox{red!15}{\textsc{post\_spec soundness}}}

\newif\ifsubmit    
\submitfalse       

\ifsubmit
  \newcommand{\todo}[1]{}
  \newcommand{\todonn}[1]{}
  
\else
  \newcommand{\todo}[1]{\ding{110}\ding{43}\textcolor{red}{#1}}
  \newcommand{\todonn}[1]{\ding{110}\ding{43}\textcolor{red}{NN: #1}}
  
\fi

\newcommand{\todoanmol}[1]{}

\newcommand{\cmark}{\textcolor{green!60!black}{\ding{51}}}
\newcommand{\xmark}{\textcolor{red!70!black}{\ding{55}}}
\newcommand{\partialmark}{\textcolor{orange!80!black}{$\bullet$}}

\newcommand{\vsb}{\textsc{Verus-SpecBench}\xspace}
\newcommand{\benchmark}{\vsb}

\newcommand{\vsg}{\textsc{Verus-SpecGym}\xspace}

\newcommand{\sweagent}{\textsc{SWE-Agent}\xspace}

\newcommand{\execspec}{\texttt{exec\_spec}\xspace}

\newcommand{\prespec}{\texttt{pre\_spec}\xspace}
\newcommand{\postspec}{\texttt{post\_spec}\xspace}
\newcommand{\presound}{\texttt{pre\_sound}\xspace}
\newcommand{\precomplete}{\texttt{pre\_complete}\xspace}
\newcommand{\postsound}{\texttt{post\_sound}\xspace}
\newcommand{\postcomplete}{\texttt{post\_complete}\xspace}

\newcommand{\numproblems}{581}

\newcommand{\gptfivethreecodex}{\texttt{gpt5.3-codex}\xspace}
\newcommand{\opusfoursix}{\texttt{opus4.6}\xspace}
\newcommand{\deepseekvfourpro}{\texttt{deepseek-v4pro}\xspace}
\newcommand{\kimiktwosix}{\texttt{kimi-k2.6}\xspace}
\newcommand{\geminithreeonepro}{\texttt{gemini-3.1pro}\xspace}
\newcommand{\glmfiveone}{\texttt{glm-5.1}\xspace}

%% file: paper_sections/main_paper/00_introduction_v2.tex
\input{paper_sections/figures_plots/tikz_images/01_hero_fig}

\section{Introduction}

AI coding agents are increasingly used to write real-world software~\citep{anthropic2025claudecode,openai2026gpt53codex}, but ensuring the correctness of AI-generated code remains a fundamental challenge~\citep{liu2023codegeneratedchatgptreally,ji2024cybersecurity,wessling2025genai}.
\emph{Verified code generation}~\citep{sunClover2024,misu2024towards,aggarwal_alphaverus_2024} offers a promising path: given a formal specification of a program's intended behavior, an agent generates code together with proofs, and a mechanical verifier checks the code against the specification.
If verification succeeds, the code is mathematically guaranteed to satisfy the specification.\looseness=-1

However, this guarantee is only useful if the specification is right.
Verified code generation assumes that a formal specification is already available, but in most settings the user starts
with an informal description of what the program should do.
For example, a programming problem states valid inputs, required outputs, edge cases, and implicit constraints in natural
language, while the verifier needs these requirements written as precise logical predicates.
Thus, as agents become better at writing code and proofs~\citep{yang2025verusage,liu2026numina}, the bottleneck shifts to
translating informal intent into the right formal specification.

To this end, our work focuses on answering the question:
\begin{quote}
\centering
    Can language-model agents write specifications\\
    that faithfully capture informal programming intent?
\end{quote}

We refer to this problem as \emph{specification autoformalization}.
A generated specification is faithful if it accepts exactly the behavior allowed by the informal problem: it should accept valid inputs and correct outputs, while rejecting invalid inputs and incorrect outputs.
If the specification is too weak, a verifier may certify an incorrect program; if it is too strong, a correct program may fail to verify.

Evaluating specification faithfulness, however, is itself a hard problem.
Existing approaches rely on either expert-written reference specifications~\citep{ye2025verina} or LLM-based  judges~\citep{sunClover2024,deng2025verifythisbench}.
Reference specifications require an expert to author a gold standard per problem and are therefore expensive to scale.
LLM judges are cheaper but only approximate, and can miss subtle errors that are precisely where unfaithful specifications tend to fail.
Neither approach gives a trustworthy, scalable signal.

We therefore introduce \vsb, a benchmark of \numproblems{} specification-writing tasks derived from Codeforces programming problems and targeting Verus~\citep{lattuada2023verus}, a verification framework for Rust, together with \vsg, an agentic environment for evaluating specification autoformalization.
In \vsg, an agent develops a formal specification for a natural-language problem by interacting with Verus, bash, and the filesystem, iteratively refining its draft based on errors and feedback returned by the verifier.
To obtain a scalable, execution-based evaluation signal, we make these generated specifications executable.
Verus specifications are logical predicates intended for the verifier rather than ordinary Rust code, so they cannot normally be run on concrete inputs.
We extend Verus's \texttt{exec\_spec} mechanism (\S\ref{ssec:evaluation}) to compile each generated specification into a Rust function, expanding the supported types and operators to cover the kinds of constraints that arise in Codeforces-style problems.
To make these evaluations practical at scale, \vsg{} additionally integrates with Harbor~\citep{harbor2026}, matching modern agent-evaluation workflows where agents interact with tools, receive feedback, and submit final answers.\looseness=-2

We test each generated specification against two sources of test cases: official Codeforces tests, and adversarial inputs extracted from Codeforces ``hacks'', inputs that competitors design to break incorrect solutions after they pass the official tests.
Since these hacks are written by humans against real solutions, they reflect the kinds of subtle edge cases that ad-hoc adversarial test-generation tends to miss.
To support continued growth of \vsb, we also provide a semi-automatic pipeline for converting new Codeforces problems and their hacks into specification-writing tasks (\S\ref{ssec:data}), so that the benchmark can grow as more Codeforces problems are released.

We evaluate six frontier and open-source language-model agents on \vsb under a fixed compute and time budget.
The strongest model, \geminithreeonepro, solves 77.8\% of tasks, while other frontier models solve 51.1--57.8\% and the open-source models we evaluate reach only 21.5--25.5\%, revealing a substantial capability gap between frontier and open systems.
Strikingly, even on problems where current agents can generate correct code, they often fail to write a faithful specification for the same problem, and our case studies reveal that the resulting failures cluster into three recurring modes: specifications that omit important input assumptions, specifications that accept incorrect outputs, and specifications that reject valid ones.

We further conduct a series of ablations to validate the key design decisions behind \vsb.
Increasing the number and diversity of test cases used in evaluation steadily lowers measured success, showing that specification faithfulness is easy to overestimate with a sparse test suite and motivating the comprehensive evaluation we provide.
Adversarial hacks turn out to be particularly valuable: they expose specification failures that the official Codeforces tests miss entirely, confirming that human-written adversarial inputs are an essential complement to ordinary test cases rather than a redundant overlay.
Finally, we find that an LLM-as-a-judge baseline misses 26\% of the failures our evaluator catches, showing that executable testing can flag specification errors that LLM judges overlook.

In summary, our contributions are:
(1) We introduce \vsg, an agentic environment for evaluating specification autoformalization, and \vsb, a benchmark of \numproblems{} specification-writing tasks derived from Codeforces problems and targeting Verus, a verifier for Rust.
(2) We develop an executable-specification evaluator by extending Verus's \texttt{exec\_spec} mechanism, enabling deterministic testing of specification faithfulness without expert-written reference specifications or LLM judges.
(3) We construct an evaluation suite from official Codeforces tests and human-written hacks (real adversarial inputs designed to break incorrect solutions), and show that hacks expose specification failures the official tests miss.
(4) We evaluate six frontier and open-source agents and find that specification autoformalization is brittle even for models that can generate correct code, with failures clustering into omitted input assumptions, accepted incorrect outputs, and rejected valid ones.

%% file: paper_sections/figures_plots/tikz_images/01_hero_fig.tex
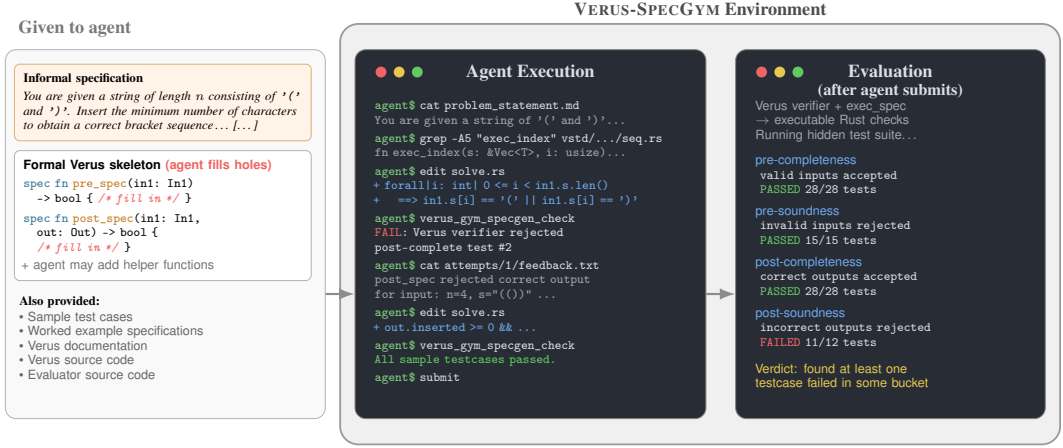
\begin{figure}[t]
\centering
\resizebox{\textwidth}{!}{%
\begin{tikzpicture}[
  font=\small,
  >=Stealth,
  arr/.style={-{Latex[length=3mm]}, line width=1.2pt, black!45},
  paneltitle/.style={font=\normalsize\bfseries},
]

\definecolor{termbg}{RGB}{40,44,52}
\definecolor{termgreen}{RGB}{100,200,100}
\definecolor{termred}{RGB}{240,100,100}
\definecolor{termyellow}{RGB}{230,200,80}
\definecolor{termblue}{RGB}{110,170,240}
\definecolor{termgray}{RGB}{160,165,175}
\definecolor{termprompt}{RGB}{130,210,130}
\definecolor{termwhite}{RGB}{220,223,228}

\node[
  draw=black!30, rounded corners=6pt,
  fill=stageProb!18,
  inner sep=0pt,
  minimum width=6.2cm, minimum height=7.15cm,
  anchor=north west
] (p1) at (0, 0) {};

\node[paneltitle, color=stageProbD, anchor=south west]
  at ([shift={(4pt,2pt)}]p1.north west) {\strut{}Given to agent};

\node[
  draw=stageSpecD!60, rounded corners=3pt,
  fill=stageSpec!25,
  text width=5.4cm,
  inner sep=5pt,
  anchor=north west,
  font=\scriptsize
] (prob) at ([shift={(5pt,-8pt)}]p1.north west) {%
  \textbf{Informal specification}\\[2pt]
  {\itshape You are given a string of length~$n$
  consisting of \texttt{'('} and \texttt{')'}. Insert
  the minimum number of characters to obtain a
  correct bracket sequence\,\ldots\,[\dots]}
};

\node[
  draw=black!30, rounded corners=3pt,
  fill=white,
  text width=5.4cm,
  inner sep=5pt,
  anchor=north west,
  font=\scriptsize\ttfamily
] (skel) at ([shift={(0,-5pt)}]prob.south west) {%
  \textbf{\textsf{Formal Verus skeleton} \textsf{\textcolor{termred}{(agent fills holes)}}}\\[2pt]
  \textcolor{stageExecD}{spec fn} \textcolor{stageSpecD}{pre\_spec}(in1: In1)\\
  \quad -> bool \{ \textcolor{termred}{\textit{/* fill in */}} \}\\[3pt]
  \textcolor{stageExecD}{spec fn} \textcolor{stageSpecD}{post\_spec}(in1: In1,\\
  \quad out: Out) -> bool \{\\
  \quad\textcolor{termred}{\textit{/* fill in */}} \}\\[2pt]
  {\textsf{\textcolor{black!55}{+ agent may add helper functions}}}
};

\node[
  anchor=north west,
  font=\scriptsize,
  text width=5.4cm,
  inner sep=3pt
] (files) at ([shift={(0,-5pt)}]skel.south west) {%
  \textbf{Also provided:}\\[1pt]
  {\color{black!60}
  \textbullet~\textsf{Sample test cases}\\
  \textbullet~\textsf{Worked example specifications}\\
  \textbullet~\textsf{Verus documentation}\\
  \textbullet~\textsf{Verus source code}\\
  \textbullet~\textsf{Evaluator source code}}
};

\node[
  draw=black!50, rounded corners=6pt,
  fill=termbg,
  inner sep=0pt,
  minimum width=6.8cm, minimum height=7.15cm,
  anchor=north west
] (p2) at ([shift={(16pt,0)}]p1.north east) {};

\node[anchor=north west] at ([shift={(8pt,-6pt)}]p2.north west) {%
  \tikz{
    \fill[termred] (0,0) circle (3pt);
    \fill[termyellow] (10pt,0) circle (3pt);
    \fill[termgreen] (20pt,0) circle (3pt);
  }%
};
\node[paneltitle, color=termwhite, anchor=north]
  at ([shift={(0,-4pt)}]p2.north) {\strut{}Agent Execution};

\node[
  anchor=north west,
  text width=6.0cm,
  inner sep=6pt,
  font=\scriptsize\ttfamily
] (term1) at ([shift={(5pt, -22pt)}]p2.north west) {%
  \textcolor{termprompt}{agent\$} \textcolor{termwhite}{cat problem\_statement.md}\\
  \textcolor{termgray}{You are given a string of '(' and ')'\ldots}\\[2pt]
  \textcolor{termprompt}{agent\$} \textcolor{termwhite}{grep -A5 "exec\_index" vstd/.../seq.rs}\\
  \textcolor{termgray}{fn exec\_index(s: \&Vec<T>, i: usize)\ldots}\\[2pt]
  \textcolor{termprompt}{agent\$} \textcolor{termwhite}{edit solve.rs}\\
  \textcolor{termblue}{+\,\,forall|i: int| 0 <= i < in1.s.len()}\\
  \textcolor{termblue}{+\,\,\quad==> in1.s[i] == '(' || in1.s[i] == ')'}\\[2pt]
  \textcolor{termprompt}{agent\$} \textcolor{termwhite}{verus\_gym\_specgen\_check}\\
  \textcolor{termred}{FAIL}\textcolor{termwhite}{: Verus verifier rejected}\\
  \textcolor{termwhite}{post-complete test \#2}\\[2pt]
  \textcolor{termprompt}{agent\$} \textcolor{termwhite}{cat attempts/1/feedback.txt}\\
  \textcolor{termgray}{post\_spec rejected correct output}\\
  \textcolor{termgray}{for input: n=4, s="(())" \ldots}\\[2pt]
  \textcolor{termprompt}{agent\$} \textcolor{termwhite}{edit solve.rs}\\
  \textcolor{termblue}{+\,\,out.inserted >= 0 \&\& \ldots}\\[2pt]
  \textcolor{termprompt}{agent\$} \textcolor{termwhite}{verus\_gym\_specgen\_check}\\
  \textcolor{termgreen}{All sample testcases passed.}\\[2pt]
  \textcolor{termprompt}{agent\$} \textcolor{termwhite}{submit}
};

\node[
  draw=black!50, rounded corners=6pt,
  fill=termbg,
  inner sep=0pt,
  minimum width=6.0cm, minimum height=7.15cm,
  anchor=north west
] (p3) at ([shift={(16pt,0)}]p2.north east) {};

\node[anchor=north west] at ([shift={(8pt,-6pt)}]p3.north west) {%
  \tikz{
    \fill[termred] (0,0) circle (3pt);
    \fill[termyellow] (10pt,0) circle (3pt);
    \fill[termgreen] (20pt,0) circle (3pt);
  }%
};
\node[paneltitle, color=termwhite, anchor=north, align=center]
  at ([shift={(0,-4pt)}]p3.north) {\strut{}Evaluation\\[-1pt]{\small(after agent submits)}};

\node[
  anchor=north west,
  text width=5.2cm,
  inner sep=6pt,
  font=\scriptsize\ttfamily
] (eval1) at ([shift={(5pt, -22pt)}]p3.north west) {%
  \textcolor{termgray}{\textsf{Verus verifier + exec\_spec}}\\
  \textcolor{termgray}{$\to$ \textsf{executable Rust checks}}\\
  \textcolor{termgray}{\textsf{Running hidden test suite\ldots}}\\[6pt]
  \textcolor{termblue}{\textsf{pre-completeness}}\\
  \textcolor{termwhite}{\,\,valid inputs accepted}\\
  \textcolor{termgreen}{\,\,PASSED}\textcolor{termwhite}{\,\,28/28 tests}\\[4pt]
  \textcolor{termblue}{\textsf{pre-soundness}}\\
  \textcolor{termwhite}{\,\,invalid inputs rejected}\\
  \textcolor{termgreen}{\,\,PASSED}\textcolor{termwhite}{\,\,15/15 tests}\\[4pt]
  \textcolor{termblue}{\textsf{post-completeness}}\\
  \textcolor{termwhite}{\,\,correct outputs accepted}\\
  \textcolor{termgreen}{\,\,PASSED}\textcolor{termwhite}{\,\,28/28 tests}\\[4pt]
  \textcolor{termblue}{\textsf{post-soundness}}\\
  \textcolor{termwhite}{\,\,incorrect outputs rejected}\\
  \textcolor{termred}{\,\,FAILED}\textcolor{termwhite}{\,\,11/12 tests}\\[7pt]
  \textcolor{termyellow}{\textsf{Verdict: found at least one}}\\
  \textcolor{termyellow}{\textsf{testcase failed in some bucket}}
};

\begin{scope}[on background layer]
\node[
  draw=black!40, rounded corners=10pt,
  fill=black!6,
  line width=1pt,
  inner xsep=8pt, inner ysep=14pt,
  fit=(p2)(p3),
] (container) {};
\end{scope}

\node[font=\normalsize\bfseries, color=black!70, anchor=south]
  at ([yshift=1pt]container.north) {\vsg{} Environment};

\draw[arr] ([yshift=-33pt]p1.east) -- ([yshift=-33pt]p2.west);
\draw[arr] ([yshift=-33pt]p2.east) -- ([yshift=-33pt]p3.west);

\end{tikzpicture}%
}
\caption{\textbf{Overview of \vsg{}.}
\textbf{Left:} Each task gives the agent an informal Codeforces problem description, a formal Verus skeleton, and the function signatures for \texttt{pre\_spec} and \texttt{post\_spec}. The \texttt{pre\_spec} function encodes input properties, while \texttt{post\_spec} encodes output properties. The task also includes sample test cases, worked examples, Verus documentation, and evaluator source code.
\textbf{Middle:} The agent reads the problem, writes formal specifications, and iteratively refines them using feedback from \texttt{verus\_gym\_specgen\_check}, which runs the Verus verifier on the sample testcases.
\textbf{Right:} After the agent submits, \vsg{} evaluates the specification on hidden tests. Verus specifications are logical predicates intended for the verifier rather than ordinary Rust code, so they cannot normally be run on concrete inputs; to get around this, we extend Verus's \texttt{exec\_spec} mechanism to compile specifications into executable Rust checks.
Command outputs are truncated due to space limits.
}
\label{fig:hero}
\end{figure}

%% file: paper_sections/main_paper/01_verus_gym_v2.tex
\section{Specification Autoformalization and Evaluation}

This section introduces the formal setup for specification autoformalization and the evaluator we use to measure whether a generated specification is faithful to the informal problem statement.
We focus on Verus~\citep{lattuada2023verus,verus-sys}, a Rust verification framework where specifications are written as logical predicates over programs.
In Verus, \texttt{requires} clauses express preconditions, \texttt{ensures} clauses express postconditions, and verification conditions are checked by the Z3 SMT solver~\citep{z3}.
We use the binary-search task in Figure~\ref{fig:binary-search-bucket-example} as a running example throughout this section.

\input{paper_sections/figures_plots/big_sample}

\subsection{Problem Setup}
\label{ssec:problem}

\textbf{Code generation.}
In \textit{code generation}, the goal is to generate a program that meets a given intent.
We are given a programming problem in the form of an \textit{informal, natural language specification} $s_{I}$ that defines the intent of the program.
This intended behavior can be represented as a relation $R_{s_I}$ over inputs and outputs, where $(x, y)\in R_{s_I}$ means that $y$ is a valid output for input $x$.
The domain $\mathrm{dom}(R_{s_I})$ is the set of valid inputs according to the informal specification.
The goal is to generate a program $p$ that produces a valid output for every valid input:
\begin{equation}
\label{eqn:codegen}
    \forall x\in \mathrm{dom}(R_{s_I}), (x, p(x))\in R_{s_I}.
\end{equation}

In practice, we do not have access to the full relation $R_{s_I}$.
Therefore, (unverified) code-generation benchmarks evaluate $p$ on a finite test set $\tau=\{(x_i,Y_i)\}$, where each $x_i$ is a valid input and $Y_i$ is the set of all correct outputs for $x_i$.
The program $p$ is considered ``correct'' if it produces a valid output for all test cases,
$\forall (x_i, Y_i)\in \tau, p(x_i)\in Y_i.$, however, it does not establish correctness beyond those test cases.

\textbf{Verified code generation.}
In \textit{verified code generation}, given an informal specification $s_I$, the goal is to generate $(s_F, p)$, where $s_F$ is a \textit{formal specification} that captures the intent of $s_I$, and $p$ is a program in a verifiable language.
The key benefit is that we can use a verifier $v(s_F, p)\in\{0,1\}$, where $v(s_F, p)=1$ means verification succeeded and $v(s_F, p)=0$ means it failed.
Letting $R_{s_F}$ be the input-output relation defined by $s_F$, a successful verification ($v(s_F, p)=1$) guarantees that for every valid input $x \in \mathrm{dom}(R_{s_F})$, the pair $(x, p(x))$ belongs to $R_{s_F}$:
\begin{equation}
\label{eqn:fcodegen}
    \forall x\in \mathrm{dom}(R_{s_F}), (x, p(x))\in R_{s_F}.
\end{equation}
Unlike test-based evaluation, this guarantee holds for every valid input rather than only a finite test set.

\textbf{Specification autoformalization.}
The underlying challenge is that the formal specification $s_F$ must be \textit{faithful} to the informal specification $s_I$.
That is, correctness with respect to the formal specification (Equation~\ref{eqn:fcodegen}) should be equivalent to correctness with respect to the informal specification (Equation~\ref{eqn:codegen}), which amounts to ensuring that $R_{s_F}$ matches $R_{s_I}$.
We refer to the problem of generating a faithful formal specification from an informal specification as \textit{specification autoformalization}.
Specifically, given an informal specification $s_I$, the goal is to generate a formal specification $s_F$ such that $R_{s_F} = R_{s_I}$.

In particular, the formal specification should be \textit{sound} and \textit{complete} with respect to the informal specification.
Soundness means that any input-output pair accepted by the formal specification is also accepted by the informal specification, i.e., $R_{s_F} \subseteq R_{s_I}$.
Completeness means that any input-output pair accepted by the informal specification is also accepted by the formal specification, i.e., $R_{s_I} \subseteq R_{s_F}$.
If soundness is violated, a program that verifies against the formal specification may not be correct with respect to the informal specification.
If completeness is violated, a program that is correct according to the informal specification may fail to verify against the formal specification.

\textbf{Pre- and post-specifications.}
A program specification decomposes into a precondition, which defines which inputs are valid, and a postcondition, which defines which outputs are correct for each valid input.
We refer to these as the \textit{pre-specification} and \textit{post-specification}.
In our benchmark skeletons, these are written as the Verus specification functions \texttt{pre\_spec} and \texttt{post\_spec}, respectively; henceforth, we use pre-specification and \texttt{pre\_spec}, and post-specification and \texttt{post\_spec}, interchangeably.

The pre-specification defines the valid inputs for the program, i.e., the domain of $R_{s_F}$.
A pre-specification is sound if it does not accept any invalid inputs, i.e., $\mathrm{dom}(R_{s_F}) \subseteq \mathrm{dom}(R_{s_I})$.
In practice, we can check soundness by taking an input $x$ that is invalid according to the informal specification and checking that \texttt{pre\_spec} does not accept $x$.
A pre-specification is complete if it accepts all valid inputs, i.e., $\mathrm{dom}(R_{s_I}) \subseteq \mathrm{dom}(R_{s_F})$.
If both hold, then $\mathrm{dom}(R_{s_F}) = \mathrm{dom}(R_{s_I})$, meaning the pre-specification accepts exactly the valid inputs.
To check completeness, we take an input $x$ that is valid according to the informal specification and check that \texttt{pre\_spec} accepts $x$.

The post-specification defines, for each valid input, which outputs are acceptable.
Intuitively, it encodes the desired functionality of the program on valid inputs.
Post-specification soundness means that for any $x\in \mathrm{dom}(R_{s_I})$, if $(x, y)\in R_{s_F}$, then $(x, y)\in R_{s_I}$.
In practice, we can check soundness by taking an output $y$ that is invalid for $x$ according to the informal specification and checking that \texttt{post\_spec} does not accept $(x, y)$.
Post-specification completeness means that for any $x\in \mathrm{dom}(R_{s_I})$, if $(x, y)\in R_{s_I}$, then $(x, y)\in R_{s_F}$.
To check completeness, we take an output $y$ that is valid for $x$ according to the informal specification and check that \texttt{post\_spec} accepts $(x, y)$.

Based on the above, evaluating a candidate specification reduces to checking it against four buckets of testcases, each designed to probe a different aspect of faithfulness:
\begin{itemize}
\item $\tau_{\mathrm{pre\text{-}comp}}$: valid inputs $x \in \mathrm{dom}(R_{s_I})$; the pre-spec should accept.
\item $\tau_{\mathrm{pre\text{-}sound}}$: invalid inputs $x \notin \mathrm{dom}(R_{s_I})$; the pre-spec should reject.
\item $\tau_{\mathrm{post\text{-}comp}}$: correct pairs $(x, y) \in R_{s_I}$; the post-spec should accept.
\item $\tau_{\mathrm{post\text{-}sound}}$: pairs $(x, y)$ with $x \in \mathrm{dom}(R_{s_I})$ but $(x, y) \notin R_{s_I}$; the post-spec should reject.
\end{itemize}
A faithful specification accepts every test in the two completeness buckets and rejects every test in the two soundness buckets.
We describe how we populate each bucket from Codeforces tests and hacks later in Section~\ref{ssec:data}.

\textbf{Example.}
Figure~\ref{fig:binary-search-bucket-example} shows one testcase per bucket for a binary-search task. Testcase~1 belongs to $\tau_{\mathrm{pre\text{-}comp}}$ and is a valid sorted input. Testcase~2 belongs to $\tau_{\mathrm{pre\text{-}sound}}$ because it violates the input contract that the array is sorted. Testcase~3 belongs to $\tau_{\mathrm{post\text{-}comp}}$: it is a correct output for the given input. Testcase~4 belongs to $\tau_{\mathrm{post\text{-}sound}}$: the output is invalid for a valid input, because the informal description asks for the position of the \emph{first} occurrence of the element.
The figure also shows four candidate specifications, each failing on one testcase. Spec~1's \texttt{pre\_spec} is incomplete: it requires a strictly increasing array, so it rejects Testcase~1's valid input with duplicates. Spec~2's \texttt{pre\_spec} is unsound: it only checks the array length, so it accepts Testcase~2's unsorted input. Spec~3's \texttt{post\_spec} is incomplete: it does not allow $\texttt{pos} = -1$, so it rejects Testcase~3's correct ``not found'' output. Spec~4's \texttt{post\_spec} is unsound: it does not require $\texttt{pos}$ to be the \emph{first} occurrence, so it accepts Testcase~4's non-leftmost match.

\subsection{Evaluation: Executable Specifications}
\label{ssec:evaluation}

The core challenge in evaluating specification autoformalization is testing whether $R_{s_F}$ matches $R_{s_I}$.
In general, we do not have access to the full relation $R_{s_I}$, so we cannot directly check equality between the two relations.
We instead assume that each informal specification is paired with test cases from the four buckets above.
This makes faithfulness evaluation an approximation: if the tests cover the relevant input and output boundary cases, then passing them provides evidence that the specification is faithful.

This evaluation has two requirements. First, given a candidate specification, the evaluator must be able to decide whether it accepts or rejects each test case across all four buckets ($\tau_{\mathrm{pre\text{-}comp}}$, $\tau_{\mathrm{pre\text{-}sound}}$, $\tau_{\mathrm{post\text{-}comp}}$, $\tau_{\mathrm{post\text{-}sound}}$). Second, each bucket must contain enough test cases to expose missing or extra constraints in the specification. This subsection addresses the first requirement; we describe how we populate each bucket later in Section~\ref{ssec:data}.

For each test case, the evaluator checks whether the generated specification returns the expected Boolean value.
The specification passes evaluation only if all test cases pass.
Verus specifications are logical predicates used by the verifier, not executable Rust functions, so they cannot always be run directly on concrete test inputs.
We therefore use a symbolic check followed by a runtime check.

In the symbolic check, the evaluator inserts each test case as a Verus assertion alongside the generated specification and runs the verifier on the resulting program.
For completeness tests, the assertion states that the specification accepts the test (e.g., \texttt{assert(pre\_spec(x))} or \texttt{assert(post\_spec(x, y))}).
For soundness tests, the assertion states that the specification rejects the test (e.g., \texttt{assert(!pre\_spec(x))} or \texttt{assert(!post\_spec(x, y))}).
If verification succeeds within the timeout, the test passes without runtime execution.
If verification fails or times out, the evaluator uses the runtime check: it compiles the specification into an executable Rust function $f_s$, runs $f_s$ on the test input, and compares the Boolean output with the expected value.
For example, in Figure~\ref{fig:binary-search-bucket-example} the runtime check is what fires the \texttt{assert\_eq!} calls inside each \texttt{main()} and reports the four specification failures.
\todoanmol{PDF note p.~5: Ask Natalie whether she has already proof-read this part.}

\textbf{Extending \texttt{exec\_spec} for Verus specifications.}
The runtime check uses Verus's \texttt{exec\_spec} mechanism, which compiles a subset of specifications into Rust code.
The original mechanism supports primitive Rust types (signed and unsigned integers, strings, booleans, characters), indexing and length operations on sequences, and quantified expressions (\texttt{forall}, \texttt{exists}) over a single variable restricted to a concrete range.
This is insufficient for many Codeforces specifications, which use richer Verus specification types and nested or multi-variable quantified expressions.

We extend \texttt{exec\_spec} to cover the core Verus specification types from the standard library, including sequences, sets, multisets, and maps.
We also support a core subset of methods on these types, such as \texttt{Seq::subrange} and \texttt{Set::contains}, and bounded quantified expressions over multiple variables.
These extensions make a larger class of generated specifications executable by the runtime check.

The original \texttt{exec\_spec} mechanism produces verified Rust code together with proof obligations showing that the executable code corresponds to the original specification.
This is useful when the executable code will be used inside a verified project, but it creates an unnecessary failure mode for our benchmark: the correspondence proof may fail even when the executable code is sufficient for testing a concrete input.
We therefore introduce \texttt{exec\_spec\_unverified}, which produces executable Rust code without the correspondence proof.
This is sufficient for our setting because the benchmark uses the executable code only for testing, not as part of a verified Rust program.
Appendix~\ref{app:exec_spec} gives the full feature list, the design of \texttt{exec\_spec\_unverified}, and examples of the symbolic and runtime checks.

%% file: paper_sections/figures_plots/big_sample.tex
\begin{figure}[tbp]
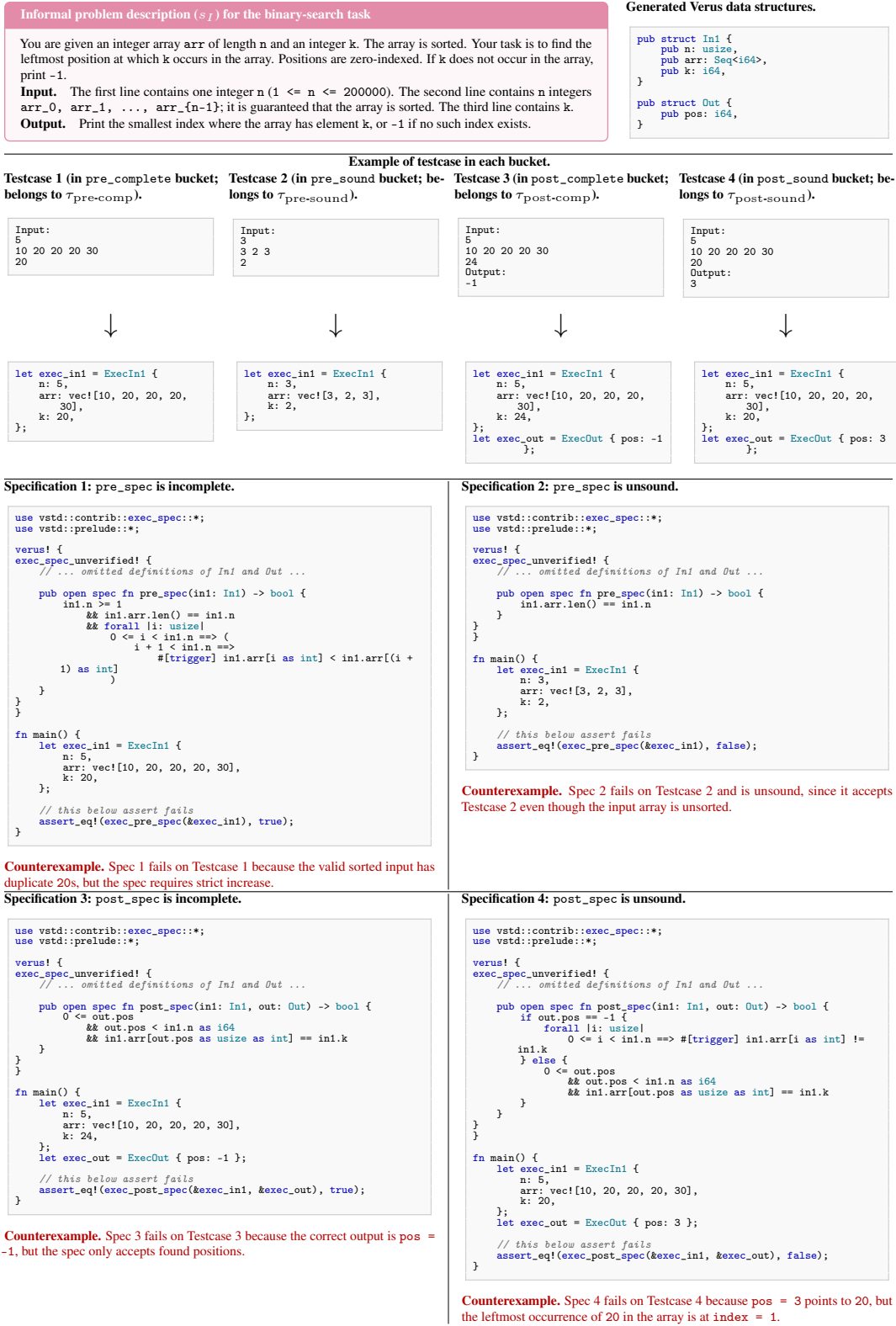

\centering
\begingroup
\tiny
\newcommand{\vgCodeStyle}{\ttfamily\fontsize{4.55}{4.75}\selectfont}
\setlength{\parskip}{0pt}

\noindent\begin{minipage}[t]{0.68\linewidth}
\vspace{0pt}
\begin{tcolorbox}[
  colback=purple!3,
  colframe=purple!45,
  title={\textbf{Informal problem description ($s_I$) for the binary-search task}},
  fonttitle=\bfseries,
  arc=2pt,
  boxrule=0.5pt,
  left=4pt,
  right=4pt,
  top=1.5pt,
  bottom=1.5pt
]
You are given an integer array \texttt{arr} of length \texttt{n} and an integer
\texttt{k}. The array is sorted. Your task is to find the leftmost position at
which \texttt{k} occurs in the array. Positions are zero-indexed. If \texttt{k}
does not occur in the array, print \texttt{-1}.

\vspace{0.15mm}
\textbf{Input.}\quad
The first line contains one integer \texttt{n} (\texttt{1 <= n <= 200000}).
The second line contains \texttt{n} integers
\texttt{arr\_0, arr\_1, ..., arr\_\{n-1\}}; it is guaranteed that the array is
sorted. The third line contains \texttt{k}.

\vspace{0.15mm}
\textbf{Output.}\quad
Print the smallest index where the array has element \texttt{k}, or \texttt{-1}
if no such index exists.
\end{tcolorbox}
\end{minipage}\hfill
\begin{minipage}[t]{0.30\linewidth}
\vspace{0pt}
\textbf{Generated Verus data structures.}
\begin{lstlisting}[language=Verus,basicstyle=\vgCodeStyle,breaklines=true,frame=single]
pub struct In1 {
    pub n: usize,
    pub arr: Seq<i64>,
    pub k: i64,
}

pub struct Out {
    pub pos: i64,
}
\end{lstlisting}
\end{minipage}

\vspace{0.25mm}
\hrule
\vspace{0.25mm}

\textbf{Example of testcase in each bucket.}

\vspace{0.3mm}
\noindent\begin{minipage}[t]{0.24\linewidth}
\textbf{Testcase 1 (in \precomplete{} bucket; belongs to
$\tau_{\mathrm{pre\text{-}comp}}$).}
\begin{lstlisting}[basicstyle=\vgCodeStyle,breaklines=true,frame=single]
Input:
5
10 20 20 20 30
20
\end{lstlisting}
\end{minipage}\hfill
\begin{minipage}[t]{0.24\linewidth}
\textbf{Testcase 2 (in \presound{} bucket; belongs to
$\tau_{\mathrm{pre\text{-}sound}}$).}
\begin{lstlisting}[basicstyle=\vgCodeStyle,breaklines=true,frame=single]
Input:
3
3 2 3
2
\end{lstlisting}
\end{minipage}\hfill
\begin{minipage}[t]{0.24\linewidth}
\textbf{Testcase 3 (in \postcomplete{} bucket; belongs to
$\tau_{\mathrm{post\text{-}comp}}$).}
\begin{lstlisting}[basicstyle=\vgCodeStyle,breaklines=true,frame=single]
Input:
5
10 20 20 20 30
24
Output:
-1
\end{lstlisting}
\end{minipage}\hfill
\begin{minipage}[t]{0.24\linewidth}
\textbf{Testcase 4 (in \postsound{} bucket; belongs to
$\tau_{\mathrm{post\text{-}sound}}$).}
\begin{lstlisting}[basicstyle=\vgCodeStyle,breaklines=true,frame=single]
Input:
5
10 20 20 20 30
20
Output:
3
\end{lstlisting}
\end{minipage}

\vspace{0.1mm}
\noindent\begin{minipage}[t]{0.24\linewidth}
\centering\large$\downarrow$
\end{minipage}\hfill
\begin{minipage}[t]{0.24\linewidth}
\centering\large$\downarrow$
\end{minipage}\hfill
\begin{minipage}[t]{0.24\linewidth}
\centering\large$\downarrow$
\end{minipage}\hfill
\begin{minipage}[t]{0.24\linewidth}
\centering\large$\downarrow$
\end{minipage}

\vspace{0.1mm}
\noindent\begin{minipage}[t]{0.24\linewidth}
\begin{lstlisting}[language=Verus,basicstyle=\vgCodeStyle,breaklines=true,frame=single]
let exec_in1 = ExecIn1 {
    n: 5,
    arr: vec![10, 20, 20, 20, 30],
    k: 20,
};
\end{lstlisting}
\end{minipage}\hfill
\begin{minipage}[t]{0.24\linewidth}
\begin{lstlisting}[language=Verus,basicstyle=\vgCodeStyle,breaklines=true,frame=single]
let exec_in1 = ExecIn1 {
    n: 3,
    arr: vec![3, 2, 3],
    k: 2,
};
\end{lstlisting}
\end{minipage}\hfill
\begin{minipage}[t]{0.24\linewidth}
\begin{lstlisting}[language=Verus,basicstyle=\vgCodeStyle,breaklines=true,frame=single]
let exec_in1 = ExecIn1 {
    n: 5,
    arr: vec![10, 20, 20, 20, 30],
    k: 24,
};
let exec_out = ExecOut { pos: -1 };
\end{lstlisting}
\end{minipage}\hfill
\begin{minipage}[t]{0.24\linewidth}
\begin{lstlisting}[language=Verus,basicstyle=\vgCodeStyle,breaklines=true,frame=single]
let exec_in1 = ExecIn1 {
    n: 5,
    arr: vec![10, 20, 20, 20, 30],
    k: 20,
};
let exec_out = ExecOut { pos: 3 };
\end{lstlisting}
\end{minipage}

\vspace{0.25mm}
\hrule
\vspace{0.25mm}

\noindent\begin{minipage}[t]{0.485\linewidth}
\textbf{Specification 1: \prespec{} is incomplete.}
\begin{lstlisting}[language=Verus,basicstyle=\vgCodeStyle,breaklines=true,frame=single]
use vstd::contrib::exec_spec::*;
use vstd::prelude::*;

verus! {
exec_spec_unverified! {
    // ... omitted definitions of In1 and Out ...

    pub open spec fn pre_spec(in1: In1) -> bool {
        in1.n >= 1
            && in1.arr.len() == in1.n
            && forall |i: usize|
                0 <= i < in1.n ==> (
                    i + 1 < in1.n ==>
                        #[trigger] in1.arr[i as int] < in1.arr[(i + 1) as int]
                )
    }
}
}

fn main() {
    let exec_in1 = ExecIn1 {
        n: 5,
        arr: vec![10, 20, 20, 20, 30],
        k: 20,
    };

    // this below assert fails
    assert_eq!(exec_pre_spec(&exec_in1), true);
}
\end{lstlisting}
\textcolor{red!70!black}{\textbf{Counterexample.} Spec 1 fails on Testcase 1
because the valid sorted input has duplicate \texttt{20}s, but the spec requires
strict increase.}
\end{minipage}\hfill\vrule width 0.35pt\hfill
\begin{minipage}[t]{0.485\linewidth}
\textbf{Specification 2: \prespec{} is unsound.}
\begin{lstlisting}[language=Verus,basicstyle=\vgCodeStyle,breaklines=true,frame=single]
use vstd::contrib::exec_spec::*;
use vstd::prelude::*;

verus! {
exec_spec_unverified! {
    // ... omitted definitions of In1 and Out ...

    pub open spec fn pre_spec(in1: In1) -> bool {
        in1.arr.len() == in1.n
    }
}
}

fn main() {
    let exec_in1 = ExecIn1 {
        n: 3,
        arr: vec![3, 2, 3],
        k: 2,
    };

    // this below assert fails
    assert_eq!(exec_pre_spec(&exec_in1), false);
}
\end{lstlisting}
\textcolor{red!70!black}{\textbf{Counterexample.} Spec 2 fails on Testcase 2
and is unsound, since it accepts Testcase 2 even though the input array is
unsorted.}
\end{minipage}

\vspace{0.25mm}
\hrule
\vspace{0.25mm}
\noindent\begin{minipage}[t]{0.485\linewidth}
\textbf{Specification 3: \postspec{} is incomplete.}
\begin{lstlisting}[language=Verus,basicstyle=\vgCodeStyle,breaklines=true,frame=single]
use vstd::contrib::exec_spec::*;
use vstd::prelude::*;

verus! {
exec_spec_unverified! {
    // ... omitted definitions of In1 and Out ...

    pub open spec fn post_spec(in1: In1, out: Out) -> bool {
        0 <= out.pos
            && out.pos < in1.n as i64
            && in1.arr[out.pos as usize as int] == in1.k
    }
}
}

fn main() {
    let exec_in1 = ExecIn1 {
        n: 5,
        arr: vec![10, 20, 20, 20, 30],
        k: 24,
    };
    let exec_out = ExecOut { pos: -1 };

    // this below assert fails
    assert_eq!(exec_post_spec(&exec_in1, &exec_out), true);
}
\end{lstlisting}
\textcolor{red!70!black}{\textbf{Counterexample.} Spec 3 fails on Testcase 3
because the correct output is \texttt{pos = -1}, but the spec only accepts found
positions.}
\end{minipage}\hfill\vrule width 0.35pt\hfill
\begin{minipage}[t]{0.485\linewidth}
\textbf{Specification 4: \postspec{} is unsound.}
\begin{lstlisting}[language=Verus,basicstyle=\vgCodeStyle,breaklines=true,frame=single]
use vstd::contrib::exec_spec::*;
use vstd::prelude::*;

verus! {
exec_spec_unverified! {
    // ... omitted definitions of In1 and Out ...

    pub open spec fn post_spec(in1: In1, out: Out) -> bool {
        if out.pos == -1 {
            forall |i: usize|
                0 <= i < in1.n ==> #[trigger] in1.arr[i as int] != in1.k
        } else {
            0 <= out.pos
                && out.pos < in1.n as i64
                && in1.arr[out.pos as usize as int] == in1.k
        }
    }
}
}

fn main() {
    let exec_in1 = ExecIn1 {
        n: 5,
        arr: vec![10, 20, 20, 20, 30],
        k: 20,
    };
    let exec_out = ExecOut { pos: 3 };

    // this below assert fails
    assert_eq!(exec_post_spec(&exec_in1, &exec_out), false);
}
\end{lstlisting}
\textcolor{red!70!black}{\textbf{Counterexample.} Spec 4 fails on Testcase 4
because \texttt{pos = 3} points to \texttt{20}, but the leftmost occurrence of
\texttt{20} in the array is at \texttt{index = 1}.}
\end{minipage}

\caption{\textbf{Binary search example.}
The top-left shows the informal problem description $s_I$ for a binary-search task. The top-right shows the data structures our pipeline produces to represent its inputs and outputs.
The middle row shows example testcases available from Codeforces, one per bucket, alongside their converted counterparts as typed Verus values.
The bottom row shows four candidate specifications $s_F$ that do not faithfully represent $s_I$, each for a different reason; each failure is caught by a testcase from a different bucket.
\textbf{Note:} \emph{this example is for illustration only; our benchmark problems are significantly more complex than this simple binary-search task.}}
\label{fig:binary-search-bucket-example}
\label{fig:binary-search-prespec-examples}
\label{fig:binary-search-postspec-examples}
\endgroup
\end{figure}

%% file: paper_sections/main_paper/01_z_data.tex
\section{Data and Agent Environment}

\subsection{From Codeforces Problems to Benchmark Tasks}
\label{ssec:data}


This section describes how we construct \vsb{} from Codeforces problems and how
agents interact with the resulting tasks.

\textbf{Data sources.}
For each source problem, we parse the Codeforces page to collect the informal
problem statement $s_I$, the official tests $\tau$, and the user-submitted hacks
$H$.
The full pipeline for constructing \vsb{} has five stages: sourcing, filtering,
hack collection, test-case conversion, and final selection.
Starting from $\tau$ and $H$, the pipeline removes problematic cases, assigns
the retained cases to the four testcase buckets, and converts each retained
testcase from raw text into typed Verus/Rust constants.
The pipeline also applies several quality filters: we remove problems with
unsupported features such as floating-point I/O, discard duplicate or truncated
tests, filter syntactically invalid hacks that would fail before a specification
is called, and require at least five testcases in every bucket.
Full pipeline details are in App.~\ref{app:additional-data-creation}.

Constructing the evaluation set from $\tau$ and $H$ requires two things:
\begin{itemize}
    \item[(A)] \textbf{Testcase conversion}: raw Codeforces text files must be
    converted into typed Verus/Rust values that can be passed to \prespec{} and
    \postspec{}.
    \item[(B)] \textbf{Bucket population}: each of the four buckets
    ($\tau_{\mathrm{pre\text{-}comp}}$,
    $\tau_{\mathrm{pre\text{-}sound}}$,
    $\tau_{\mathrm{post\text{-}comp}}$,
    $\tau_{\mathrm{post\text{-}sound}}$) must contain enough testcases to catch
    both overly strong and overly weak specifications.
\end{itemize}
We describe the conversion pipeline first, and then explain how we use official
tests and hacks to populate the four buckets.

\textbf{(A) Converting Codeforces Tests to Rust Code.}
Codeforces tests are raw text files; the evaluator needs typed Verus/Rust values.
To check whether a \prespec{} accepts an input, the evaluator calls it on a
typed value of the problem's input type.
To check a \postspec{}, it calls the specification on both an input value and an
output value.
For each problem, we therefore need a parser $R$ that maps each raw testcase
$t$ to typed executable values $E = R(t)$.
Table~\ref{tab:data-1027c-test-case-format} shows a Codeforces 1027C testcase in
the original text format, alongside the Verus types selected by the construction
agent and the executable Rust values produced by the parser.

However, we also need to ensure this conversion is lossless.
If the parser drops part of the input, reorders fields, or maps two different
raw tests to the same typed value, then the evaluator may test the wrong
concrete testcase.
To ensure the conversion is lossless, we ask a benchmark-construction agent
(\gptfivethreecodex{} inside \sweagent) to write both the parser $R$ and a
printer $P$ that maps typed values back to raw text.
The parser is accepted only if every retained testcase round-trips:
\[
    T_{\mathrm{reproduced}} = P(R(t)) \mathrel{==} t.
\]
For postcondition buckets, we also run the same check on the corresponding
output.
If any round-trip check fails, we feed the failure back to the agent and ask it
to revise $R$ and $P$.
This loop continues until all retained testcases round-trip successfully, or the
attempt is discarded.
Figure~\ref{fig:lossless-conversion-1027c} shows this process in detail for
Codeforces 1027C.
The left column shows raw text testcases from Codeforces.
The parser $R$ maps each raw testcase $t$ to typed values $E=R(t)$, shown in the
middle column.
The printer $P$ then maps those typed values back to text, producing
$T_{\mathrm{reproduced}}=P(E)=P(R(t))$.
The conversion is accepted only if $T_{\mathrm{reproduced}}$ matches the
original testcase $t$.
The bottom row shows the parser and printer code that finally passed this
round-trip check for Codeforces 1027C.
Figure~\ref{fig:binary-search-bucket-example} gives a simpler binary-search
example.

The data types chosen by the construction agent are fixed for the benchmark task.
They must be compatible with Verus's \texttt{exec\_spec} feature so the
evaluator can produce both specification-level values and executable Rust values.
Fixing the data types ahead of time lets the benchmark agent focus on writing
faithful \prespec{} and \postspec{} bodies.
Full details of the conversion pipeline are available in App.~\ref{app:additional-data-creation}.

\input{paper_sections/figures_plots/1027c_lossless_conversion}

\textbf{(B) Collecting Evaluation Test Cases.}
Our testcases come from two Codeforces sources.
First, each problem has \emph{official tests}: valid inputs paired with accepted
outputs that Codeforces uses to judge submissions.
These naturally contribute to $\tau_{\mathrm{pre\text{-}comp}}$ and
$\tau_{\mathrm{post\text{-}comp}}$: the input should be accepted by \prespec{},
and the input-output pair should be accepted by \postspec{}.


Second, we use Codeforces's hack system.
In this system, after a participant's solution $p$ passes all official tests,
another participant may propose an adversarial input designed to break
$p$.\footnote{\url{https://codeforces.com/blog/entry/107753}}
These hacks become valuable additions to the different buckets because they
often target edge cases and implicit constraints missing from ordinary official
tests.

Figure~\ref{fig:hack-journey} summarizes how a hack is routed into buckets.
If Codeforces rejects the hack input as invalid, we add it to
$\tau_{\mathrm{pre\text{-}sound}}$.
If Codeforces accepts the input as valid, we add it to
$\tau_{\mathrm{pre\text{-}comp}}$.
For valid inputs, we also observe the output produced by the program being
hacked when run on that input.
If the Codeforces checker accepts this output, we add the input-output pair to
$\tau_{\mathrm{post\text{-}comp}}$; otherwise, we add the pair to
$\tau_{\mathrm{post\text{-}sound}}$.
In this way, hacks become a useful source of realistic testcases for all four
buckets.
This is especially important because LLMs struggle to create counterexamples
for almost-correct but subtly wrong submissions to programming
problems~\citep{sinha2025can}, whereas Codeforces hacks are human-generated
adversarial inputs written against real submitted solutions.

\input{paper_sections/figures_plots/tikz_images/02_hack_journey}

\begin{figure}[!tbp]
    \centering
    \includegraphics[width=\linewidth]{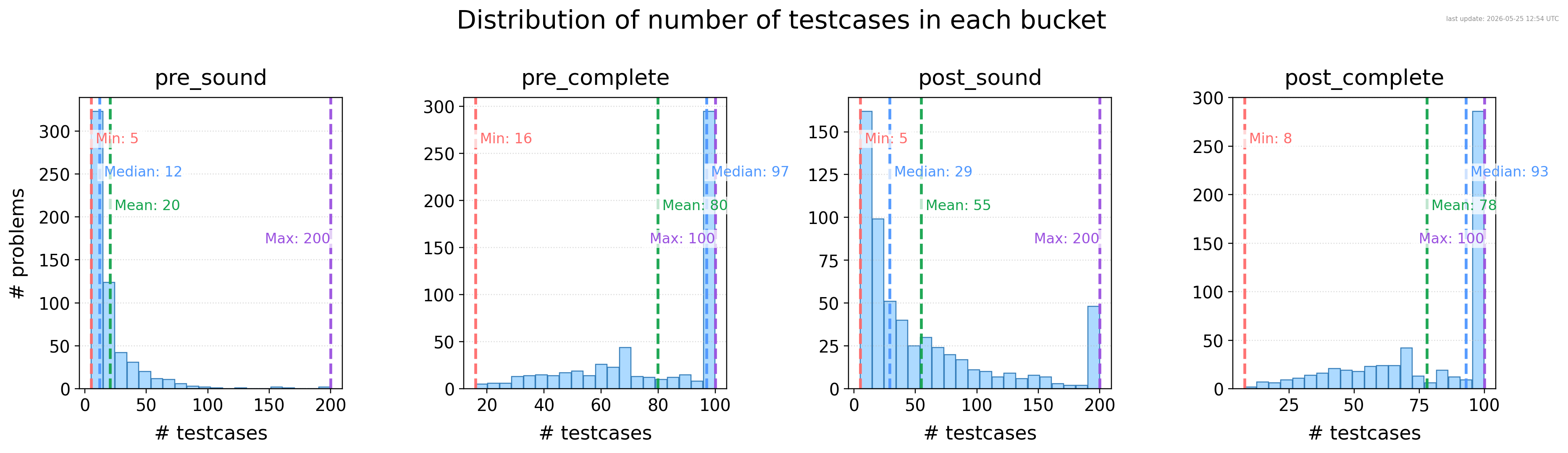}
    \caption{Distribution of the number of test cases per problem in each evaluation bucket. Median counts are 12 (pre-sound), 97 (pre-complete), 29 (post-sound), and 93 (post-complete). The completeness buckets are larger because they include all official Codeforces tests, while the soundness buckets are sourced primarily from hacks.}
    \label{fig:dataset-testcase-count-distribution}
\end{figure}

\textbf{Dataset statistics.}
\vsb{} contains \numproblems{} problems spanning a wide range of topics and
difficulty ratings.
Figure~\ref{fig:dataset-testcase-count-distribution} shows the distribution of
test cases per problem in each bucket.
On average, problems have 21 pre-sound, 80 pre-complete, 55 post-sound, and 78
post-complete test cases.
All four buckets have at least 5 test cases per problem.
App.~\ref{app:additional-dataset-statistics} reports additional statistics,
including the rating distribution (Figure~\ref{fig:dataset-rating-distribution})
and tag distribution (Figure~\ref{fig:dataset-tag-distribution}).

\subsection{Agent Environment}
\label{ssec:data-format}
Each benchmark task is a problem directory exposed to an AI agent.
The agent reads and edits files, fills the specification holes in the provided
skeleton, and submits its final specification for evaluation.
We summarize the task files, tools, and expected output below.


\textbf{Pre- and post-specifications and proofs.}
Each problem contains a skeleton \texttt{solve.rs} file that the agent must fill
in.
The file contains standard Verus library imports, along with fixed structs that
define the problem's input and output types.
These types are produced by the data-conversion pipeline described above.
The main placeholders in the skeleton code are for the pre-specification and
post-specification:
\begin{lstlisting}[language=Verus]
  pub open spec fn pre_spec(in1: In1) -> bool { }
  pub open spec fn post_spec(in1: In1, out: Out) -> bool { }
\end{lstlisting}
The agent must fill in the bodies of these specification functions.
It is free to write helper functions and use them in the specifications.
Listing~\ref{lst:case-study-1027c-template} in App. \ref{app:case-study-1027c} shows the complete skeleton for Codeforces 1027C.
%

The file also contains additional code that is necessary for evaluation, such
as placeholders where testcases will be injected, and annotations for
\texttt{exec\_spec}.
We show the skeleton for an example problem in App.~\ref{app:skeleton}, and walk
through the symbolic proving and \texttt{exec\_spec} testing flow on a concrete
example in App.~\ref{app:example-symbolic-eval} and
App.~\ref{app:example-exec-spec-eval}.
At a higher level, Figure~\ref{fig:testcase-journey} summarizes the process used
to decide whether a candidate specification $s_F$ produced by the agent accepts
or rejects a testcase.

\textbf{Provided and hidden testcases.}
Test cases are organized into the four categories described in
\S\ref{ssec:evaluation} and \S\ref{ssec:data}.
For each problem, we provide the agent with three sample testcases drawn from
the completeness buckets, and no sample testcases from the soundness buckets.
When the agent runs the local evaluation command, the specification is tested
against these provided sample testcases and the agent receives feedback,
including Verus error messages for failing cases.
A separate, larger set of testcases is hidden from the agent and used only
during final evaluation.
A specification is considered correct only if it passes all testcases across all
four categories, both provided and hidden.
App.~\ref{app:skeleton} walks through the full evaluation mechanism on a
concrete example, showing how testcases are injected into the skeleton and how
the symbolic and \texttt{exec\_spec} checks are applied.

\textbf{Agent tools.}
The agent framework provides standard file-system tools
(reading, writing, editing files) and shell access.
The agent can evaluate its current specification at any time by running
a special shell command (\texttt{verus\_gym\_specgen\_check}), which tests the
specification against the provided sample testcases and writes detailed
feedback, including Verus error messages, to a log directory.
The agent iterates on the specification based on this feedback.
The agent also has access to a submit tool.
A rollout ends when the agent submits a final specification, exhausts its
budget, or exceeds the 75-minute timeout.

\textbf{Additional materials.}
In addition to the skeleton code and provided sample testcases, the agent's
environment contains:
(1)~the Codeforces problem statement,
(2)~a worked example of a completed specification for a different problem,
(3)~Verus documentation, including the \texttt{exec\_spec} guide;
(4)~the evaluator source code, which the agent may inspect to understand the
evaluation flow;
(5)~the Verus source code; and
(6)~a detailed prompt written by Verus experts that describes the task format,
the semantics of soundness and completeness, how test-case snippets are used,
and a recommended solving strategy.
The full prompt is shown in App.~\ref{app:full-prompt}.
The materials and prompt are intended to increase the probability that
benchmark failures reflect the difficulty of specification autoformalization
rather than unfamiliarity with Verus syntax.

%% file: paper_sections/figures_plots/1027c_lossless_conversion.tex
\begin{figure}[p]
\centering
\begingroup
\scriptsize
\setlength{\tabcolsep}{2.5pt}
\renewcommand{\arraystretch}{1.05}

\newcommand{\roundtriparrow}{\(\xrightarrow{\;\;\text{Printer }P\;\;}\)}
\newcommand{\parsearrow}{\(\xrightarrow{\;\;\text{Parser }R\;\;}\)}
\newcommand{\roundtripcheck}{%
{\Large\textcolor{red}{?}}\\[-1mm]
\resizebox{0.88\linewidth}{!}{\(T_{\mathrm{reproduced}}\)}\\[-0.25mm]
\resizebox{0.88\linewidth}{!}{\(= P(E) = P(R(t))\)}\\[0.5mm]
{\color{red}\bfseries Is}\\[-0.4mm]
{\color{red}\bfseries\resizebox{0.88\linewidth}{!}{\(T_{\mathrm{reproduced}} \mathrel{==} t\)?}}%
}
\lstdefinestyle{losslessConversionCode}{
  language=Verus,
  basicstyle=\ttfamily\fontsize{5.1}{5.45}\selectfont,
  numbers=left,
  numberstyle=\tiny\color{gray},
  numbersep=3pt,
  breaklines=true,
  breakatwhitespace=false,
  columns=flexible,
  keepspaces=true,
  frame=single,
  framesep=2pt,
  xleftmargin=1.2em,
  xrightmargin=0pt,
  backgroundcolor=\color{gray!4},
  rulecolor=\color{gray!40},
  showstringspaces=false,
  tabsize=2,
  aboveskip=0pt,
  belowskip=0pt,
}
\lstdefinestyle{losslessTestcaseText}{
  basicstyle=\ttfamily\fontsize{5.0}{5.25}\selectfont,
  breaklines=true,
  breakatwhitespace=true,
  columns=flexible,
  keepspaces=true,
  showstringspaces=false,
  frame=none,
  aboveskip=1pt,
  belowskip=0pt,
}

\begin{tabular}{p{0.305\textwidth}c p{0.315\textwidth}c p{0.125\textwidth}}
\toprule
\textbf{Original Codeforces test \(t\)}
& & \textbf{Generated executable values \(E=R(t)\)}
& & \textbf{Reproduced testcase} \\
\midrule
\begin{minipage}[t]{\linewidth}
\begin{tcolorbox}[colback=red!3,colframe=red!35,boxrule=0.45pt,arc=1.5mm,left=2pt,right=2pt,top=2pt,bottom=2pt]
\textbf{A testcase in \(\tau_{\mathrm{pre\text{-}sound}}\)}
\begin{lstlisting}[style=losslessTestcaseText]
1
6
67114656 67114656 67114657 67114657 67114658 67114658
\end{lstlisting}
\end{tcolorbox}
\end{minipage}
&
\parsearrow
&
\begin{minipage}[t]{\linewidth}
\begin{tcolorbox}[colback=blue!3,colframe=blue!35,boxrule=0.45pt,arc=1.5mm,left=2pt,right=2pt,top=2pt,bottom=2pt]
\begin{Verbatim}[fontsize=\tiny]
let exec_in1 = ExecIn1 {
    ns: vec![6],
    sticks: vec![
        vec![67114656, 67114656,
             67114657, 67114657,
             67114658, 67114658],
    ],
};
\end{Verbatim}
\end{tcolorbox}
\end{minipage}
&
\roundtriparrow
&
\begin{minipage}[t]{\linewidth}
\begin{tcolorbox}[colback=orange!4,colframe=orange!50!black,boxrule=0.45pt,arc=1.5mm,left=2pt,right=2pt,top=2pt,bottom=2pt]
\centering
\roundtripcheck
\end{tcolorbox}
\end{minipage}
\\[1mm]
\begin{minipage}[t]{\linewidth}
\begin{tcolorbox}[colback=red!3,colframe=red!35,boxrule=0.45pt,arc=1.5mm,left=2pt,right=2pt,top=2pt,bottom=2pt]
\textbf{A testcase in \(\tau_{\mathrm{post\text{-}sound}}\)}
\begin{lstlisting}[style=losslessTestcaseText]
Input:
1
4
1 1 10000 10000
Output:
0 0 0 0
\end{lstlisting}
\end{tcolorbox}
\end{minipage}
&
\parsearrow
&
\begin{minipage}[t]{\linewidth}
\begin{tcolorbox}[colback=blue!3,colframe=blue!35,boxrule=0.45pt,arc=1.5mm,left=2pt,right=2pt,top=2pt,bottom=2pt]
\begin{Verbatim}[fontsize=\tiny]
let exec_in1 = ExecIn1 {
    ns: vec![4],
    sticks: vec![vec![1, 1, 10000, 10000]],
};
let exec_out = ExecOut {
    rectangles: vec![
        ExecRectangle { s1: 0, s2: 0,
                        s3: 0, s4: 0 },
    ],
};
\end{Verbatim}
\end{tcolorbox}
\end{minipage}
&
\roundtriparrow
&
\begin{minipage}[t]{\linewidth}
\begin{tcolorbox}[colback=orange!4,colframe=orange!50!black,boxrule=0.45pt,arc=1.5mm,left=2pt,right=2pt,top=2pt,bottom=2pt]
\centering
\roundtripcheck
\end{tcolorbox}
\end{minipage}
\\[1mm]
\begin{minipage}[t]{\linewidth}
\begin{tcolorbox}[colback=green!3,colframe=green!45!black,boxrule=0.45pt,arc=1.5mm,left=2pt,right=2pt,top=2pt,bottom=2pt]
\textbf{A testcase in \(\tau_{\mathrm{post\text{-}comp}}\)}
\begin{lstlisting}[style=losslessTestcaseText]
Input:
1
4
1 1 10000 10000
Output:
1 1 10000 10000
\end{lstlisting}
\end{tcolorbox}
\end{minipage}
&
\parsearrow
&
\begin{minipage}[t]{\linewidth}
\begin{tcolorbox}[colback=blue!3,colframe=blue!35,boxrule=0.45pt,arc=1.5mm,left=2pt,right=2pt,top=2pt,bottom=2pt]
\begin{Verbatim}[fontsize=\tiny]
let exec_in1 = ExecIn1 {
    ns: vec![4],
    sticks: vec![vec![1, 1, 10000, 10000]],
};
let exec_out = ExecOut {
    rectangles: vec![
        ExecRectangle { s1: 1, s2: 1,
                        s3: 10000, s4: 10000 },
    ],
};
\end{Verbatim}
\end{tcolorbox}
\end{minipage}
&
\roundtriparrow
&
\begin{minipage}[t]{\linewidth}
\begin{tcolorbox}[colback=orange!4,colframe=orange!50!black,boxrule=0.45pt,arc=1.5mm,left=2pt,right=2pt,top=2pt,bottom=2pt]
\centering
\roundtripcheck
\end{tcolorbox}
\end{minipage}
\\
\bottomrule
\end{tabular}

\vspace{1.5mm}
\begin{tabular}{p{0.485\textwidth}p{0.485\textwidth}}
\toprule
\textbf{Parser \(R\): parse raw text into executable values}
&
\textbf{Printer \(P\): print executable values back to text} \\
\midrule
\begin{minipage}[t]{\linewidth}
\begin{lstlisting}[style=losslessConversionCode]
pub fn read_input_from_path(path: &str) -> Inputs {
    let content = fs::read_to_string(path).expect("failed to read .in");
    let mut lines = content.lines();
    let t: usize = lines.next().expect("missing T line")
        .trim().parse().expect("bad T");
    let mut ns = Vec::with_capacity(t);
    let mut sticks = Vec::with_capacity(t);
    for _ in 0..t {
        let n: i64 = lines.next().expect("missing n line")
            .trim().parse().expect("bad n");
        let arr: Vec<i64> = lines.next().expect("missing sticks line")
            .split_whitespace()
            .map(|tok| tok.parse::<i64>().expect("bad stick length"))
            .collect();
        ns.push(n);
        sticks.push(arr);
    }
    ExecIn1 { ns, sticks }
}

pub fn read_output_from_path(path: &str) -> ExecOut {
    let content = fs::read_to_string(path).expect("failed to read .out");
    let mut iter = content.split_whitespace();
    let mut rectangles = Vec::new();
    loop {
        let s1 = match iter.next() {
            Some(tok) => tok.parse::<i64>().expect("bad output int"),
            None => break,
        };
        let s2 = iter.next().expect("missing second side").parse().unwrap();
        let s3 = iter.next().expect("missing third side").parse().unwrap();
        let s4 = iter.next().expect("missing fourth side").parse().unwrap();
        rectangles.push(ExecRectangle { s1, s2, s3, s4 });
    }
    ExecOut { rectangles }
}
\end{lstlisting}
\end{minipage}
&
\begin{minipage}[t]{\linewidth}
\begin{lstlisting}[style=losslessConversionCode]
pub fn pre_spec_print(in1: &ExecIn1) {
    let mut stdout = io::BufWriter::new(io::stdout());
    let t = in1.ns.len();
    write!(stdout, "{}", t).expect("failed to write T");
    if t > 0 { writeln!(stdout).expect("failed to write newline"); }
    for i in 0..t {
        writeln!(stdout, "{}", in1.ns[i]).expect("failed to write n");
        let arr = &in1.sticks[i];
        if !arr.is_empty() {
            write!(stdout, "{}", arr[0]).expect("failed to write first stick");
            for j in 1..arr.len() {
                write!(stdout, " {}", arr[j]).expect("failed to write stick");
            }
        }
        if i + 1 < t { writeln!(stdout).expect("failed to write newline"); }
    }
    stdout.flush().expect("failed to flush stdout");
}

pub fn post_spec_print(in1: &ExecIn1, out: &ExecOut) {
    let mut stdout = io::BufWriter::new(io::stdout());
    let m = out.rectangles.len();
    for i in 0..m {
        let r = &out.rectangles[i];
        write!(stdout, "{} {} {} {}", r.s1, r.s2, r.s3, r.s4)
            .expect("failed to write rectangle");
        if i + 1 < m { writeln!(stdout).expect("failed to write newline"); }
    }
    stdout.flush().expect("failed to flush stdout");
}
\end{lstlisting}
\end{minipage}
\\
\bottomrule
\end{tabular}
\caption{\textbf{Lossless conversion for Codeforces 1027C.}
For a raw Codeforces testcase \(t\), the parser \(R\) produces executable Rust
values \(E=R(t)\). The printer \(P\) then maps those values back to text,
yielding \(T_{\mathrm{reproduced}}=P(E)=P(R(t))\). The conversion is lossless
only if \(T_{\mathrm{reproduced}}\mathrel{==}t\) byte-for-byte; otherwise the benchmark
might evaluate the wrong concrete testcase.}
\label{fig:lossless-conversion-1027c}
\endgroup
\end{figure}

%% file: paper_sections/figures_plots/tikz_images/02_hack_journey.tex
\begin{figure}[H]
\centering
\newcommand{\hackbucket}[1]{{\sffamily\bfseries\boldmath #1}}
\resizebox{\textwidth}{!}{%
\begin{tikzpicture}[
  font=\large,
  >=Stealth,
  arr/.style={-{Latex[length=2.4mm]}, semithick, black!65},
  dashedarr/.style={-{Latex[length=2.4mm]}, semithick, dashed, black!50},
  source/.style={
    draw=black!55,
    rounded corners=4pt,
    fill=stageProb!25,
    align=center,
    text width=4.0cm,
    minimum height=1.45cm,
    inner sep=5pt
  },
  process/.style={
    draw=black!55,
    rounded corners=4pt,
    fill=stageEval!25,
    align=center,
    text width=3.8cm,
    minimum height=1.15cm,
    inner sep=5pt
  },
  decision/.style={
    draw=black!60,
    diamond,
    aspect=2.05,
    fill=stageExec!22,
    align=center,
    inner sep=1pt,
    text width=2.65cm,
    font=\normalsize
  },
  bucket/.style={
    draw=black!60,
    rounded corners=4pt,
    align=center,
    text width=3.35cm,
    minimum height=1.35cm,
    inner sep=5pt,
    font=\large\bfseries
  },
  prebucket/.style={bucket, fill=stageSpec!25},
  soundbucket/.style={bucket, fill=red!12},
  completebucket/.style={bucket, fill=green!14},
  note/.style={
    draw=black!30,
    rounded corners=3pt,
    fill=white,
    align=left,
    text width=4.05cm,
    inner sep=5pt,
    font=\normalsize
  },
  lbl/.style={
    fill=white,
    fill opacity=0.96,
    text opacity=1,
    align=center,
    inner sep=2pt,
    font=\normalsize,
    text=black!70
  }
]

\node[source] (hacker) at (0,0) {
  \includegraphics[width=0.65cm]{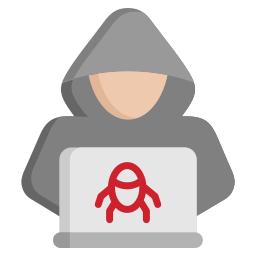}\\[-1pt]
  \textbf{Hacker proposes}\\
  input $i$ against\\
  candidate program $p$
};

\node[decision] (valid) at (5.5,0) {Does Codeforces\\accept $i$ as\\valid input?};

\node[prebucket] (precomplete) at (11.0,0) {
  \hackbucket{pre\_complete}\\[2pt]
  {\normalfont valid input $i$}\\
  {\normalfont should be accepted by \prespec{}}
};

\node[process] (run) at (16.0,0) {Run candidate\\program $p(i)$\\to produce $o_p$};

\node[decision] (checker) at (21.5,0) {Does the\\Codeforces checker\\accept $o_p$?};

\node[prebucket] (presound) at (5.5,-4.0) {
  \hackbucket{pre\_sound}\\[2pt]
  {\normalfont invalid input $i$}\\
  {\normalfont should be rejected by \prespec{}}
};

\node[completebucket] (postcomplete) at (27.0,3.0) {
  \hackbucket{post\_complete}\\[2pt]
  {\normalfont correct pair $(i,o)$}\\
  {\normalfont should be accepted by \postspec{}}
};

\node[soundbucket] (postsound) at (27.0,-3.0) {
  \hackbucket{post\_sound}\\[2pt]
  {\normalfont wrong pair $(i,o_p)$}\\
  {\normalfont should be rejected by \postspec{}}
};

\node[note] (gt) at (16.0,3.8) {
  If the hack protocol exposes a ground-truth \texttt{Answer:} $o^\star$,
  then $(i,o^\star)$ is also added to \hackbucket{post\_complete}.
};

\draw[arr] (hacker) -- (valid);

\draw[arr] (valid) -- (presound)
  node[lbl, pos=0.45, right=3pt] {invalid};

\draw[arr] (valid) -- (precomplete)
  node[lbl, pos=0.5, above=2pt] {valid};

\draw[arr] (precomplete) -- (run);
\draw[arr] (run) -- (checker);

\draw[arr] (checker) -- (postcomplete)
  node[lbl, pos=0.5, above=2pt, sloped] {accepted};

\draw[arr] (checker) -- (postsound)
  node[lbl, pos=0.5, below=2pt, sloped] {rejected};

\draw[dashedarr] (gt.east) to[out=0,in=160] (postcomplete.west);

\end{tikzpicture}%
}
\caption{\textbf{How Codeforces hacks become \vsg{} testcase buckets.}
A hack begins when a participant proposes an input $i$ against a candidate program $p$.
The diagram shows how we use artifacts produced by the hack for buckets in our test suite, depending on the result of the hack.}
\label{fig:hack-journey}
\end{figure}

%% file: paper_sections/main_paper/02_experiments.tex
\section{Experiments}
\label{sec:experiments}

\subsection{Experimental setup.}
We evaluate language-model agents using \sweagent~\citep{yang2024sweagent}.
Within \sweagent, we evaluate the six models reported in Table~\ref{tab:main}: \geminithreeonepro, \gptfivethreecodex~\citep{openai2026gpt53codex}, \opusfoursix~\citep{anthropic2025claude4}, \deepseekvfourpro~\citep{deepseekai2026deepseekv4}, \glmfiveone~\citep{glm5team2026glm5}, and \kimiktwosix~\citep{kimiteam2026kimik25}. Each agent is given a budget of \$2.5 and 75 minutes per problem. We primarily evaluate models using Pass@1, i.e., the fraction of problems for which the agent-generated specification passes all test cases across all four buckets. We additionally report the average fraction of testcases passed in each bucket. Full hyperparameters are in App.~\ref{app:eval-hyperparams}.

\subsection{Main results.}
Table~\ref{tab:main} shows the end-to-end Pass@1 performance of each agent on our benchmark.
\geminithreeonepro{} achieves the highest Pass@1 (0.778), followed by \gptfivethreecodex{} (0.578) and \opusfoursix{} (0.511).
The open-source models remain lower, with Pass@1 between 0.215 and 0.255 under the same cost cap, indicating substantial room for improvement in open-source agents for this task.
The bucket-level columns report the average fraction of testcases passed by each model in each bucket. A larger summary figure is shown in App.~\ref{app:additional-insights} (Figure~\ref{fig:score-summary}).

\input{paper_sections/tables/00_main_table}

\paragraph{Specification generation remains difficult even when code generation succeeds.}
We also compare specification generation against ordinary code generation for \gptfivethreecodex{} on problems where testcase-based code evaluation is well-defined.
Among the problems where \gptfivethreecodex{} wrote an incorrect specification and every input has a unique correct output, 187 have available code-generation runs.
The same model solves 153 of these 187 programming tasks in Python, corresponding to a code-generation success rate of 81.8\% on this subset.
Thus, many failures are not explained by an inability to solve the underlying Codeforces problem: the model can often write correct executable code while still failing to write a faithful formal specification.
Details are in App.~\ref{app:insights-codegen-aspect}.

\subsection{Analysis.}
We next analyze the main design choices in the evaluator and the main failure modes in generated specifications.
We first study whether soundness tests are necessary for measuring specification correctness, and whether adding more testcases changes the measured success rates.
We then study how often \execspec{} resolves testcases that symbolic checking leaves unknown.
Finally, we compare our \texttt{exec\_spec}-based evaluation against LLM-based evaluation of specification faithfulness, and summarize recurring failure modes from the qualitative analysis.

\paragraph{Soundness testcases are necessary for evaluation.}
Specifications are often unsound, so soundness testcases extracted from Codeforces hacks are essential for measuring specification correctness.
Table~\ref{tab:main} supports this: if we evaluate only on completeness testcases, pass@1-completeness is substantially higher, but once we add soundness testcases to the evaluation, pass@1 drops from 77\% to 58\% for \gptfivethreecodex{}, 82\% to 78\% for \geminithreeonepro{}, and 59\% to 51\% for \opusfoursix{}.
These drops show that models can write specifications that accept the provided valid examples while still accepting invalid inputs or incorrect outputs, making soundness tests crucial.

\paragraph{More testcases improve evaluation with diminishing returns.}
Given that soundness and completeness tests expose real specification failures, the next question is how much additional coverage we gain from adding more testcases.
For our benchmark, we find that adding more testcases does increase coverage, but the marginal benefit of each additional testcase decreases once the budget is large enough.
In App.~\ref{app:insights-testcase-coverage}, we estimate the expected probability of catching at least one specification failure under smaller testcase budgets and find that increasing the number of testcases has high marginal value initially, before the curves flatten as additional tests increasingly repeat already-covered failure patterns; this suggests that the current test budget captures most failures exposed by these test buckets.

\paragraph{\execspec{} resolves testcases left unknown by symbolic checking.}
The evaluator first tries to resolve the testcase symbolically; if both acceptance and rejection are inconclusive, it falls back to \execspec{}, compiles the specification into executable Rust, and runs it on the testcase to obtain a concrete accept/reject verdict.
Figure~\ref{fig:testcase-journey} shows the evaluator's decision tree for deciding whether a model-written specification accepts or rejects a concrete testcase.
App.~\ref{app:insights-failed-testcase-verdicts} shows the resulting distribution of testcase resolution outcomes across evaluated runs (Figure~\ref{fig:resolution-distribution}).
The first pattern is that high-performing models do not only produce fewer wrong verdicts; they also produce specifications that the evaluator can analyze.
\geminithreeonepro{} and \gptfivethreecodex{} have relatively small compile/syntax-error fractions across all four buckets, while \deepseekvfourpro{}, \glmfiveone{}, and \kimiktwosix{} are dominated by compile/syntax errors in several buckets.
This suggests that part of the gap between stronger and weaker models is the ability to stay within the Verus and \execspec{} fragment needed by the benchmark.

Among the analyzable testcases, \execspec{} is especially important for postcondition checking.
In the \postcomplete{} and \postsound{} buckets, \geminithreeonepro{} and \gptfivethreecodex{} resolve a large fraction of testcases via the executable fallback, whereas purely symbolic resolution accounts for a smaller share.
Without \execspec{}, these cases would have stopped at the symbolically unknown node in Figure~\ref{fig:testcase-journey}, and we would not know whether the specification accepted or rejected those concrete examples.
By contrast, the \precomplete{} bucket contains a larger share of symbolically resolved correct verdicts, indicating that many input-validity constraints can be proved symbolically.
Moreover, for at least 86\% of benchmark problems, at least one evaluated model writes an \execspec{}-compatible specification that satisfies all testcases.
This suggests that \execspec{} has enough feature coverage for models to write compatible faithful specifications for most benchmark problems, rather than failing because the problems lack a correct Verus specification that is also compatible with \execspec{}.

\input{paper_sections/figures_plots/tikz_images/00_testcase_journey}
\mainFigureBarrier

\paragraph{LLM-based evaluation misses specification errors.}
We compare our \texttt{exec\_spec}-based evaluation metric against LLM-based evaluation of specification faithfulness.
When we used \gptfivethreecodex{} as an LLM judge on its own specifications, it marked 49 of 191 incorrect but compilable specifications as correct.
Thus, the LLM judge failed to identify errors in 25.7\% of the incorrect specifications for which our benchmark found a verifiable failing testcase.
These are cases where the executable testcases expose a concrete error in the generated specification, but the LLM judge does not identify the specification as incorrect.
\todoanmol{myabe more apt to say: testcases can find issues where LLMs are not able to..}
Details are in App.~\ref{app:insights-llm-judge}.

\paragraph{Failure modes.}
Our qualitative analysis reveals three recurring failure modes.
First, models may miss important input properties, such as omitting a global structural promise or accepting inputs that violate existence guarantees (App.~\ref{app:case-study-1028c},~\ref{app:case-study-1027c}).
These failures can prevent a prover from using assumptions needed to verify correct code.
Second, models may reject correct outputs.
For example, \geminithreeonepro{} writes an overly complex interval-union postcondition that rejects valid answers, while \opusfoursix{} succeeds with a simpler characterization (App.~\ref{app:case-study-2074d}).
Third, models may accept incorrect outputs, such as non-coprime pairs or suboptimal rectangles (App.~\ref{app:case-study-1051b},~\ref{app:case-study-1027c}).
These failures could allow incorrect code to be verified.
Details are in App.~\ref{app:insights-unsuccessful-cases}. We also find that agents are less likely to generate correct specifications for harder problems, as measured by Codeforces rating (App.~\ref{app:insights-difficulty}), and are brittle across repeated attempts, with substantially lower pass$^k$ than pass@$k$ (App.~\ref{app:insights-pass-at-k}).

%% file: paper_sections/tables/00_main_table.tex
\begin{table}[t]
\centering
\scriptsize
\setlength{\tabcolsep}{3pt}
\renewcommand{\arraystretch}{1.15}
\caption{Main results on \vsb tasks in \vsg environments. Pass@1 is the fraction of problems where the generated specification passes all test cases. Pass@1-completeness is the fraction of problems where the generated specification passes all completeness tests. Each agent is given a budget of \$2.5 per problem.}
\label{tab:main}
\resizebox{\textwidth}{!}{%
\begin{tabular}{lcccccc}
\toprule
\textbf{Model}
& \textbf{Pass@1}
& \textbf{Pass@1-Comp.}
& \multicolumn{4}{c}{\textbf{Average Testcase Fraction Passed}} \\
\cmidrule(lr){4-7}
&
&
& \textbf{Pre-Sound}
& \textbf{Pre-Complete}
& \textbf{Post-Complete}
& \textbf{Post-Sound} \\
\midrule
\rowcolor{gray!15}
\multicolumn{7}{l}{\textit{Closed-source models; \$2.5 cost cap}} \\
\geminithreeonepro{} & 0.778 & 0.824 & 0.909 & 0.926 & 0.892 & 0.891 \\
\gptfivethreecodex{} & 0.578 & 0.766 & 0.826 & 0.910 & 0.878 & 0.766 \\
\opusfoursix{} & 0.511 & 0.587 & 0.671 & 0.705 & 0.658 & 0.655 \\
\midrule
\rowcolor{gray!15}
\multicolumn{7}{l}{\textit{Open-source models; max 400 steps, \$2.5 cost cap}} \\
\deepseekvfourpro{} & 0.243 & 0.318 & 0.380 & 0.422 & 0.378 & 0.349 \\
\glmfiveone{} & 0.215 & 0.248 & 0.350 & 0.375 & 0.297 & 0.288 \\
\kimiktwosix{} & 0.255 & 0.291 & 0.329 & 0.350 & 0.315 & 0.301 \\
\bottomrule
\end{tabular}%
}
\end{table}

%% file: paper_sections/figures_plots/tikz_images/00_testcase_journey.tex
\begin{figure}[!htbp]
\centering
\resizebox{0.84\linewidth}{!}{%
\begin{tikzpicture}[
  font=\small,
  >=Stealth,
  node distance=8mm and 10mm,
  start/.style={
    draw=black!55,
    rounded corners=3pt,
    fill=stageProb!35,
    align=center,
    text width=4.3cm,
    minimum height=8mm
  },
  question/.style={
    draw=black!60,
    diamond,
    aspect=2.15,
    fill=stageExec!25,
    align=center,
    inner sep=1pt,
    text width=3.1cm
  },
  action/.style={
    draw=black!60,
    rounded corners=3pt,
    fill=stageEval!30,
    align=center,
    text width=3.9cm,
    minimum height=8mm
  },
  terminal/.style={
    draw=black!55,
    rounded corners=3pt,
    align=center,
    text width=2.6cm,
    minimum height=8mm
  },
  accept/.style={terminal, fill=green!18},
  reject/.style={terminal, fill=red!13},
  unknown/.style={terminal, fill=gray!16},
  note/.style={
    draw=black!35,
    rounded corners=3pt,
    fill=white,
    align=left,
    text width=4.2cm,
    font=\footnotesize
  },
  arr/.style={-{Latex[length=2mm]}, semithick, black!65},
  stagebox/.style={draw=black!25, rounded corners=4pt, inner sep=5pt},
  numbadge/.style={circle, draw=black!70, fill=white, inner sep=0pt,
    font=\tiny\bfseries, minimum size=3.8mm, line width=0.5pt}
]

\node[start] (start) {Concrete testcase $t$\\+ submitted specification $s$};
\node[note, right=14mm of start, text width=5.0cm] (legend) {
  \textbf{Six resolution categories}\\
  compile-or-syntax-error\\
  accept-via-symbolic\\
  reject-via-symbolic\\
  accept-via-exec\\
  reject-via-exec\\
  indeterminate-during-exec
};

\node[question, below=9mm of start] (proveaccept) {Can Verus prove\\$s(t)$?};
\node[accept, right=18mm of proveaccept, text width=3.0cm] (symaccept) {\textbf{accept-via-}\\\textbf{symbolic}};

\node[question, below=11mm of proveaccept] (provereject) {Can Verus prove\\$\neg s(t)$?};
\node[reject, right=18mm of provereject, text width=3.0cm] (symreject) {\textbf{reject-via-}\\\textbf{symbolic}};

\node[unknown, below=11mm of provereject] (symunknown) {Symbolically\\\textbf{unknown}};

\node[action, below=9mm of symunknown] (compile) {Compile executable\\counterpart via \execspec{}};
\node[question, below=10mm of compile] (run) {Run executable\\spec on $t$};
\node[unknown, right=18mm of compile, text width=3.1cm] (compileerr) {\textbf{compile-or-}\\\textbf{syntax-error}};
\node[reject, left=15mm of run, text width=2.9cm] (execreject) {\textbf{reject-via-}\\\textbf{exec}};
\node[accept, right=15mm of run, text width=2.9cm] (execaccept) {\textbf{accept-via-}\\\textbf{exec}};
\node[unknown, below=9mm of run, text width=3.4cm] (execunknown) {\textbf{indeterminate-}\\\textbf{during-exec}};

\node[action, below=14mm of execunknown, text width=4.8cm] (classify) {Classify using the\\testcase bucket label};
\node[terminal, below=8mm of classify, fill=stageSpec!25, text width=6.8cm] (final) {
  resolution category:\\
  compile-or-syntax-error; accept-via-symbolic; reject-via-symbolic;\\
  accept-via-exec; reject-via-exec; indeterminate-during-exec
};

\node[numbadge, anchor=south east] at (start.north west) {0};
\node[numbadge, anchor=south east] at (proveaccept.north west) {1};
\node[numbadge, anchor=south east] at (symaccept.north west) {2};
\node[numbadge, anchor=south east] at (provereject.north west) {3};
\node[numbadge, anchor=south east] at (symreject.north west) {4};
\node[numbadge, anchor=south east] at (symunknown.north west) {5};
\node[numbadge, anchor=south east] at (compile.north west) {6};
\node[numbadge, anchor=south east] at (compileerr.north west) {7};
\node[numbadge, anchor=south east] at (run.north west) {8};
\node[numbadge, anchor=south east] at (execreject.north west) {9};
\node[numbadge, anchor=south east] at (execaccept.north west) {10};
\node[numbadge, anchor=south east] at (execunknown.north west) {11};
\node[numbadge, anchor=south east] at (classify.north west) {12};
\node[numbadge, anchor=south east] at (final.north west) {13};

\draw[arr] (start) -- (proveaccept);
\draw[arr] (legend.south west) to[out=-120,in=25] (proveaccept.north east);

\draw[arr] (proveaccept) -- node[above, font=\footnotesize] {yes} (symaccept);
\draw[arr] (proveaccept) -- node[right, font=\footnotesize] {no} (provereject);
\draw[arr] (provereject) -- node[above, font=\footnotesize] {yes} (symreject);
\draw[arr] (provereject) -- node[right, font=\footnotesize] {no} (symunknown);

\draw[arr] (symunknown) -- (compile);
\draw[arr] (compile) -- node[above, font=\footnotesize] {fails} (compileerr);
\draw[arr] (compile) -- node[right, font=\footnotesize] {succeeds} (run);
\draw[arr] (run) -- node[above, font=\footnotesize] {false} (execreject);
\draw[arr] (run) -- node[above, font=\footnotesize] {true} (execaccept);
\draw[arr] (run) -- node[right, font=\footnotesize] {error/timeout} (execunknown);

\draw[arr] (symaccept.south) |- (classify.east);
\draw[arr] (symreject.south) |- (classify.east);
\draw[arr] (compileerr.south) |- (classify.east);
\draw[arr] (execreject.south) |- (classify.west);
\draw[arr] (execaccept.south) |- (classify.east);
\draw[arr] (execunknown) -- (classify);
\draw[arr] (classify) -- (final);

\begin{scope}[on background layer]
  \node[stagebox, fill=stageExec!10, fit=(proveaccept) (provereject) (symaccept) (symreject) (symunknown)] (symbox) {};
  \node[stagebox, fill=stageEval!10, fit=(compile) (run) (compileerr) (execreject) (execaccept) (execunknown)] (execbox) {};
\end{scope}
\node[font=\footnotesize\bfseries, color=stageExecD, anchor=south west] at (symbox.north west) {Symbolic Verus checks};
\node[font=\footnotesize\bfseries, color=stageEvalD, anchor=south west] at (execbox.north west) {\execspec{} fallback};

\end{tikzpicture}%
}
\caption{\textbf{Decision tree for resolving one testcase.}
For each testcase, the evaluator checks whether the submitted specification accepts or rejects the concrete input-output example.
It first tries to prove acceptance or rejection symbolically in Verus.
If both symbolic checks are inconclusive, it compiles the specification to executable Rust via \execspec{} and runs it on the concrete testcase.
The observation is assigned to one of six operational resolution categories: compile-or-syntax-error, accept-via-symbolic, reject-via-symbolic, accept-via-exec, reject-via-exec, or indeterminate-during-exec.
For \precomplete{} and \postcomplete{} testcases, we count accept-via-symbolic and accept-via-exec as correct categories, while reject-via-symbolic and reject-via-exec indicate that the specification rejects a valid input or correct output.
For \presound{} and \postsound{} testcases, we count reject-via-symbolic and reject-via-exec as correct categories, while accept-via-symbolic and accept-via-exec indicate that the specification accepts an invalid input or incorrect output.
}
\label{fig:testcase-journey}
\end{figure}
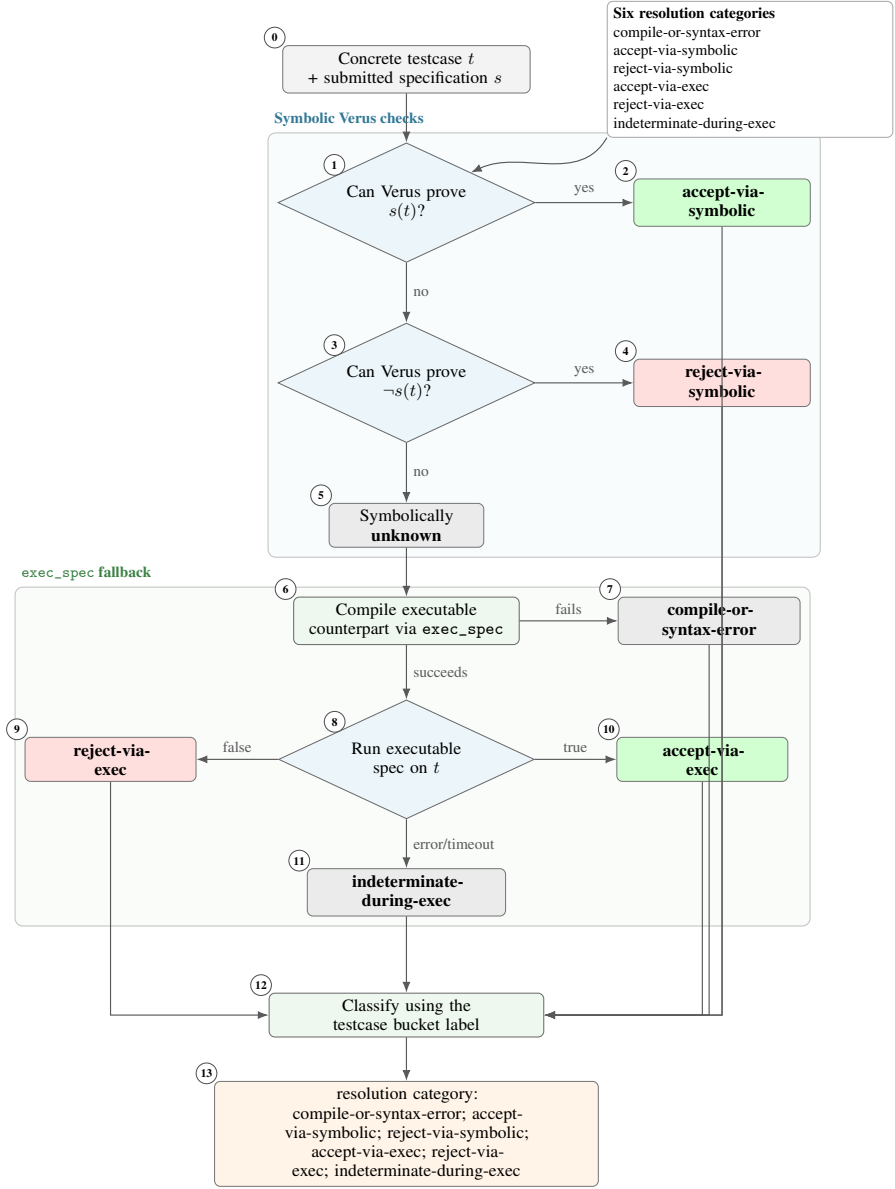

%% file: paper_sections/main_paper/03_related_work.tex
\section{Related Work}

\begin{table*}[t]
\centering
\renewcommand{\arraystretch}{1.15}
\resizebox{\textwidth}{!}{%
\begin{tabular}{lcccccccc}
\toprule
& \textbf{Mainstream} & \textbf{NL$\to$} & \textbf{Gold-Spec} & \textbf{Non-LLM} & \textbf{Pre/Post} & \textbf{Human-} & \textbf{Auto-} & \textbf{Agentic} \\
\textbf{Work} & \textbf{Lang?} & \textbf{Spec} & \textbf{Free} & \textbf{Verifier} & \textbf{$\times$ S/C} & \textbf{Adv.} & \textbf{Curated} & \textbf{Eval} \\
\midrule
\multicolumn{9}{l}{\textit{Without specification-faithfulness evaluation}} \\
AutoVerus~\citep{yangAutoVerusAutomatedProof2025} & \cmark~(Rust) & \xmark & --- & --- & --- & --- & \xmark & \partialmark \\
VeruSAGE~\citep{yang2025verusage} & \cmark~(Rust) & \xmark & --- & --- & --- & --- & \partialmark & \cmark \\
AlphaVerus~\citep{aggarwal_alphaverus_2024} & \cmark~(Rust) & \xmark & --- & --- & --- & --- & \cmark & \xmark \\
PSV~\citep{wilf2025psv} & \cmark~(Rust) & \xmark & --- & --- & --- & --- & \cmark & \xmark \\
\citet{misu2024towards} & \xmark~(Dafny) & \cmark & --- & --- & --- & --- & \partialmark & \xmark \\
SpecGen~\citep{ma2025specgen} & \cmark~(Java) & \xmark & --- & --- & --- & --- & \xmark & \xmark \\
AutoSpec~\citep{wen2024autospec} & \cmark~(C) & \xmark & --- & --- & --- & --- & \xmark & \xmark \\
MSG~\citep{fu2025msg} & \xmark~(Move) & \xmark & --- & --- & --- & --- & \xmark & \cmark \\
\midrule
\multicolumn{9}{l}{\textit{Specification-faithfulness evaluation}} \\
\citet{endres2024postconditions} & \cmark~(Java) & \cmark & \cmark & \cmark & \partialmark & \xmark & \xmark & \xmark \\
\citet{lahiri2024evaluating} & \xmark~(Dafny) & \cmark & \cmark & \cmark & \partialmark & \xmark & \xmark & \xmark \\
Clover~\citep{sunClover2024} & \xmark~(Dafny) & \partialmark & \cmark & \partialmark & \xmark & \xmark & \xmark & \xmark \\
VERINA~\citep{ye2025verina} & \xmark~(Lean) & \cmark & \xmark & \cmark & \cmark & \xmark & \xmark & \partialmark \\
VeriAct~\citep{misu2025veriact} & \cmark~(Java) & \xmark & \cmark & \cmark & \cmark & \xmark & \xmark & \cmark \\
\midrule
\rowcolor{gray!15}
\textbf{\benchmark~(Ours)} & \cmark~\textbf{(Rust)} & \cmark & \cmark & \cmark & \cmark & \cmark & \cmark & \cmark \\
\bottomrule
\end{tabular}%
}
\caption{
\textbf{Comparison with closest related works on verified code generation, specification generation, and specification-faithfulness evaluation.}
\textit{Mainstream Language?}: target language is widely used in production (vs.\ research/specialized).
\textit{NL$\to$Spec}: takes natural-language problem descriptions as input.
\textit{Gold-Spec-Free}: evaluation does not require expert-written reference formal specifications.
\textit{Non-LLM Verifier}: spec evaluation uses a deterministic oracle (verifier, tests) rather than an LLM judge.
\textit{Pre/Post~$\times$~S/C}: evaluates both preconditions and postconditions along soundness and completeness axes.
\textit{Human-Adv.}: evaluation uses human-written adversarial test inputs (in \benchmark{}, Codeforces hacks).
\textit{Auto-Curated}: the benchmark or dataset can be scaled via automation (e.g., extraction, translation, or self-play) rather than requiring manual curation per instance.
\textit{Agentic Eval}: evaluation operates within an interactive agent environment.
\cmark~= yes, \xmark~= no, \partialmark~= partial or indirect, --- = not applicable.
}
\label{tab:related_work_comparison}
\end{table*}
Table~\ref{tab:related_work_comparison} provides a structured comparison of our work with the closest related efforts across eight dimensions; we summarize the key contrasts below.

\textbf{Verified Code Generation.}
Recent projects use LLMs for formally verified program synthesis, either generating proofs from code and specifications~\citep{yangAutoVerusAutomatedProof2025, chen2024safe, yang2025verusage} or generating code and proofs from specifications, sometimes synthesizing the specification as well~\citep{aggarwal_alphaverus_2024, misu2024towards, wilf2025psv, sunClover2024}.
However, verifier guarantees apply only relative to the specification: if the specification misses intent, verified code can still be wrong.
\citet{aggarwal_alphaverus_2024} identify this as specification misalignment; we isolate specification autoformalization and automate evaluation of faithfulness to program intent.

\textbf{Evaluating Formal Specifications.}
Recent work recognizes specification evaluation as distinct from checking whether code verifies~\citep{lahiri2024evaluating, deng2025verifythisbench,misu2025veriact}.
Existing methods compare against reference specifications or semantic metrics~\citep{wen2024autospec,ma2025specgen}, test rejection of buggy mutants~\citep{endres2024postconditions, lahiri2024evaluating}, or reconstruct code to detect weak specifications~\citep{sunClover2024}.
These approaches rely on existing artifacts or reference specifications rather than benchmarking autoformalization from natural language.
VERINA~\citep{ye2025verina} also separates soundness and completeness, but relative to reference formal specifications; concurrent VeriAct~\citep{misu2025veriact} similarly targets multi-axis evaluation via Hoare-triple harnesses.
Our approach needs no reference formal specification: we use Verus's \texttt{exec\_spec} to test generated specifications directly against Codeforces hacks, adversarial inputs written by competitors to break accepted solutions.


\textbf{Agentic Evaluation.}
Agentic benchmarks such as SWE-bench~\citep{jimenez2024swebench} evaluate autonomous software-engineering agents, but on unverified code with test-based evaluation.
For formal verification, VeruSAGE~\citep{yang2025verusage} studies agentic Verus proof generation, recent benchmarks add iterative refinement for code and proofs~\citep{miranda2025veribench, zhao2026algoveri}, and MSG~\citep{fu2025msg} uses a multi-agent architecture for specification generation.
These works either assume specifications are given or use task-specific agent architectures.
Our benchmark evaluates general-purpose agents on natural-language-to-specification autoformalization in an interactive environment where they invoke the verifier, observe feedback, and refine specifications.

%% file: paper_sections/main_paper/04_conclusion.tex
\section{Conclusion}

We introduced \vsb{} and \vsg{}, a benchmark and agentic environment for evaluating specification autoformalization: translating informal programming intent into faithful formal specifications.
The benchmark pairs Codeforces problems with official tests and human-written hacks, and the evaluator uses executable specifications to check whether a generated \prespec{} and \postspec{} accept and reject the right concrete cases.
This gives a scalable faithfulness signal without requiring expert-written reference specifications or relying only on LLM judgment.
The data-creation pipeline can also be extended to new Codeforces problems as they appear, making continued benchmark growth possible~\citep{jain2024livecodebench}.

Our experiments show that specification autoformalization is a distinct bottleneck.
Even when frontier models can generate correct code for a problem, they often fail to write a faithful specification for that same problem.
Adversarial hacks are especially useful in this setting because they expose specification failures that official tests miss.
We also find that an LLM-as-judge baseline misses 26\% of the failures caught by our evaluator, suggesting that executable testing is a more reliable signal for the subtle errors that arise in this task.

\textbf{Limitations.}
This work focuses on single-file competition-style problems.
Repository-level specifications for multi-file, real-world software systems are more realistic and may expose additional weaknesses in LLMs' ability to autoformalize specifications in practical settings.
Faithfulness evaluation also remains an approximation: finite test suites can expose many specification errors, but they cannot rule out all possible errors.

%% file: paper_sections/main_paper/acknowledgements.tex
\section*{Acknowledgments}

This material is based upon work supported by the National Science Foundation CISE Graduate Fellowships under Grant No. 2313998, the National Science Foundation under Grant Nos. DMS-2434614 and DMS-2502281, AFRL and DARPA under Agreement FA8750-24-9-1000, the CyLab Future Enterprise initiative, and gifts from Amazon, the Beneficial AI Foundation, Convergent Research, as well as a grant of API credits from Microsoft Azure and Gemini. Pranjal is supported by a SoftBank Group-Arm Fellowship. Any opinions, findings, conclusions, or recommendations expressed in this material are those of the author(s) and do not necessarily reflect the views of these entities.

%% file: paper_sections/appendix/appendix_verus_binary_search.tex
\section{Example of Formally Verified Code}
\label{app:binary-search-verus-example}

Figure~\ref{fig:binary-search-verus-example} shows a complete Verus-verified
implementation for the first-occurrence search problem used in the running example,
whose informal description $s_I$ is shown in Figure~\ref{fig:binary-search-bucket-example}.
The blue region is the formal specification $s_F$ (the \texttt{pre\_spec} and
\texttt{post\_spec} predicates, with the helper \texttt{sorted\_nondecreasing}),
written over the data model in yellow (the \texttt{In1}/\texttt{Out} types, declared
with \texttt{exec\_spec\_unverified!}, which generates the executable
\texttt{ExecIn1}/\texttt{ExecOut} types).
The purple region is the executable code, and the green region is the proof
annotations that help Verus check the code against $s_F$, including loop
invariants, decreases clauses, and \texttt{assert} proof hints; \texttt{deep\_view()}
connects the executable values back to the specification.

\definecolor{binaryDataColor}{RGB}{120,80,0}
\definecolor{binarySpecColor}{RGB}{0,82,150}
\definecolor{binaryExecColor}{RGB}{115,55,135}
\definecolor{binaryProofColor}{RGB}{35,110,55}
\definecolor{binaryNeutralBg}{RGB}{248,248,248}
\definecolor{binaryDataBg}{RGB}{255,244,214}
\definecolor{binarySpecBg}{RGB}{224,239,255}
\definecolor{binaryExecBg}{RGB}{245,232,250}
\definecolor{binaryProofBg}{RGB}{225,244,225}

\lstdefinestyle{binaryFigureBase}{
  language=Verus,
  basicstyle=\fontfamily{lmtt}\fontsize{5.2}{5.6}\selectfont,
  frame=none,
  breaklines=true,
  columns=fullflexible,
  keepspaces=true,
  aboveskip=0pt,
  belowskip=0pt,
  xleftmargin=0.4em,
  xrightmargin=0.2em,
  framesep=1.2pt,
  alsoletter=_,
  emph={sorted_nondecreasing,pre_spec,post_spec,rust_codegen_solve},
  emphstyle=\bfseries
}
\lstdefinestyle{binaryNeutralListing}{style=binaryFigureBase,backgroundcolor=\color{binaryNeutralBg}}
\lstdefinestyle{binaryDataListing}{style=binaryFigureBase,backgroundcolor=\color{binaryDataBg}}
\lstdefinestyle{binarySpecListing}{style=binaryFigureBase,backgroundcolor=\color{binarySpecBg}}
\lstdefinestyle{binaryExecListing}{style=binaryFigureBase,backgroundcolor=\color{binaryExecBg}}
\lstdefinestyle{binaryProofListing}{style=binaryFigureBase,backgroundcolor=\color{binaryProofBg}}

\begin{figure}[H]
\centering
\setlength{\topsep}{0pt}\setlength{\partopsep}{0pt}\setlength{\parskip}{0pt}
\lstset{alsoletter=_,
  emph={sorted_nondecreasing,pre_spec,post_spec,rust_codegen_solve},
  emphstyle=\bfseries}
\begin{lstlisting}[style=binaryNeutralListing]
use vstd::contrib::exec_spec::*;
use vstd::prelude::*;

verus! {
\end{lstlisting}
\begin{lstlisting}[style=binaryDataListing]
exec_spec_unverified! {
  pub struct In1 { pub n: usize, pub arr: Seq<i64>, pub k: i64 }
  pub struct Out { pub pos: i64 }
}
\end{lstlisting}
\begin{lstlisting}[style=binarySpecListing]
pub open spec fn sorted_nondecreasing(arr: Seq<i64>) -> bool {
  forall |i: int, j: int| 0 <= i <= j < arr.len() ==> arr[i] <= arr[j]
}

pub open spec fn pre_spec(in1: In1) -> bool {
  &&& in1.arr.len() == in1.n
  &&& in1.n <= 200_000
  &&& sorted_nondecreasing(in1.arr)
}

pub open spec fn post_spec(in1: In1, out: Out) -> bool {
  if out.pos == -1 {
    forall |i: int| 0 <= i < in1.n as int ==> #[trigger] in1.arr[i] != in1.k
  } else {
    &&& 0 <= out.pos
    &&& out.pos < in1.n as i64
    &&& in1.arr[out.pos as int] == in1.k
    &&& forall |i: int| 0 <= i < out.pos ==> #[trigger] in1.arr[i] != in1.k
  }
}
\end{lstlisting}
\begin{lstlisting}[style=binaryExecListing]
pub fn rust_codegen_solve(in1: &ExecIn1) -> (out: ExecOut)
  requires pre_spec(in1.deep_view()),
  ensures post_spec(in1.deep_view(), out.deep_view()),
{
  let mut lo: usize = 0;
  let mut hi: usize = in1.arr.len();

  while lo < hi
\end{lstlisting}
\begin{lstlisting}[style=binaryProofListing]
    invariant
      lo <= hi,
      hi <= in1.arr.len(),
      pre_spec(in1.deep_view()),
      forall |j: int| 0 <= j < lo as int ==> #[trigger] in1.arr@[j] < in1.k,
      forall |j: int| hi as int <= j < in1.arr.len() as int ==> #[trigger] in1.arr@[j] >= in1.k,
    decreases hi - lo,
\end{lstlisting}
\begin{lstlisting}[style=binaryExecListing]
  {
    let mid = lo + (hi - lo) / 2;
\end{lstlisting}
\begin{lstlisting}[style=binaryProofListing]
    assert(lo <= mid);
    assert(mid < hi);
\end{lstlisting}
\begin{lstlisting}[style=binaryExecListing]

    if in1.arr[mid] < in1.k {
\end{lstlisting}
\begin{lstlisting}[style=binaryProofListing]
      assert forall |j: int| 0 <= j < (mid + 1) as int implies #[trigger] in1.arr@[j] < in1.k by {
        if j < lo as int {
        } else {
          assert(j <= mid as int);
          assert(in1.deep_view().arr == in1.arr@);
          assert(sorted_nondecreasing(in1.arr@));
          assert(in1.arr@[j] <= in1.arr@[mid as int]);
        }
      }
\end{lstlisting}
\begin{lstlisting}[style=binaryExecListing]
      lo = mid + 1;
    } else {
\end{lstlisting}
\begin{lstlisting}[style=binaryProofListing]
      assert forall |j: int| mid as int <= j < in1.arr.len() as int implies #[trigger] in1.arr@[j] >= in1.k by {
        assert(in1.deep_view().arr == in1.arr@);
        assert(sorted_nondecreasing(in1.arr@));
        assert(in1.arr@[mid as int] <= in1.arr@[j]);
      }
\end{lstlisting}
\begin{lstlisting}[style=binaryExecListing]
      hi = mid;
    }
  }

  if lo < in1.arr.len() && in1.arr[lo] == in1.k {
    return ExecOut { pos: lo as i64 };
  }
\end{lstlisting}
\begin{lstlisting}[style=binaryProofListing]
  assert(lo == hi);
  assert forall |j: int| 0 <= j < in1.arr.len() as int implies #[trigger] in1.arr@[j] != in1.k by {
    if j < lo as int {
      assert(in1.arr@[j] < in1.k);
    } else if lo < in1.arr.len() {
      assert(lo as int <= j);
      assert(in1.arr@[lo as int] != in1.k);
      assert(in1.arr@[lo as int] >= in1.k);
      assert(in1.arr@[lo as int] > in1.k);
      assert(in1.deep_view().arr == in1.arr@);
      assert(sorted_nondecreasing(in1.arr@));
      assert(in1.arr@[lo as int] <= in1.arr@[j]);
    } else {
      assert(j < lo as int);
    }
  }
\end{lstlisting}
\begin{lstlisting}[style=binaryExecListing]
  ExecOut { pos: -1 }
}

} // verus!
\end{lstlisting}
\caption{\textbf{Example of formally verified code: binary search for first occurrence.}
Blue highlights the formal specification $s_F$ (the \texttt{pre\_spec} and
\texttt{post\_spec} predicates), written over the data model in yellow (the
\texttt{In1}/\texttt{Out} types, declared with \texttt{exec\_spec\_unverified!}).
Purple highlights the executable code, and green highlights the proof annotations
that help Verus verify the code against $s_F$, including loop invariants,
decreases clauses, and \texttt{assert} proof hints. The proof connects executable
values to the specification using \texttt{deep\_view()}.}
\label{fig:binary-search-verus-example}
\end{figure}

%% file: paper_sections/appendix/related_work_table.tex

%% file: paper_sections/appendix/appendix_skeleton.tex
\section{Task Format and Evaluation: A Worked Example}
\label{app:skeleton}

We walk through a complete example to illustrate the agent's task
and the evaluation mechanism.

\subsection{Problem Statement}
\label{app:skeleton-problem-statement}
Along with the specification-generation background prompt (App.~\ref{app:full-prompt}),
each task gives the agent the informal Codeforces problem statement.
For example, for Codeforces 1197D, the agent receives the statement shown below.

\begin{tcolorbox}[colback=blue!3,colframe=blue!40,fontupper=\scriptsize,title={\textbf{Informal specification (natural language) $s_I$ for Codeforces 1197-D}}]
You are given an array $a_1, a_2, \dots, a_n$ and two integers $m$ and $k$.
You can choose some subarray $a_l, a_{l+1}, \dots, a_{r-1}, a_r$.
The cost of subarray $a_l, a_{l+1}, \dots, a_{r-1}, a_r$ is equal to
$\sum\limits_{i=l}^{r} a_i - k \left\lceil \frac{r - l + 1}{m} \right\rceil$,
where $\left\lceil x \right\rceil$ is the least integer greater than or equal to $x$.
The cost of empty subarray is equal to zero.

For example, if $m = 3$, $k = 10$ and $a = [2, -4, 15, -3, 4, 8, 3]$,
then the cost of some subarrays are:

\begin{itemize}
    \item $a_3 \dots a_3: 15 - k \left\lceil \frac{1}{3} \right\rceil = 15 - 10 = 5$;
    \item $a_3 \dots a_4: (15 - 3) - k \left\lceil \frac{2}{3} \right\rceil = 12 - 10 = 2$;
    \item $a_3 \dots a_5: (15 - 3 + 4) - k \left\lceil \frac{3}{3} \right\rceil = 16 - 10 = 6$;
    \item $a_3 \dots a_6: (15 - 3 + 4 + 8) - k \left\lceil \frac{4}{3} \right\rceil = 24 - 20 = 4$;
    \item $a_3 \dots a_7: (15 - 3 + 4 + 8 + 3) - k \left\lceil \frac{5}{3} \right\rceil = 27 - 20 = 7$.
\end{itemize}

Your task is to find the maximum cost of some subarray (possibly empty) of array $a$.

\medskip
\textbf{Input.}
The first line contains three integers $n$, $m$, and $k$
($1 \le n \le 3 \cdot 10^5, 1 \le m \le 10, 1 \le k \le 10^9$).
The second line contains $n$ integers $a_1, a_2, \dots, a_n$
($-10^9 \le a_i \le 10^9$).

\medskip
\textbf{Output.}
Print the maximum cost of some subarray of array $a$.

\medskip
\textbf{Examples.}

\textbf{Input}
\begin{flushleft}
\ttfamily
7 3 10\\
2 -4 15 -3 4 8 3
\end{flushleft}
\textbf{Output}
\begin{flushleft}
\ttfamily
7
\end{flushleft}

\textbf{Input}
\begin{flushleft}
\ttfamily
5 2 1000\\
-13 -4 -9 -20 -11
\end{flushleft}
\textbf{Output}
\begin{flushleft}
\ttfamily
0
\end{flushleft}
\end{tcolorbox}

\subsection{Skeleton File}

When starting the task, the agent is asked to write specifications in a file called
\texttt{solve.rs}.
For the problem above, the file is initialized from the skeleton shown below.
The agent fills in the bodies of \texttt{pre\_spec} and \texttt{post\_spec}, and may
optionally fill in the four proof helpers (\texttt{pre\_spec\_soundness\_proof}, etc.) to give
Verus additional facts for discharging testcase assertions.

The input and output types (\texttt{In1}, \texttt{Out}), declared inside
\texttt{exec\_spec\_unverified!}, are fixed for the problem and encode how test inputs and
outputs are represented in Verus; the agent cannot change them.

The yellow-highlighted \texttt{check\_*} and \texttt{main\_exec\_*} functions are also fixed;
the agent cannot modify them.
They contain the red \texttt{\_\_PASTE\_\_} markers, which are not Verus code — they are
template placeholders.
The skeleton is a template, not a standalone file: \texttt{exec\_in1} and \texttt{exec\_out} are
undefined until the evaluator injects them, so the skeleton does not compile as shown.

Once the model submits its filled \texttt{solve.rs} with \texttt{pre\_spec} and
\texttt{post\_spec}, the evaluator runs it against every testcase in the four buckets
($\tau_{\mathrm{pre\text{-}comp}}$, $\tau_{\mathrm{pre\text{-}sound}}$,
$\tau_{\mathrm{post\text{-}comp}}$, $\tau_{\mathrm{post\text{-}sound}}$).
For each testcase, it substitutes the \texttt{\_\_PASTE\_\_} markers with concrete values:
\texttt{// \_\_PASTE\_out.input\_defn\_\_} is replaced with a concrete \texttt{exec\_in1}
definition, and, for post buckets, \texttt{// \_\_PASTE\_out.gt\_output\_defn\_\_} is also
replaced with a concrete \texttt{exec\_out} definition.
The evaluator then follows the decision tree in Figure~\ref{fig:testcase-journey} on the
resulting file to determine whether the specification accepts or rejects that testcase.

\definecolor{skeletonFixedBg}{RGB}{255,244,214}
\definecolor{skeletonNeutralBg}{RGB}{248,248,248}
\newcommand{\skeletonPasteInput}{\textcolor{red}{\ttfamily // \_\_PASTE\_out.input\_defn\_\_}}
\newcommand{\skeletonPasteGtOutput}{\textcolor{red}{\ttfamily // \_\_PASTE\_out.gt\_output\_defn\_\_}}
\lstdefinestyle{skeletonListingBase}{
  language=Verus,
  escapeinside={(*@}{@*)},
  numbers=left,
  numberstyle=\tiny\color{gray},
  stepnumber=1,
  numbersep=5pt,
  breaklines=true,
  columns=fullflexible,
  keepspaces=true,
  aboveskip=0pt,
  belowskip=0pt,
}
\lstdefinestyle{skeletonNeutralListing}{style=skeletonListingBase,backgroundcolor=\color{skeletonNeutralBg}}
\lstdefinestyle{skeletonFixedListing}{style=skeletonListingBase,backgroundcolor=\color{skeletonFixedBg}}

\begin{lstlisting}[style=skeletonNeutralListing,name=skeleton1197,firstnumber=1]
use vstd::prelude::*;
use vstd::contrib::exec_spec::*;

verus! {

    exec_spec_unverified! {
        pub struct In1 {
            pub n: i64,
            pub m: i64,
            pub k: i64,
            pub a: Seq<i64>,
        }

        pub struct Out {
            pub answer: i64,
        }

        pub open spec fn pre_spec(in1: In1) -> bool {
            // <-- agent has to fill this body and is free to define additional helper functions/imports -->

}

        pub open spec fn post_spec(in1: In1, out: Out) -> bool {
            // <-- agent has to fill this body and is free to define additional helper functions/imports -->

}

    }

pub open proof fn pre_spec_soundness_proof(in1: In1) -> bool {
    // <-- agent may optionally fill in -->

}

pub open proof fn pre_spec_completeness_proof(in1: In1) -> bool {
    // <-- agent may optionally fill in -->

}

pub open proof fn post_spec_soundness_proof(in1: In1, out: Out) -> bool {
    // <-- agent may optionally fill in -->

}

pub open proof fn post_spec_completeness_proof(in1: In1, out: Out) -> bool {
    // <-- agent may optionally fill in -->

}
\end{lstlisting}

\begin{lstlisting}[style=skeletonFixedListing,name=skeleton1197,firstnumber=last]
fn check_pre_spec_completeness() {
        (*@\skeletonPasteInput@*)

        proof {
            pre_spec_completeness_proof(exec_in1.deep_view());
        }

        assert(pre_spec(exec_in1.deep_view()));

    }
\end{lstlisting}

\begin{lstlisting}[style=skeletonFixedListing,name=skeleton1197,firstnumber=last]
fn check_pre_spec_soundness() {
        (*@\skeletonPasteInput@*)

        proof {
            pre_spec_soundness_proof(exec_in1.deep_view());
        }

        assert(!pre_spec(exec_in1.deep_view()));

    }
\end{lstlisting}

\begin{lstlisting}[style=skeletonFixedListing,name=skeleton1197,firstnumber=last]
fn check_post_spec_completeness() {
        (*@\skeletonPasteInput@*)
        (*@\skeletonPasteGtOutput@*)

        proof {
            post_spec_completeness_proof(exec_in1.deep_view(), exec_out.deep_view());
        }

        assert(post_spec(exec_in1.deep_view(), exec_out.deep_view()));

    }
\end{lstlisting}

\begin{lstlisting}[style=skeletonFixedListing,name=skeleton1197,firstnumber=last]
fn check_post_spec_soundness() {
        (*@\skeletonPasteInput@*)
        (*@\skeletonPasteGtOutput@*)

        proof {
            post_spec_soundness_proof(exec_in1.deep_view(), exec_out.deep_view());
        }

        assert(!post_spec(exec_in1.deep_view(), exec_out.deep_view()));

    }
\end{lstlisting}

\begin{lstlisting}[style=skeletonFixedListing,name=skeleton1197,firstnumber=last]
pub fn main_exec_pre_spec_check() {
        (*@\skeletonPasteInput@*)
        let result = exec_pre_spec(&exec_in1);
        // The evaluator may rewrite this assertion polarity for soundness cases.
        assert!(result);
    }
\end{lstlisting}

\begin{lstlisting}[style=skeletonFixedListing,name=skeleton1197,firstnumber=last]
pub fn main_exec_post_spec_check() {
        (*@\skeletonPasteInput@*)
        (*@\skeletonPasteGtOutput@*)
        let result = exec_post_spec(&exec_in1, &exec_out);
        // The evaluator may rewrite this assertion polarity for soundness cases.
        assert!(result);
    }
\end{lstlisting}

\begin{lstlisting}[style=skeletonNeutralListing,name=skeleton1197,firstnumber=last]
fn main() {
        unimplemented!("main() body is replaced by evaluator with a direct function call");
    }

}
\end{lstlisting}

\subsection{Evaluation via Symbolic Verification}
\label{app:example-symbolic-eval}

The four \texttt{check\_*} functions in the skeleton implement the symbolic path (nodes~1--5 in
Figure~\ref{fig:testcase-journey}).
For each testcase, the evaluator builds two derived files from the submitted \texttt{solve.rs}:
a \emph{completeness file} and a \emph{soundness file}.
Each derived file keeps only the relevant \texttt{check\_*} function for that file's role
(completeness or soundness), together with the model's \texttt{pre\_spec}/\texttt{post\_spec} and
any helper functions; all other \texttt{check\_*} and \texttt{main\_exec\_*} functions are
dropped.
The completeness file contains \texttt{check\_*\_completeness} with
\texttt{assert(pre\_spec(...))} or \texttt{assert(post\_spec(...))}, depending on the bucket;
the soundness file contains \texttt{check\_*\_soundness} with the negated assertion.
Both have the \texttt{\_\_PASTE\_\_} markers replaced with the concrete testcase values.
\texttt{deep\_view()} converts the injected \texttt{ExecIn1}/\texttt{ExecOut} values into the
specification-level \texttt{In1}/\texttt{Out} types consumed by \texttt{pre\_spec} and
\texttt{post\_spec}.

For a testcase in $\tau_{\mathrm{post\text{-}comp}}$, the completeness file looks like:

\begin{lstlisting}[language=Verus]
fn check_post_spec_completeness() {
    // injected by the evaluator:
    let exec_in1 = ExecIn1 {
        n: 7, m: 3, k: 10,
        a: vec![2, -4, 15, -3, 4, 8, 3],
    };
    let exec_out = ExecOut { answer: 7 };

    proof {
        post_spec_completeness_proof(
            exec_in1.deep_view(), exec_out.deep_view());
    }
    assert(post_spec(
        exec_in1.deep_view(), exec_out.deep_view()));
}
\end{lstlisting}

The evaluator runs Verus on both files.
If the completeness file verifies, the testcase resolves as accept-via-symbolic (node~2).
If the soundness file verifies instead, it resolves as reject-via-symbolic (node~4).
If neither verifies, the testcase is symbolically unknown (node~5) and falls back to the
executable path.

The proof helpers are optional: if the model adds lemmas to
\texttt{post\_spec\_completeness\_proof}, Verus uses those facts when discharging the
\texttt{assert} in the completeness wrapper.

\subsection{Evaluation via Executable Specifications (\texttt{exec\_spec})}
\label{app:example-exec-spec-eval}

When both symbolic checks are inconclusive (node~5), the evaluator falls back to running the
specification as executable Rust (nodes~6--11 in Figure~\ref{fig:testcase-journey}).
The macro \texttt{exec\_spec\_unverified!} generates \texttt{exec\_pre\_spec} and
\texttt{exec\_post\_spec} from the Verus specification functions.
The evaluator builds two exec files — a completeness file and a soundness file — using the same
\texttt{\_\_PASTE\_\_} injection as the symbolic path.
If compilation fails, the testcase resolves as compile-or-syntax-error (node~7).
Otherwise the compiled binary is run on the concrete testcase (node~8).

The completeness exec file expects the predicate to return \texttt{true}; the soundness exec
file expects \texttt{false}.
For a post-completeness testcase, the completeness exec file looks like:

\begin{lstlisting}[language=Verus]
pub fn main_exec_post_spec_check() {
    // injected by the evaluator:
    let exec_in1 = ExecIn1 {
        n: 7, m: 3, k: 10,
        a: vec![2, -4, 15, -3, 4, 8, 3],
    };
    let exec_out = ExecOut { answer: 7 };

    let result = exec_post_spec(&exec_in1, &exec_out);
    assert!(result);
}
\end{lstlisting}


The result determines the resolution: \texttt{true} gives accept-via-exec (node~10),
\texttt{false} gives reject-via-exec (node~9), and an error or timeout gives
indeterminate-during-exec (node~11).
Unlike the SMT \texttt{assert} in the symbolic path, this is an ordinary Rust \texttt{assert!}.

%% file: paper_sections/appendix/appendix_exec_spec.tex
\section{Specification to Executable Code Translation in Verus with \texttt{exec\_spec}}

Here, we describe the \texttt{exec\_spec} feature and the subset of the Verus specification language that it supports.
\label{app:exec_spec}

\subsection{Implementation Details}

The \texttt{exec\_spec} feature is implemented as a macro that the user applies to their specification code. Any specification code contained within the macro is automatically compiled to executable equivalents which can be referenced by other executable code.

This benchmark uses the \texttt{exec\_spec\_unverified!} macro, which generates executable code from specification code. Verus also ships with the \texttt{exec\_spec\_verified!} macro, which uses the same procedure as \texttt{exec\_spec\_unverified!} to generate executable code from specification code, but also produces Verus-checked \textit{proofs of equivalence} between the executable code and the original specification code.
While the executable code produced by the two macros is identical, \texttt{exec\_spec\_verified!} is intended to support \textit{verified clients} that consume the generated executable code. The proofs of equivalence generated by \texttt{exec\_spec\_verified!} ensure the absence of arithmetic overflow, infinite loops, and precondition violations in the generated executable code. However, due to incompleteness in the underlying verifier, the proof generation procedure can sometimes fail to generate a successful proof, leading to difficult-to-debug verification errors in the generated code. In contrast, \texttt{exec\_spec\_unverified!} skips the proof generation step to avoid these difficulties. Errors such as arithmetic overflow, infinite loops, and precondition violations will still materialize in panics (i.e., runtime errors) when the generated code is run against error-triggering input. In this benchmark, the code generated by \texttt{exec\_spec\_unverified!} is only run by the harness to execute test cases, so the proofs of equivalence are not needed by our evaluation methodology.

\subsection{Example}

In the example below, two \texttt{struct}s (\texttt{Point} and \texttt{Polygon}) and two \texttt{spec fn}s (\texttt{on\_line} and \texttt{is\_rect}) are defined within the \texttt{exec\_spec\_unverified!} macro.

\begin{lstlisting}[language=Verus]
use vstd::contrib::exec_spec::*;
use vstd::prelude::*;

verus! {

exec_spec_unverified! {

  struct Point {
    x: i64,
    y: i64,
  }

  struct Polygon {
    points: Seq<Point>
  }

  spec fn on_line(points: Seq<Point>) -> bool {
    forall |i: usize| #![auto] 0 <= i < points.len() ==> points[i as int].y == points[i as int].x
  }

  spec fn is_rect(poly: Polygon) -> bool {
    poly.points.len() == 4
  }
}

} // verus!
\end{lstlisting}
The \texttt{exec\_spec\_unverified!} macro generates executable code equivalent to the following (the macro-generated code has been simplified for readability). Each struct is compiled to a new struct whose name is prefixed with \texttt{Exec-} and whose field types have been converted to their executable equivalents (e.g., \texttt{Seq} is changed to \texttt{Vec}). Each \texttt{spec fn} is compiled to a new \texttt{exec fn} where quantified expressions (\texttt{forall}) have been converted to loops and Verus specification functions (\texttt{Seq<T>::len}) have been converted to Rust executable equivalents (\texttt{Vec<T>::len} or \texttt{[T]::len}).
\begin{lstlisting}[language=Verus]
struct ExecPoint {
  x: i64,
  y: i64,
}

struct ExecPolygon {
  points: Vec<Point>,
}

fn exec_on_line(points: &[Point]) -> bool {
  ({
    let mut _res = true;
    {
      let _lower_i = 0;
      let _upper_i = points.len();
      let mut i = _lower_i;
      if _lower_i < _upper_i {
        while i < _upper_i {
          if (!(points.index(i).y == points.index(i).x)) {
            _res = false;
            break;
          }
          i += 1;
        }
      }
    }
    _res
  })
}

fn exec_is_rect(poly: &Polygon) -> bool {
    poly.points.len() == 4
}
\end{lstlisting}

The generated executable code can then be used in Rust executable code. The following code runs without panics.

\begin{lstlisting}[language=Verus]
fn main() {
  let p1 = ExecPoint { x: 1, y: 1 };
  let p2 = ExecPoint { x: 2, y: 2 };
  let points = vec![p1, p2];
  let b1 = exec_on_line(&points);
  assert_eq!(b1, true);
  let poly = ExecPolygon { points: points };
  let b2 = exec_is_rect(&poly);
  assert_eq!(b2, false);
}
\end{lstlisting}

\subsection{Supported Specification Constructs}

At the time of writing, the \texttt{exec\_spec\_unverified!} macro supports the following Verus specification constructs.
\begin{itemize}
    \item Arithmetic operations
    
    \item Logical operators (\texttt{\&\&}, \texttt{||}, \texttt{\&\&\&}, \texttt{|||}, \texttt{!}, \texttt{==>})
    
    \item \texttt{if}, \texttt{match}, and \texttt{matches} expressions
    
    \item \texttt{spec fn} calls, including recursion
    
    \item Rust primitive types: integers (\texttt{i8}, \texttt{i16}, ~\dots, \texttt{isize}, \texttt{u8}, \texttt{u16}, ~\dots, \texttt{usize}), \texttt{bool}, and \texttt{char}. Note that Verus types \texttt{int} and \texttt{nat} are not supported as arguments or fields in user-defined types.

    \item Verus \texttt{SpecString} (an alias to \texttt{Seq<char>} which compiles to Rust \texttt{String}/\texttt{\&str}) and string literals: equality, indexing, \texttt{len}

    \item Rust \texttt{Option<T>}: equality, \texttt{unwrap}

    \item Verus \texttt{Seq<T>} (compiled to Rust \texttt{Vec<T>} or \texttt{\&[T]} depending on the context) and \texttt{seq!} literals: equality, \texttt{len}, indexing, \texttt{subrange}, \texttt{add}, \texttt{push}, \texttt{update}, \texttt{empty}, \texttt{to\_multiset}, \texttt{drop\_first}, \texttt{drop\_last}, \texttt{take}, \texttt{skip}, \texttt{first}, \texttt{last}, \texttt{is\_suffix\_of}, \texttt{is\_prefix\_of}, \texttt{contains}, \texttt{index\_of}, \texttt{index\_of\_first}, \texttt{index\_of\_last}

    \item Verus \texttt{Map<K, V>} (compiled to Rust \texttt{HashMap<K, V>}): equality, \texttt{len}, indexing, \texttt{empty}, \texttt{dom}, \texttt{insert}, \texttt{remove}, \texttt{get}. Note that indexing is only supported on \texttt{Map<K, V>} where \texttt{K} is a primitive type (e.g. \texttt{usize}); for other types \texttt{K}, use \texttt{get} instead.

    \item Verus \texttt{Set<T>} (compiled to Rust \texttt{HashSet<T>}): equality, \texttt{len}, \texttt{empty}, \texttt{contains}, \texttt{insert}, \texttt{remove}, \texttt{union}, \texttt{intersect}, \texttt{difference}

    \item Verus \texttt{Multiset<T>} (compiled to \texttt{ExecMultiset<T>}, a type implemented in \texttt{vstd::contrib::exec\_spec} whose internal representation is a \texttt{HashMap}): equality, \texttt{len}, \texttt{count}, \texttt{empty}, \texttt{singleton}, \texttt{add}, \texttt{sub}

    \item User-defined structs and enums. These user-defined types should be defined using \texttt{exec\_spec\_unverified!}-compatible types for the fields (e.g. \texttt{Seq}). Such user-defined types are then compiled to executable versions whose name is prefixed with \texttt{Exec-}, with executable versions of each field’s type (e.g. \texttt{Vec<T>}/\texttt{[T]}).

    \item Bounded universally quantified expressions of the form: \texttt{forall |x1: <type1>, x2: <type2>, ..., xN: <typeN>| <guard1> \&\& <guard2> \&\& ... \&\& <guardN> ==> <body>}, and existentially quantified expressions of the form: \texttt{exists |x1: <type1>, x2: <type2>, ..., xN: <typeN>| <guard1> \&\& <guard2> \&\& ... \&\& <guardN> \&\& <body>}, where:
    \begin{itemize}
        \item \texttt{<guardI>} is of the form \texttt{<lowerI> <op1> xI <op2> <upperI>}, where \texttt{<op1>}, \texttt{<op2>} are either \texttt{<=} or \texttt{<}, and \texttt{<lowerI>} and \texttt{<upperI>} can mention \texttt{xJ} for all \texttt{J < I}
        \item \texttt{<typeI>} is a Rust primitive integer (\texttt{i8}, \texttt{i16}, ~\dots, \texttt{isize}, \texttt{u8}, \texttt{u16}, ~\dots, \texttt{usize}) or \texttt{char}
    \end{itemize}
\end{itemize}

%% file: paper_sections/appendix/additional_data_creation.tex
\section{Additional Details About Data Creation}
\label{app:additional-data-creation}

\subsection{Data Collection Pipeline}

We describe the full pipeline for constructing \vsb from Codeforces contest
problems. The pipeline has five stages: sourcing, filtering, hack collection,
test-case conversion, and final selection. We collect official Codeforces tests
and user-submitted hacks, route them into the four testcase buckets from
\S\ref{ssec:data}, remove problematic or duplicate cases, and convert the
remaining raw text testcases into typed Verus/Rust constants for evaluation.
We now describe each stage in turn.

Figure~\ref{fig:setdiagram} gives a compact view of why these four buckets are
needed: a candidate specification can disagree with the informal intent either
on the valid-input domain or on the input-output relation, and each asymmetric
difference corresponds to one completeness or soundness bucket.

\input{paper_sections/figures_plots/teaser_figure/spec_faithful}

\textbf{Stage 1: Sourcing.}
We collect 10k problems from all Codeforces contests held up to
Dec 2025.

\textbf{Stage 2: Initial filtering.}
We apply several filters to retain a broad set of problems suitable for our
benchmark:
\begin{enumerate}
    \item \textbf{Hack availability.} We remove problems from early contests
    that predate the Codeforces hack system, since hacks are essential for our
    soundness test cases.
    \item \textbf{No floating point.} We exclude problems that involve
    floating-point arithmetic, as Verus does not currently support
    floating-point reasoning. 
\end{enumerate}

\textbf{Stage 3: Hack collection and categorization.}
For each remaining problem, we collect official Codeforces tests and all
user-submitted hacks. As described in \S\ref{ssec:data} and summarized in
Figure~\ref{fig:hack-journey}, a hack produces a test case whose bucket
(pre-$\ast$ or post-$\ast$, soundness or completeness) depends on the
Codeforces verdict:
\begin{enumerate}
    \item \textbf{Invalid input} (hack rejected by the Codeforces validator):
    the input is added to $\tau_{\mathrm{pre\text{-}sound}}$.
    \item \textbf{Valid input, incorrect output} (the hacked solution produces
    output rejected by the Codeforces checker): the input contributes to
    $\tau_{\mathrm{pre\text{-}comp}}$, and the input-output pair is added to
    $\tau_{\mathrm{post\text{-}sound}}$.
    \item \textbf{Valid input, correct output} (the hacked solution happens to
    produce accepted output): the input contributes to
    $\tau_{\mathrm{pre\text{-}comp}}$, and the pair is added to
    $\tau_{\mathrm{post\text{-}comp}}$.
\end{enumerate}

Several cleaning steps follow bucket assignment. Different attackers often propose
identical hacks targeting different submissions; we de-duplicate test cases by
their raw input-output content. Some test cases are only partially visible on
the platform, and very large cases may be truncated by Codeforces when displayed
in contest logs or hack metadata. We discard any test case whose raw text or
metadata is incomplete or truncated.

\textbf{Catch: semantic vs.\ syntactic invalid hacks.}
Not all invalid-input hacks are useful. A hack input may be rejected for two
reasons. A \emph{syntactic} violation means the text itself is malformed: a
non-integer token where an integer is expected, or the wrong number of values on
a line. Any parser rejects such inputs before \prespec{} is ever invoked, so
they reveal nothing about whether the specification captures the problem's
constraints. A \emph{semantic} violation means the text is syntactically valid
but breaks a stated constraint---a value outside the allowed range, or a
configuration that violates a problem guarantee. These are the inputs \prespec{}
must reject. For each rejected hack, Codeforces provides a machine-generated statement
describing why the input was invalid. We apply regex-based filters over this
statement to identify and discard syntactic violations, preventing them from
entering $\tau_{\mathrm{pre\text{-}sound}}$. Figure~\ref{fig:case-study-1027c-invalid-testcases} shows
concrete examples for Codeforces 1027C (App.~\ref{app:case-study-1027c}).
Test case 1 is syntactically invalid: the first line contains the token
\texttt{hello} instead of an integer $T$, violating the input grammar.
Test cases 2 and 3 are semantically invalid: test case 2 contains a stick
length of $-1$, which violates the constraint $a_j \ge 1$; test case 3 has
all sticks with distinct lengths, so no rectangle can be formed, violating
the problem's guarantee. Only test cases 2 and 3 are kept for
$\tau_{\mathrm{pre\text{-}sound}}$.

\begingroup
\newcommand{\badtoken}[1]{\textcolor{red!70!black}{\textbf{#1}}}
\begin{figure}[H]
    \centering
    \begingroup
    \setlength{\fboxsep}{8pt}
    \noindent\begin{minipage}[t]{0.32\linewidth}
    \textbf{Invalid Test Case 1} {\small\textit{(Syntactically Incorrect)}}
    \begin{Verbatim}[commandchars=\\\{\},fontsize=\scriptsize,frame=single]
\badtoken{hello}
4
7 2 2 7
8
2 8 1 4 8 2 1 5
5
5 5 5 5 5
\end{Verbatim}
    \textcolor{red!70!black}{\textbf{Invalid.} The first line must be an
    integer $T$; the token \texttt{hello} violates the input grammar.}
    \end{minipage}\hfill
    \begin{minipage}[t]{0.32\linewidth}
    \textbf{Invalid Test Case 2} {\small\textit{(Semantically Incorrect)}}
    \begin{Verbatim}[commandchars=\\\{\},fontsize=\scriptsize,frame=single]
3
4
7 2 2 7
8
2 8 1 4 8 2 1 5
5
5 5 5 5 \badtoken{-1}
\end{Verbatim}
    \textcolor{red!70!black}{\textbf{Invalid.} Stick lengths must satisfy
    $a_j \ge 1$; the value $-1$ violates this domain constraint.}
    \end{minipage}\hfill
    \begin{minipage}[t]{0.32\linewidth}
    \textbf{Invalid Test Case 3} {\small\textit{(Semantically Incorrect)}}
    \begin{Verbatim}[commandchars=\\\{\},fontsize=\scriptsize,frame=single]
3
4
7 2 2 7
8
2 8 1 4 8 2 1 5
5
\badtoken{5 2 1 3 4}
\end{Verbatim}
    \textcolor{red!70!black}{\textbf{Invalid.} All sticks have distinct
    lengths, so no rectangle can be formed, violating the problem's
    guarantee.}
    \end{minipage}
    \endgroup

    \caption{Representative invalid test cases for Codeforces 1027C. Syntactically incorrect cases violate the input grammar itself, while semantically incorrect cases are well-formed inputs that violate problem constraints or guarantees. Each box shows the raw input snippet and highlights why the instance must be rejected by \prespec.}
    \label{fig:case-study-1027c-invalid-testcases}
    \todoanmol{todo, connect to main paper and explain the pipeline}
\end{figure}
\endgroup

After de-duplication and semantic/syntactic filtering, we apply the final test
coverage filter: we remove any problem with fewer than 5 test cases remaining in
any required bucket.

\textbf{Stage 4: Test-case conversion.}
Section~\ref{ssec:data} describes the conversion pipeline in detail, including the agentic loop used to construct the parser $R$ and printer $P$ for each problem and the lossless round-trip check $T_{\mathrm{reproduced}} = P(R(t)) \mathrel{==} t$.
Table~\ref{tab:data-1027c-test-case-format} shows a Codeforces 1027C testcase in
the original text format, alongside the Verus types selected by the construction
agent and the executable Rust values produced by the parser.
Figure~\ref{fig:lossless-conversion-1027c} shows a concrete example for Codeforces 1027C.

\begin{table}[H]
\centering
\small
\setlength{\tabcolsep}{3pt}
\begin{tabular}{p{0.30\textwidth}p{0.35\textwidth}p{0.30\textwidth}}
\toprule
Example Codeforces Test Case & Generated Verus Types & Executable Rust Values \\
\midrule
\begin{minipage}[t]{\linewidth}\raggedright
\textbf{Input:}\\
\texttt{3}\\
\texttt{4}\\
\texttt{7 2 2 7}\\
\texttt{8}\\
\texttt{2 8 1 4 8 2 1 5}\\
\texttt{5}\\
\texttt{5 5 5 5 5}

\medskip
\textbf{Expected Output:}\\
\texttt{2 7 7 2}\\
\texttt{2 2 1 1}\\
\texttt{5 5 5 5}
\end{minipage}
&
\begin{minipage}[t]{\linewidth}\ttfamily\raggedright
pub struct In1 \{\\
\ \ pub ns: Seq<i64>,\\
\ \ pub sticks: Seq<Seq<i64>{}>\\
\}\\
\medskip
pub struct Rectangle \{\\
\ \ pub s1: i64,\\
\ \ pub s2: i64,\\
\ \ pub s3: i64,\\
\ \ pub s4: i64\\
\}\\
\medskip
pub struct Out \{\\
\ \ pub rectangles: \\
\ \ \ \ Seq<Rectangle>\\
\}
\end{minipage}
&
\begin{minipage}[t]{\linewidth}\ttfamily\raggedright
let exec\_in1 = ExecIn1 \{\\
\ \ ns: vec![4, 8, 5],\\
\ \ sticks: vec![\\
\ \ \ \ vec![7, 2, 2, 7],\\
\ \ \ \ ...\\
\ \ ],\\
\};\\

\medskip
let exec\_out = ExecOut \{\\
\ \ rectangles: vec![\\
\ \ \ \ ExecRectangle \{ \\
\ \ \ \ \ \ s1:~2, s2:~7, \\
\ \ \ \ \ \ s3:~7, s4:~2 \\
\ \ \ \ \},\\
\ \ \ \ ...\\
\ \ ],\\
\};
\end{minipage}\\
\bottomrule
\end{tabular}
\caption{From Codeforces 1027C, the left column shows an example test case in the original input and output format. The middle column shows the specification-only Verus data types generated by our conversion pipeline to represent the input and output data; these types are all compatible with \texttt{exec\_spec}. The right column shows the well-typed Rust values generated by our parser from the original test case data (the names of executable Rust types generated by \texttt{exec\_spec} begin with \texttt{Exec}-).}
\label{tab:data-1027c-test-case-format}
\end{table}

Not every test case converts successfully:
the LLM agent may fail to produce a correct parser for some problem formats.
We retain only problems for which at least 5 test cases are successfully
converted and pass the round-trip check in each bucket
($\tau_{\mathrm{pre\text{-}sound}}$, $\tau_{\mathrm{pre\text{-}comp}}$,
$\tau_{\mathrm{post\text{-}sound}}$, $\tau_{\mathrm{post\text{-}comp}}$).
If any bucket exceeds 200 test cases, we randomly sample 200 from it.

\textbf{Stage 5: Final selection.}
From the problems that survive all preceding stages, we sample
\numproblems{} problems for the evaluation set, balancing coverage across
difficulty ratings and topic tags.

\textbf{Final dataset.}
The resulting \vsb dataset contains \numproblems{} problems spanning a range of
difficulty ratings and topics. Additional dataset statistics, including rating,
tag, and test-case-count distributions, are reported in
App.~\ref{app:additional-dataset-statistics}.

%% file: paper_sections/figures_plots/teaser_figure/spec_faithful.tex
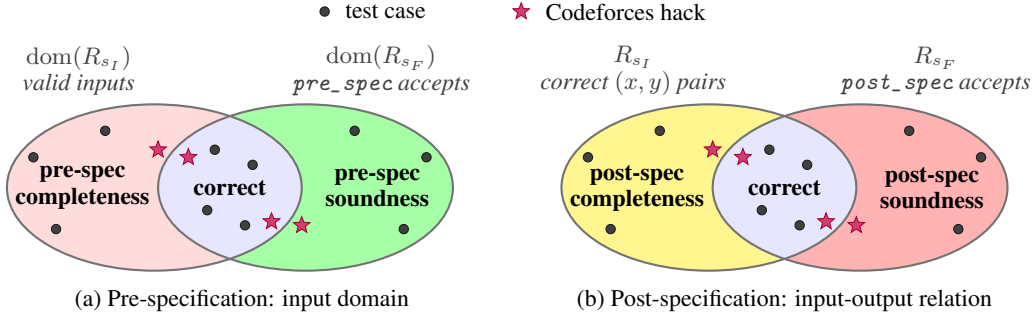
\begin{figure}[H]
    \centering
    \tikzset{
      testpt/.style={circle, fill=black!75, draw=black!90, inner sep=0pt, minimum size=3.5pt},
      hackpt/.style={star, star points=5, star point ratio=2.3, fill=purple!75, draw=purple!90!black, inner sep=0pt, minimum size=6.5pt},
      setname/.style={align=center, font=\small\itshape, color=black!75},
      rlabel/.style={align=center, font=\small\bfseries}
    }

    \begin{tikzpicture}[font=\small, baseline]
      \node[testpt] at (0,0) {};
      \node[anchor=west] at (0.15,0) {test case};
      \node[hackpt] at (2.6,0) {};
      \node[anchor=west] at (2.8,0) {Codeforces hack};
    \end{tikzpicture}\\[2pt]

    \begin{minipage}[t]{0.48\textwidth}
    \centering
    \begin{tikzpicture}[font=\small]
    \fill[pink!55] (-1.0, 0) ellipse (1.95 and 1.10);
    \fill[green!35] ( 1.0, 0) ellipse (1.95 and 1.10);
    \begin{scope}
      \clip (-1.0, 0) ellipse (1.95 and 1.10);
      \clip ( 1.0, 0) ellipse (1.95 and 1.10);
      \fill[blue!10] (-5,-3) rectangle (5,3);
    \end{scope}
    \draw[thick, black!55] (-1.0, 0) ellipse (1.95 and 1.10);
    \draw[thick, black!55] ( 1.0, 0) ellipse (1.95 and 1.10);

    \node[setname] at (-2.0, 1.55) {$\mathrm{dom}(R_{s_I})$\\valid inputs};
    \node[setname] at ( 2.0, 1.55) {$\mathrm{dom}(R_{s_F})$\\\texttt{pre\_spec} accepts};

    \node[rlabel] at (-1.95, 0.0) {pre-spec\\completeness};
    \node[rlabel] at ( 1.95, 0.0) {pre-spec\\soundness};
    \node[rlabel] at (0, 0.0) {correct};

    \foreach \x/\y in {-2.60/0.40, -2.30/-0.55, -1.65/0.75} {
      \node[testpt] at (\x, \y) {};
    }
    \foreach \x/\y in {-0.20/0.50, 0.30/0.30, -0.30/-0.30, 0.20/-0.50} {
      \node[testpt] at (\x, \y) {};
    }
    \foreach \x/\y in {2.60/0.40, 2.30/-0.55, 1.65/0.75} {
      \node[testpt] at (\x, \y) {};
    }

    \node[hackpt] at (-0.95, 0.50) {};   
    \node[hackpt] at (-0.55, 0.40) {};   
    \node[hackpt] at ( 0.55, -0.45) {};  
    \node[hackpt] at ( 0.95, -0.50) {};  
    \end{tikzpicture}\\[2pt]
    {\small (a) Pre-specification: input domain}
    \end{minipage}\hfill
    \begin{minipage}[t]{0.48\textwidth}
    \centering
    \begin{tikzpicture}[font=\small]
    \fill[yellow!55] (-1.0, 0) ellipse (1.95 and 1.10);
    \fill[red!30]    ( 1.0, 0) ellipse (1.95 and 1.10);
    \begin{scope}
      \clip (-1.0, 0) ellipse (1.95 and 1.10);
      \clip ( 1.0, 0) ellipse (1.95 and 1.10);
      \fill[blue!10] (-5,-3) rectangle (5,3);
    \end{scope}
    \draw[thick, black!55] (-1.0, 0) ellipse (1.95 and 1.10);
    \draw[thick, black!55] ( 1.0, 0) ellipse (1.95 and 1.10);

    \node[setname] at (-2.0, 1.55) {$R_{s_I}$\\correct $(x,y)$ pairs};
    \node[setname] at ( 2.0, 1.55) {$R_{s_F}$\\\texttt{post\_spec} accepts};

    \node[rlabel] at (-1.95, 0.0) {post-spec\\completeness};
    \node[rlabel] at ( 1.95, 0.0) {post-spec\\soundness};
    \node[rlabel] at (0, 0.0) {correct};

    \foreach \x/\y in {-2.60/0.40, -2.30/-0.55, -1.65/0.75} {
      \node[testpt] at (\x, \y) {};
    }
    \foreach \x/\y in {-0.20/0.50, 0.30/0.30, -0.30/-0.30, 0.20/-0.50} {
      \node[testpt] at (\x, \y) {};
    }
    \foreach \x/\y in {2.60/0.40, 2.30/-0.55, 1.65/0.75} {
      \node[testpt] at (\x, \y) {};
    }

    \node[hackpt] at (-0.95, 0.50) {};
    \node[hackpt] at (-0.55, 0.40) {};
    \node[hackpt] at ( 0.55, -0.45) {};
    \node[hackpt] at ( 0.95, -0.50) {};
    \end{tikzpicture}\\[2pt]
    {\small (b) Post-specification: input-output relation}
    \end{minipage}
    \caption{\textbf{Conceptual picture of faithfulness evaluation.} A faithful specification matches informal intent on \textbf{(a)}~the input domain and \textbf{(b)}~the input-output relation. The asymmetric differences correspond to the four evaluation categories (\PreSpecCompleteness{}, \PreSpecSoundness{}, \PostSpecCompleteness{}, \PostSpecSoundness{}). Hacks ($\star$) add boundary cases where permissive or restrictive specifications can fail.
    }
\todoanmol{PDF note p.~5: Make sure this figure is referred to somewhere; it is not super clear what the figure is doing.}
    \label{fig:setdiagram}
    \end{figure}

%% file: paper_sections/appendix/additional_dataset_statistics.tex
\section{Additional Dataset Statistics}
\label{app:additional-dataset-statistics}


We summarize key statistics of the \vsb dataset below.

\begin{figure}[H]
    \centering
    \includegraphics[width=\linewidth]{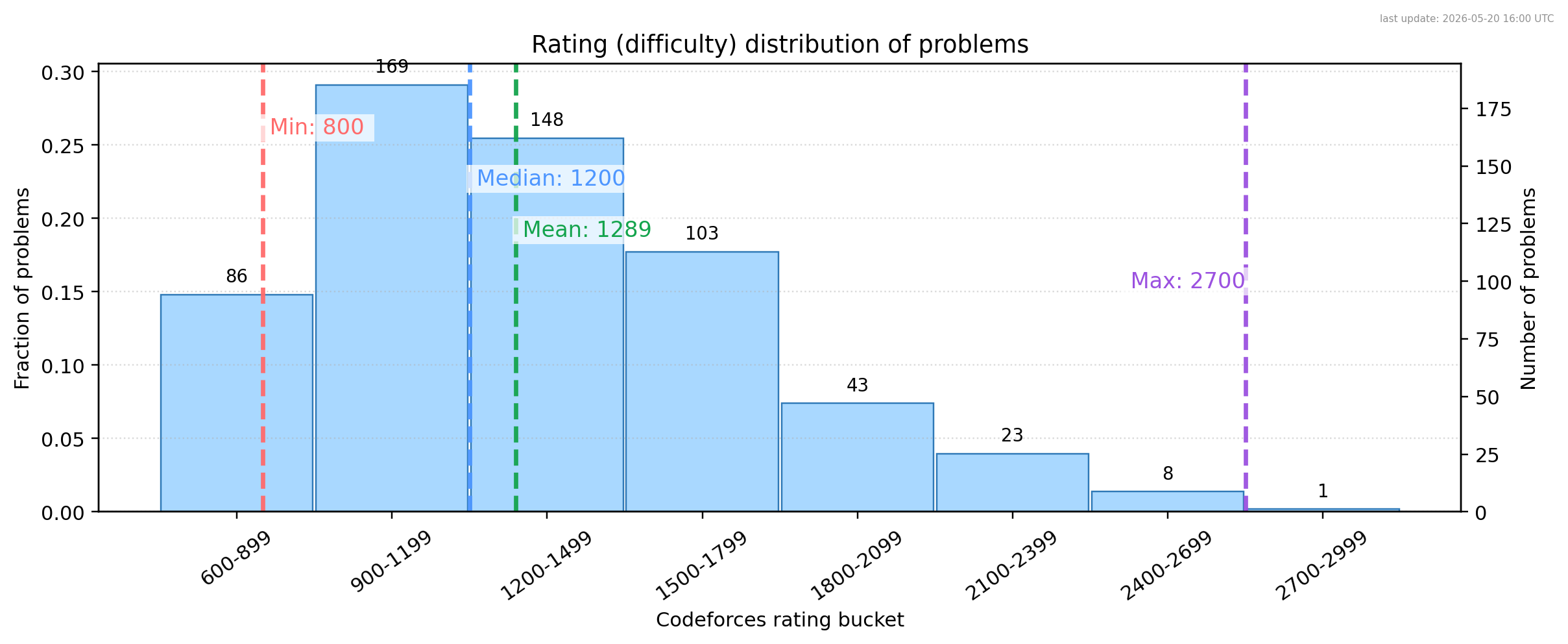}
    \caption{Codeforces rating distribution of \vsb problems. Ratings range from 800 to 2700, with a median of 1200 and a mean of 1289. The majority of problems fall in the 900--1499 range, with diminishing but non-trivial representation at higher difficulty levels.}
    \label{fig:dataset-rating-distribution}
\end{figure}

\textbf{Difficulty.}
Figure~\ref{fig:dataset-rating-distribution} shows the Codeforces rating
distribution. Problems span ratings from 800 (easiest) to 2700 (hardest), with
a median of 1200 and a mean of 1289.

\begin{figure}[H]
    \centering
    \includegraphics[width=\linewidth]{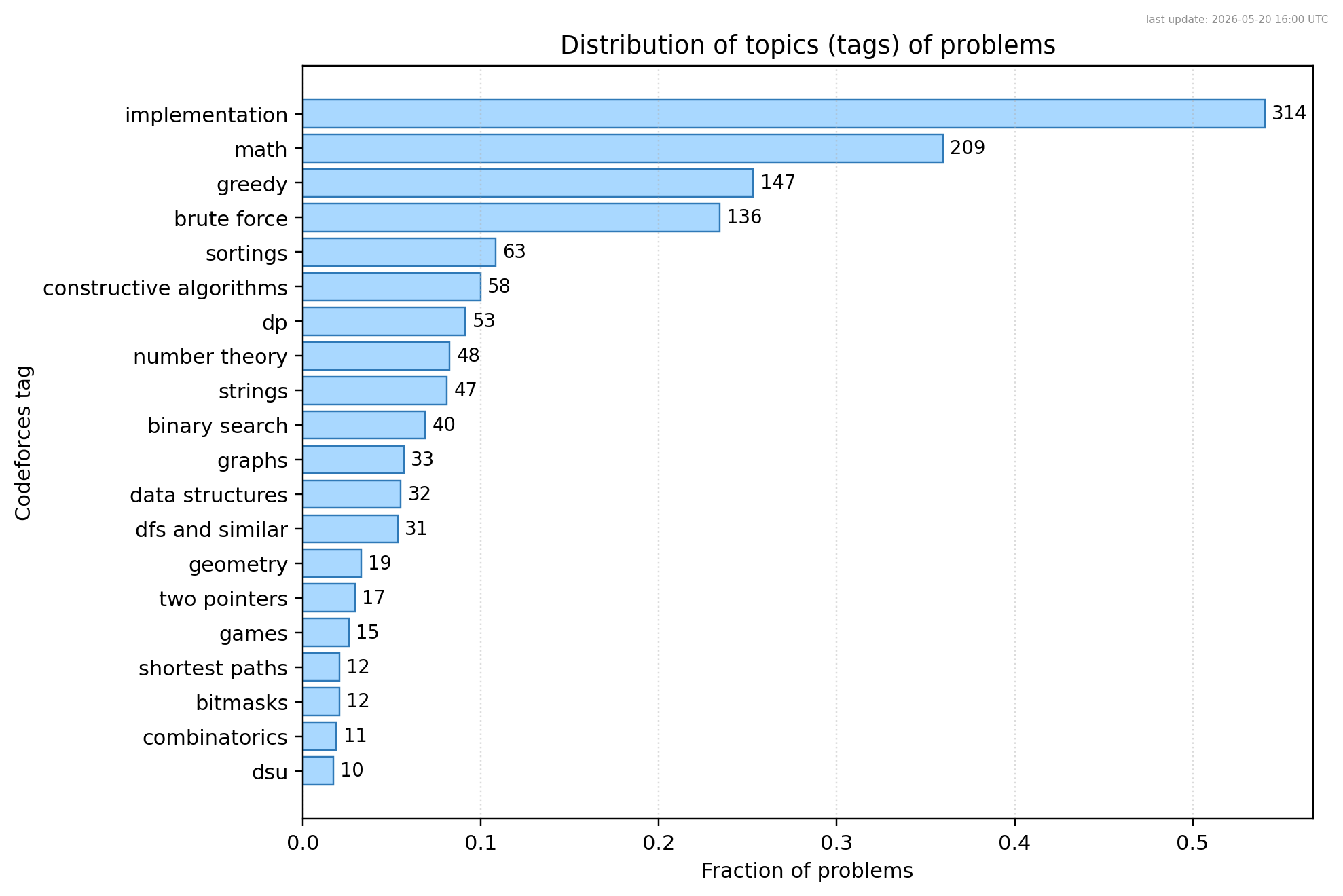}
    \caption{Distribution of Codeforces topic tags across \vsb problems (a problem may carry multiple tags). The benchmark covers a broad range of competitive programming topics, from implementation-heavy tasks to algorithmic problems involving dynamic programming, graph search, and number theory.}
    \label{fig:dataset-tag-distribution}
\end{figure}

\textbf{Topic coverage.}
Figure~\ref{fig:dataset-tag-distribution} shows the distribution of Codeforces
topic tags (problems may carry multiple tags).

\input{paper_sections/figures_plots/1027c_informal_spec}

%% file: paper_sections/figures_plots/1027c_informal_spec.tex
\begin{figure}[t]
\begin{tcolorbox}[colback=blue!3,colframe=blue!40,fontupper=\scriptsize,title={\textbf{Informal specification (natural language) $s_I$ for Codeforces 1027-C}}]
\textbf{Problem link.}\quad \url{https://codeforces.com/problemset/problem/1027/C}
\hfill{\scriptsize\textit{2s / 256MB}}

\medskip
You have $n$ sticks of the given lengths.
Your task is to choose exactly four of them in such a way that they can form a rectangle.
No sticks can be cut to pieces, each side of the rectangle must be formed by a single stick.
No stick can be chosen multiple times.
It is guaranteed that it is always possible to choose such sticks.

Let $S$ be the area of the rectangle and $P$ be the perimeter of the rectangle.
The chosen rectangle should have the value $\frac{P^2}{S}$ minimal possible.
The value is taken without any rounding.
If there are multiple answers, print any of them.
Each testcase contains several lists of sticks, for each of them you are required to solve the problem separately.

\medskip
\textbf{Input.}\quad
The first line contains a single integer $T$ ($T \ge 1$) — the number of lists of sticks in the testcase.
Then $2T$ lines follow — lines $(2i - 1)$ and $2i$ of them describe the $i$-th list.
The first line of the pair contains a single integer $n$ ($4 \le n \le 10^6$) — the number of sticks in the $i$-th list.
The second line of the pair contains $n$ integers $a_1, a_2, \dots, a_n$ ($1 \le a_j \le 10^4$) — lengths of the sticks in the $i$-th list.
It is guaranteed that for each list there exists a way to choose four sticks so that they form a rectangle.
The total number of sticks in all $T$ lists doesn't exceed $10^6$ in each testcase.

\medskip
\textbf{Output.}\quad
Print $T$ lines. The $i$-th line should contain the answer to the $i$-th list of the input.
That is the lengths of the four sticks you choose from the $i$-th list, so that they form a rectangle and the value $\frac{P^2}{S}$ of this rectangle is minimal possible.
You can print these four lengths in arbitrary order.
If there are multiple answers, print any of them.

\medskip
\textbf{Example.}

\textbf{Input}
\begin{flushleft}\ttfamily
3\\
4\\
7 2 2 7\\
8\\
2 8 1 4 8 2 1 5\\
5\\
5 5 5 5 5
\end{flushleft}
\textbf{Output}
\begin{flushleft}\ttfamily
2 7 7 2\\
2 2 1 1\\
5 5 5 5
\end{flushleft}

\medskip
\textbf{Note.}\quad
There is only one way to choose four sticks in the first list, they form a rectangle with sides $2$ and $7$, its area is $2 \cdot 7 = 14$, perimeter is $2(2 + 7) = 18$. $\frac{18^2}{14} \approx 23.143$.
The second list contains subsets of four sticks that can form rectangles with sides $(1, 2)$, $(2, 8)$ and $(1, 8)$. Their values are $\frac{6^2}{2} = 18$, $\frac{20^2}{16} = 25$ and $\frac{18^2}{8} = 40.5$, respectively. The minimal one of them is the rectangle $(1, 2)$.
You can choose any four of the $5$ given sticks from the third list, they will form a square with side $5$, which is still a rectangle with sides $(5, 5)$.
\end{tcolorbox}

\centering
\caption{Informal specification $s_I$, i.e., the natural-language description, for Codeforces 1027C (\url{https://codeforces.com/problemset/problem/1027/C}).}
\label{fig:data-1027c-informal-desc}
\end{figure}

%% file: paper_sections/appendix/additional_insights.tex
\section{Additional Insights}
\label{app:additional-insights}

\textbf{Evaluation hyperparameters.}
\label{app:eval-hyperparams}
Each agent is given a budget of \$2.5 per problem.
For \sweagent, we additionally impose a limit of 400 API calls per problem.
The base agent timeout is 75 minutes per problem.
All experiments run on a combination of Modal cloud and Docker containers, with 4~CPUs and 8\,GB RAM per task.
Budgeted agent evaluations also depend on practical API-level details.
First, API latency affects how many interaction rounds an agent can complete
before the 75-minute wall-clock timeout.
Second, dollar-budget consumption depends partly on prompt-cache behavior:
providers differ in how they price cached tokens, and cache-hit rates can vary
depending on how long a cache remains warm.

\begin{figure}[H]
    \centering
    \includegraphics[width=\linewidth]{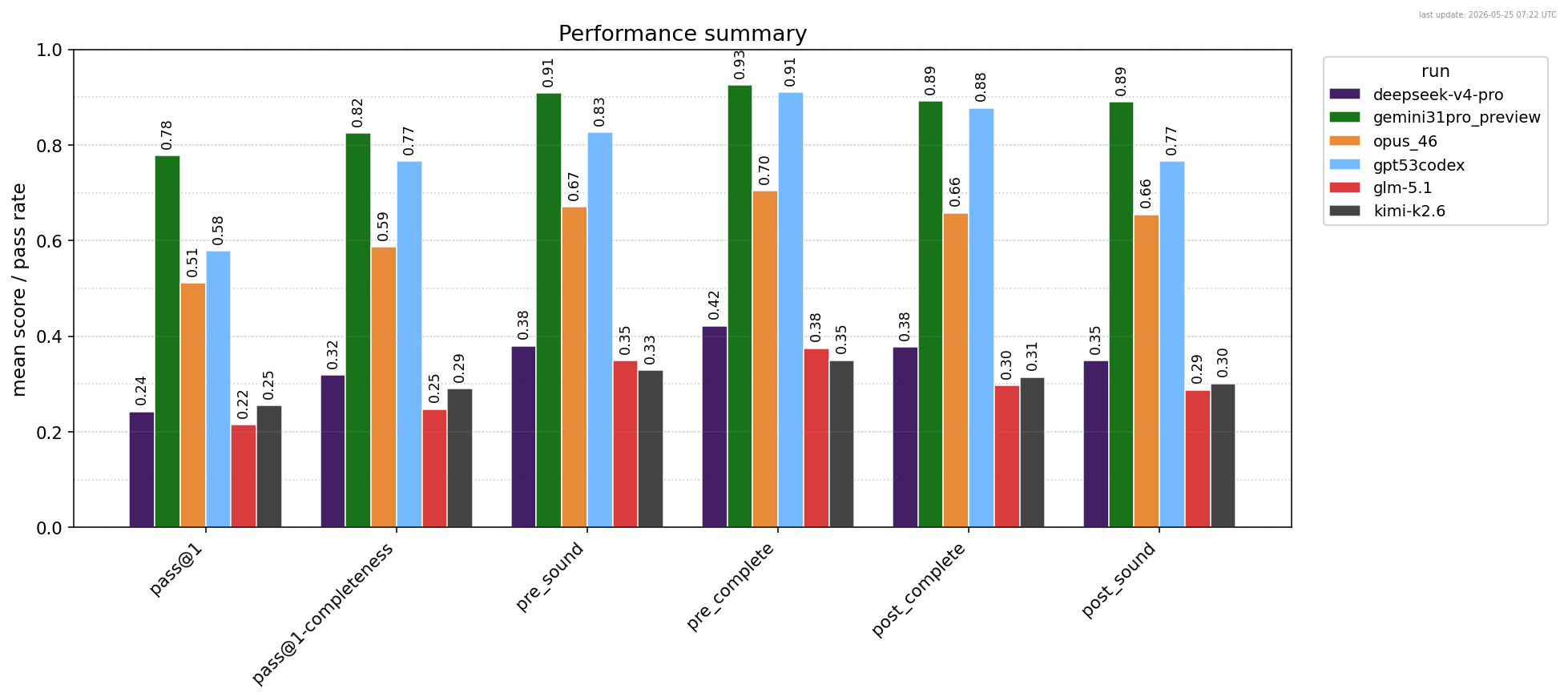}
    \caption{Summary of Pass@1 and testcase-bucket performance across evaluated models.}
    \label{fig:score-summary}
\end{figure}

\subsection{Does problem difficulty affect specification-generation success?}
\label{app:insights-difficulty}

\begin{figure}[H]
    \centering
    \includegraphics[width=\linewidth]{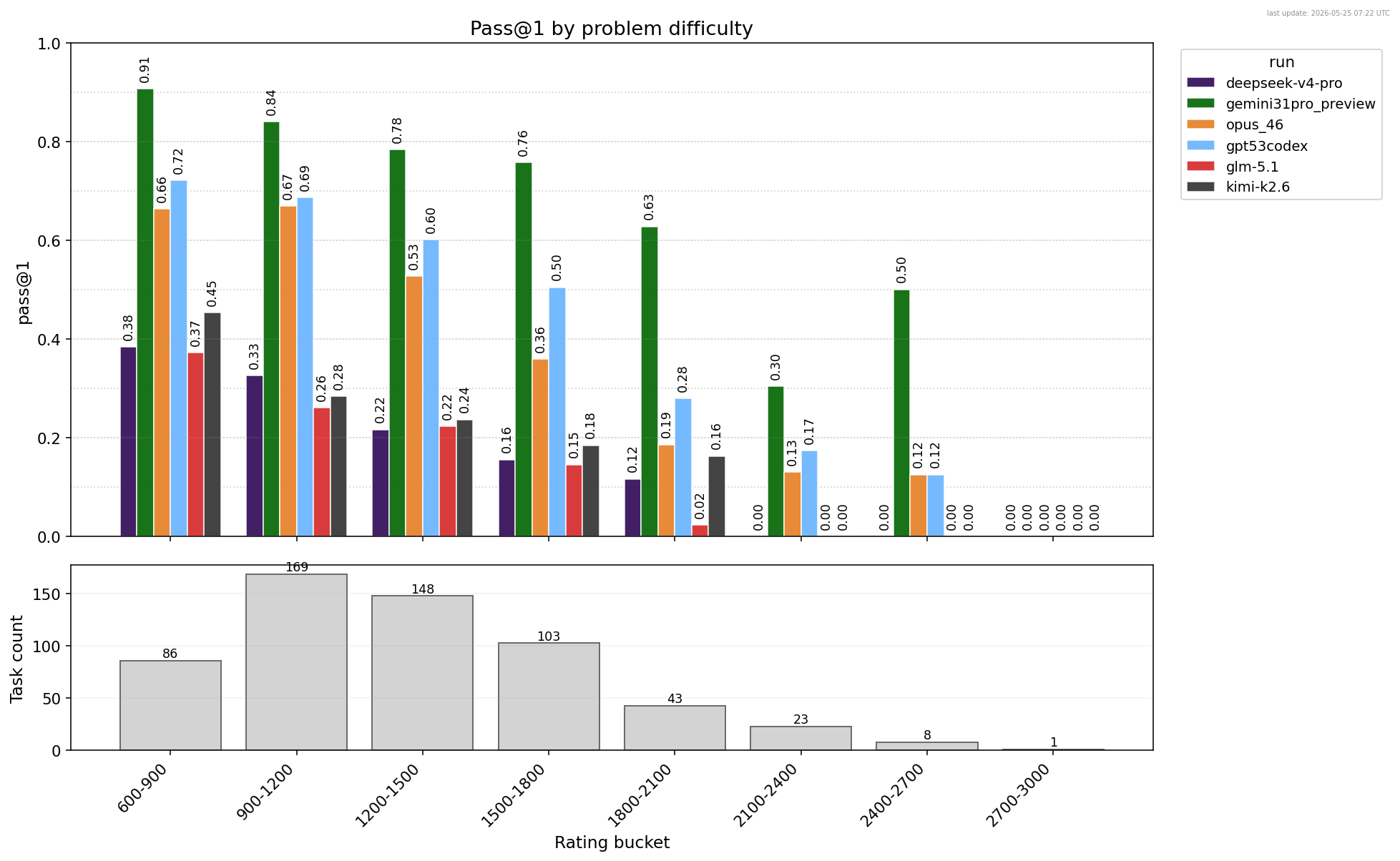}
    \caption{Pass@1 by Codeforces problem difficulty rating (top) and number of problems per rating bucket (bottom). Performance degrades consistently with difficulty for all models. Even on easy problems (600--900), the best model (\geminithreeonepro) solves only 90\%.}
    \label{fig:pass-at-one-by-rating}
\end{figure}

Figure~\ref{fig:pass-at-one-by-rating} breaks down Pass@1 by Codeforces
difficulty rating. Performance degrades consistently as problems become harder:
\geminithreeonepro drops from 0.90 on the easiest bucket (600--900) to 0.62 at
1800--2100 and 0.50 at 2400--2700, while \gptfivethreecodex falls from 0.73 to
0.27 over the same range. Open-source models follow the same trend but at a
lower baseline, with all models scoring near zero on the hardest problems
(2100+). The bottom panel shows the task count per bucket; the hardest
buckets contain few problems, so their estimates are noisier.

Notably, even the easiest problems are far from solved: the best model reaches
0.90 at 600--900 and 0.84 at 900--1200, indicating that roughly 10--16\% of
easy problems still defeat frontier agents. This suggests that specification
autoformalization poses challenges beyond algorithmic difficulty---even
straightforward problems can have subtle input constraints or edge cases that
are hard to formalize completely.

\subsection{How are testcases resolved, and how does \texorpdfstring{\execspec}{exec\_spec} help?}
\label{app:insights-failed-testcase-verdicts}

Figure~\ref{fig:testcase-journey} shows the decision tree used to resolve one
testcase.
For each testcase, the evaluator first asks Verus to prove the relevant
assertion symbolically.
If symbolic reasoning is inconclusive, the evaluator falls back to \execspec{}:
the specification is compiled into executable Rust and run on the concrete
testcase.
This process assigns each testcase to one of six categories:
\emph{compile/syntax error}, \emph{symbolic correct},
\emph{symbolic incorrect}, \emph{exec-spec correct},
\emph{exec-spec incorrect}, or \emph{indeterminate}.
The ``correct'' categories mean the resolved accept/reject verdict matches the
testcase bucket; the ``incorrect'' categories mean it disagrees.
An \emph{indeterminate} testcase is one where neither symbolic reasoning nor
runtime execution produced a conclusive verdict, for example because execution
timed out or exceeded memory limits.

\begin{figure}[H]
    \centering
    \includegraphics[width=\linewidth]{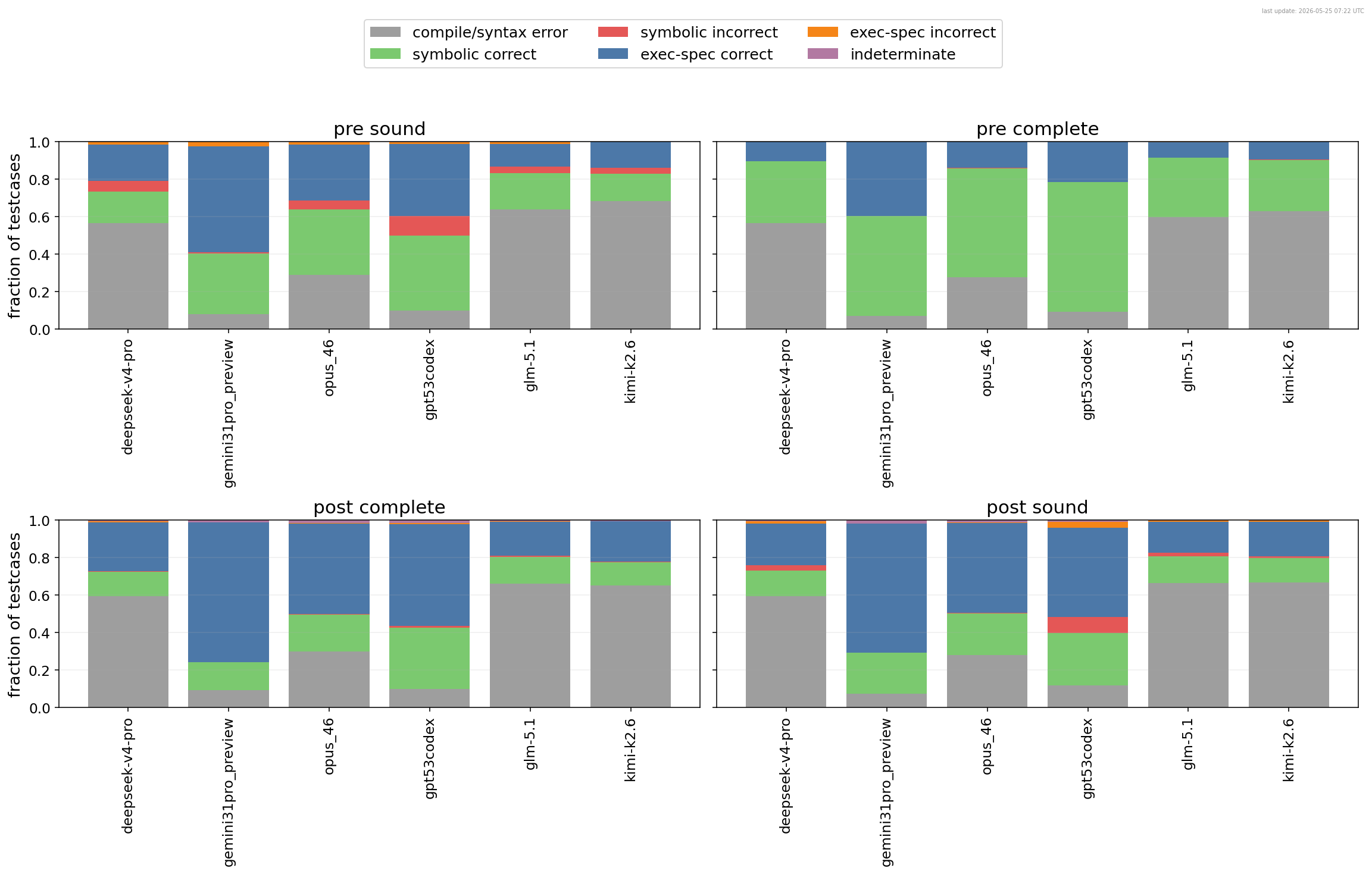}
    \caption{Distribution of testcase resolution outcomes across evaluated runs, broken down by evaluation bucket. Blue regions show cases where symbolic reasoning was inconclusive but \execspec{} produced a deterministic verdict. Compile/syntax errors (gray) dominate weaker models, while stronger models resolve most testcases via \execspec{} or symbolic checking. The \execspec{} fallback is especially important for postcondition buckets, where output relations are too complex for symbolic reasoning alone.}
    \label{fig:resolution-distribution}
\end{figure}

Figure~\ref{fig:resolution-distribution} shows the resulting breakdown for
each bucket.
The first pattern is syntactic: stronger models produce specifications that the
evaluator can actually analyze.
\geminithreeonepro{} and
\gptfivethreecodex{} have relatively small compile/syntax-error fractions
across all four buckets, while \deepseekvfourpro{}, \glmfiveone{}, and
\kimiktwosix{} are dominated by compile/syntax errors in several buckets.
Thus, part of the gap between models is not only semantic precision, but also
the ability to stay inside the Verus and \execspec{} fragment used by the
benchmark.

Among the analyzable testcases, \execspec{} is especially important for
postcondition checking. In the \postcomplete{} and \postsound{} buckets,
\geminithreeonepro{} and \gptfivethreecodex{} resolve a large fraction of
testcases via the executable fallback, whereas purely symbolic resolution
accounts for a smaller share. Without \execspec{}, those blue segments would
largely have remained indeterminate under symbolic checking alone. This is
consistent with the intuition that postconditions often involve richer output
relations that are hard for SMT to prove directly, but can still be evaluated
deterministically on concrete testcases. By contrast, the \precomplete{} bucket
contains a larger share of symbolic-correct verdicts, especially for
\geminithreeonepro{}, \opusfoursix{}, and \gptfivethreecodex{}, indicating that
many input-validity constraints are simple enough for symbolic checking to
discharge directly.

Incorrect resolved verdicts are visible but generally smaller than the correct
portions.
The clearest exception is \gptfivethreecodex{} on soundness buckets, where a
nontrivial red band indicates symbolically resolved but incorrect verdicts.
These cases are useful diagnostically.
They mean Verus could prove what the model's specification does on the testcase,
and that proven behavior disagrees with the benchmark's expected verdict.
Thus, these are concrete semantic failures in the generated specification, not
cases where the evaluator failed to resolve the testcase.

\subsection{Examples of unsuccessful cases}
\label{app:insights-unsuccessful-cases}

We highlight representative failure modes surfaced by our case studies
(App.~\ref{app:case-study-1027c}--\ref{app:case-study-2074d}).

\textbf{Excessively weak preconditions (pre-soundness).}
On Problem~1028C (App.~\ref{app:case-study-1028c}), both \gptfivethreecodex{}
and \geminithreeonepro{} check per-rectangle coordinate constraints but omit the
global promise that at least $n{-}1$ of the $n$ input rectangles share a common
point. This missing property is not merely a surface-level omission: it is the
key property needed to prove that the output is correct, so its absence
would block any downstream verification effort.
On Problem~1027C (App.~\ref{app:case-study-1027c}), \kimiktwosix{} accepts
inputs that violate the guarantee that the given side lengths can form
rectangles, again weakening the precondition from what is needed to solve the problem.

\textbf{Excessively weak postconditions (post-soundness).}
On Problem~1051B (App.~\ref{app:case-study-1051b}), \gptfivethreecodex{}
checks that paired elements are not both even rather than verifying
$\gcd = 1$, so the specification incorrectly accepts pairs such as $(3,6)$.
On Problem~1027C (App.~\ref{app:case-study-1027c}), \gptfivethreecodex{} checks that the output rectangle is
feasible (valid side lengths) but omits the optimality criterion (minimizing
$P^2/S$), allowing suboptimal outputs to pass. \kimiktwosix{} on the same
problem only checks the output shape (two pairs of equal values) without
verifying that the lengths originate from the input or that the rectangle is
optimal.
Such weak postconditions are particularly dangerous: they could allow incorrect code to be formally verified, silently undermining the guarantees that formal verification is meant to provide.

\textbf{Excessively strong postconditions (post-completeness).}
On Problem~2074D (App.~\ref{app:case-study-2074d}), \geminithreeonepro{}
writes an overly complex interval-union specification for the output that
rejects correct solutions. In contrast, \opusfoursix{} succeeds on the same
problem with a simpler column-wise characterization, suggesting that
over-specification is itself a failure mode distinct from under-specification.


\subsection{\texorpdfstring{pass@$k$ and pass$^k$}{pass@k and pass\^k}}
\label{app:insights-pass-at-k}
We evaluated \gptfivethreecodex{} three times on the full set of
\numproblems{} problems.
Here, a problem is ``solved'' if the generated specification passes all
testcase buckets, not merely if the model can solve the original programming
problem.
Across the three independent runs, the model succeeds on 0.578, 0.559, and
0.566 fraction of problems, respectively.
The union across runs is substantially larger: pass@3 is 0.756, with 439 of
\numproblems{} problems solved by at least one attempt.
This gap shows that repeated sampling can recover many additional correct
specifications.

Conversely, only 202 of \numproblems{} problems (34.8\%) are solved by
\emph{all three} attempts (pass$^3$).
The wide spread between pass@3 (75.6\%) and pass$^3$ (34.8\%) shows that
specification generation with \gptfivethreecodex{} is still brittle: even when
the model can solve a problem, it often does so unreliably across attempts.

\subsection{Do models solve the same problems or different problems?}
\label{app:insights-same-problems}

\begin{figure}[H]
    \centering
    \includegraphics[width=\linewidth]{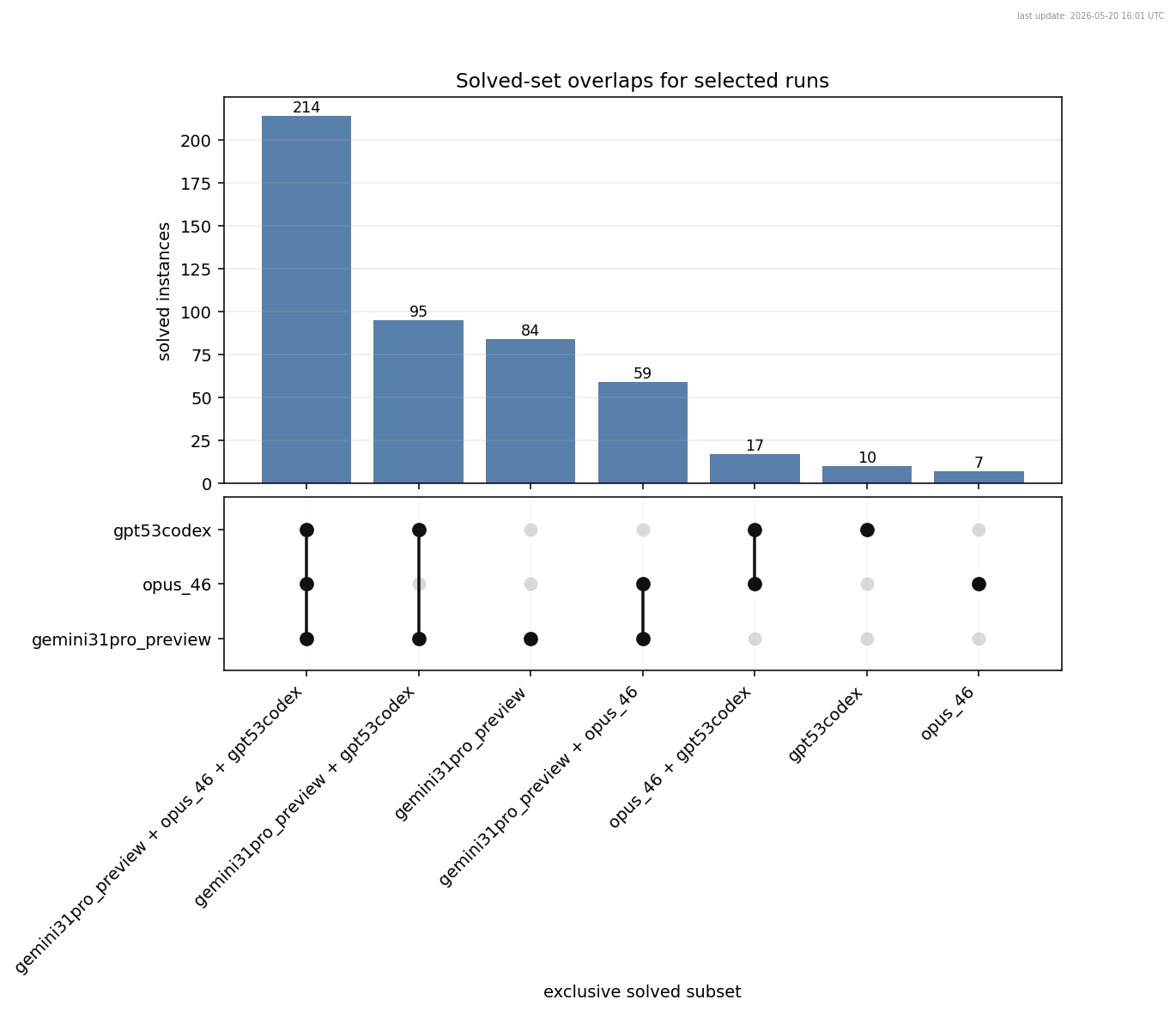}
    \caption{UpSet plot of solved-problem overlaps among the three closed-source models. Each bar shows the number of problems solved \emph{exclusively} by the indicated model subset. 214 problems are solved by all three models; the remaining bars reveal substantial complementarity, with \geminithreeonepro contributing the most unique solves (84).}
    \label{fig:solved-overlap}
\end{figure}

Figure~\ref{fig:solved-overlap} shows an UpSet plot of the solved-problem sets
for the three closed-source models, where each bar counts problems solved
\emph{exclusively} by the indicated subset.
A shared core of 214 problems is solved by all three models, but
beyond this core the models diverge: \geminithreeonepro contributes the most
unique solves, while \opusfoursix's solved set is largely subsumed by the other
two. The union across all three models covers 486 of \numproblems{} problems,
substantially more than any single model, suggesting that
best-of-$k$ selection across models could yield meaningful gains.

\subsection{Do models find code-generation easier than specification generation?}
\label{app:insights-codegen-aspect}
\todoanmol{PDF note p.~30: Anmol to add examples.}
\todoanmol{PDF note p.~32: Add examples about which testcases were used to evaluate the code-generation part.}
This analysis asks whether failed specifications are simply a byproduct of the
model not understanding the underlying programming problem.
They often are not.
Our data collection pipeline lets us measure code-generation performance only
on problems for which each input has a unique correct output.
This uniqueness condition is important: when multiple outputs are valid,
testcase-based evaluation may incorrectly reject a program that returns a
different valid output.
Overall, 489 problems in our benchmark satisfy this condition.

In our benchmark, \gptfivethreecodex{} writes incorrect specifications for 245
problems.
Of these, 197 have exactly one correct output for every input, so we can test
whether the same model can solve the underlying programming task by generating
Python code and checking it against the available testcases.
Among the 187 uniquely evaluable failed-specification problems with available
code-generation runs, \gptfivethreecodex{} solves 153, a code-generation
success rate of 81.8\%.
Thus, many failures are not caused by the model being unable to solve the
programming problem.
In a large fraction of cases, the model can produce correct executable code
while still failing to write a correct formal specification.

\subsection{Can LLMs evaluate specifications in a judge setting?}
\label{app:insights-llm-judge}
\todoanmol{PDF note p.~31: Anmol to add examples of LLM judge failures of the reasoning.}
We asked \gptfivethreecodex{} to act as a judge for specifications generated by
the same model.
For each judgment, we prompted the model with the informal Codeforces problem
statement, a few original Codeforces testcases together with their
representations in Verus data structures, and the generated specification to be
evaluated.
The judge then classified the specification as correct or incorrect.

The results of the LLM-as-a-judge experiment are shown in Table~\ref{tab:codex-judge-confusion}.
We focus on the 527 problems where \gptfivethreecodex{} produced a
compile-clean specification.
Our evaluator accepts 336 of these specifications and rejects the remaining
191.
Among the benchmark-accepted specifications, the \gptfivethreecodex{} judge
marks 310 as correct and 26 as incorrect.
Among the benchmark-rejected specifications, the judge marks 49 as correct and
142 as incorrect.
The main failure mode is false acceptance: the judge accepts 49 specifications
for which our evaluator finds at least one concrete testcase that disagrees
with the informal problem intent.
This is why testcase-based evaluation is useful even when the judge model is
strong.

\begin{table}[t]
\centering
\small
\setlength{\tabcolsep}{7pt}
\renewcommand{\arraystretch}{1.15}
\caption{Confusion matrix for \gptfivethreecodex{} used as an LLM judge on compile-clean semantic cases. Columns indicate correctness as approximated by benchmark testcases; rows indicate correctness according to the Codex judge. The 54 benchmark-incorrect specifications that did not compile were excluded from this LLM-judge analysis.}
\label{tab:codex-judge-confusion}
\resizebox{\linewidth}{!}{%
\begin{tabular}{lcc}
\toprule
& \textbf{Correct as per Benchmark testcases}
& \textbf{Incorrect as per Benchmark testcases} \\
\midrule
\textbf{Correct as per Codex judge} & 310 (92.3\%) & 49 (25.7\%) \\
\textbf{Incorrect as per Codex judge} & 26 (7.7\%) & 142 (74.3\%) \\
\bottomrule
\end{tabular}%
}
\end{table}

\subsection{Could more testcases have helped?}
\label{app:insights-testcase-coverage}

To study whether our test-case budget is sufficient, we estimate what would
happen under a smaller budget that keeps at most $m$ test cases per bucket.
This is an exact retrospective calculation under uniform subsampling, rather
than a new benchmark run.
For each bucket, this corresponds to uniformly sub-sampling
$k=\min(m,T)$ test cases without replacement, where $T$ is the total number of
available tests in the bucket.  If $P$ of the $T$ tests pass, the probability
that the sub-sample catches at least one failure is
$1-\binom{P}{k}/\binom{T}{k}$, with the natural boundary cases when $k=0$ or
$P<k$.  For an entire problem, we combine the bucket-level probabilities under
the assumption that the bucket samples are independent.

\begin{figure}[H]
    \centering
    \includegraphics[width=\linewidth]{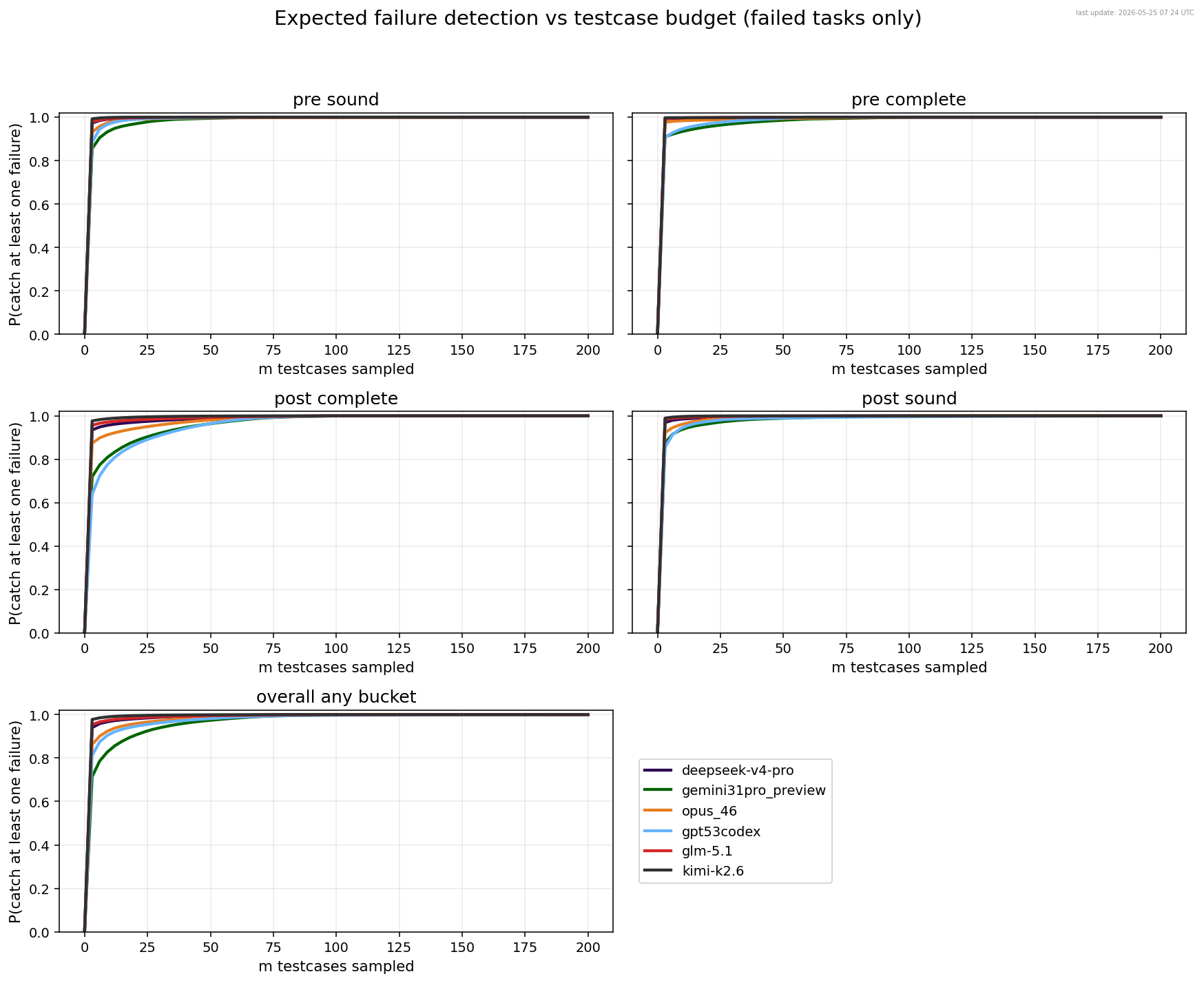}
    \caption{Estimated probability of catching at least one specification failure as a function of the test-case budget $m$, restricted to tasks that fail at the full test-case budget. Conditional on a failure existing, a small number of tests usually detects it; postcondition completeness has the slowest saturation, suggesting a broader diversity of edge cases in that bucket.}
    \label{fig:test-budget-catch-probability-failed-only}
\end{figure}

Figure~\ref{fig:test-budget-catch-probability-failed-only} shows this quantity
conditioned on tasks that fail under the full test suite.
Given that a specification fails under the full suite, the probability of
detecting a failure is already very high with a small budget: most buckets
reach near-saturation within roughly the first few dozen tests.
This is expected: the first few sampled testcases have high marginal value,
because different incorrect specifications often fail for different edge-case
patterns, and a small number of diverse tests can already cover many distinct
failure modes.  After this initial region, the curves flatten: additional tests
are often variations of failure modes already represented in the sample, so each
extra test case contributes less new discriminatory power.
The slowest panel is postcondition completeness, where some models continue to
gain until around $m \approx 50$--$75$; this suggests that postcondition
completeness contains a wider variety of edge cases than the other buckets.
Even there, the curve is mostly flat long before the largest available
test-case counts.

\begin{figure}[H]
    \centering
    \includegraphics[width=\linewidth]{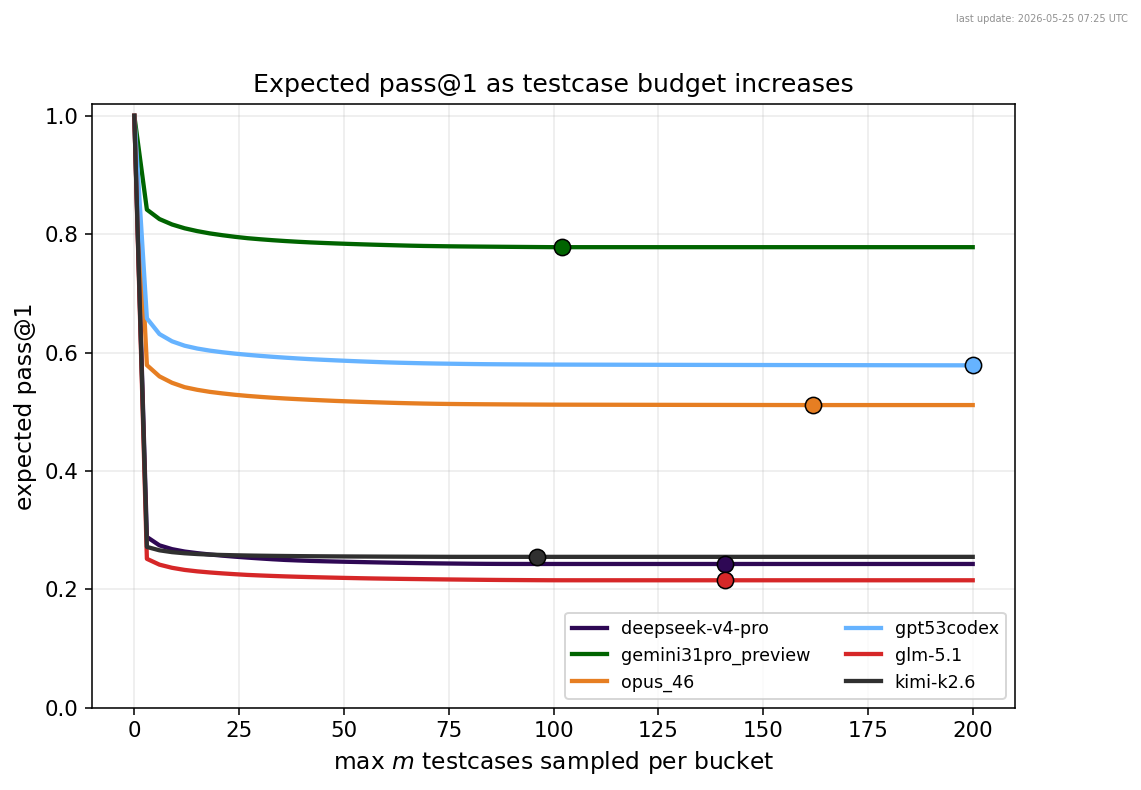}
    \caption{Expected Pass@1 as a function of the maximum test-case budget $m$ per bucket. With very few test cases, all models appear artificially strong; scores drop sharply as the first diverse tests expose distinct failure modes, then flatten as additional tests become increasingly redundant. Dots mark the first sampled $m$ where each curve has converged to its terminal plotted value.}
    \label{fig:pass-at-one-vs-test-budget}
\end{figure}

Figure~\ref{fig:pass-at-one-vs-test-budget} shows the same calculation from
the model-score perspective.
At $m=0$, every submission would pass vacuously, so all curves start at 1.
Increasing $m$ produces a sharp initial drop because the first few tests expose
many distinct bugs.
After roughly $m=25$--$50$, the curves enter a long, shallow tail: adding more
tests still catches a few additional failures, but the marginal effect is small
because many remaining tests repeat already-covered failure patterns.
The dots mark the first sampled budget $m$ at which the plotted curve has
converged to its terminal value in our grid.
They are therefore empirical convergence points for this retrospective
calculation, not the actual number of tests available in the benchmark.


%% file: paper_sections/appendix/failure_modes.tex
\caseStudyBarrier

\section{Case studies of LLM performance on specific tasks}
\label{app:failure-modes}

\input{paper_sections/appendix/case_study_latex/problem_1051_B}

\caseStudyBarrier

\input{paper_sections/appendix/case_study_latex/problem_1027__C}

\caseStudyBarrier

\input{paper_sections/appendix/case_study_latex/problem_1028_C}

\caseStudyBarrier

\input{paper_sections/appendix/case_study_latex/problem_2074_D}

\caseStudyBarrier

%% file: paper_sections/appendix/case_study_latex/problem_1051_B.tex
\subsection{Case Study: Problem 1051B}
\label{app:case-study-1051b}

On this problem, \gptfivethreecodex{} fails to write a faithful specification.
The task is simple to check from a programming-contest perspective: a candidate
output should say \texttt{YES} and then provide a perfect matching of all integers
in the interval, with every pair relatively prime. The failures below show a
generated specification that captures a plausible necessary condition for
relative primality, but misses the actual gcd condition.

\begin{tcolorbox}[colback=blue!3,colframe=blue!40,fontupper=\scriptsize,title={\textbf{Informal specification (natural language) $s_I$ for Codeforces 1051-B}}]
\textbf{Problem link.}\quad
\url{https://codeforces.com/problemset/problem/1051/B}

\medskip
You are given a set of all integers from $l$ to $r$ inclusive, $l < r$,
$(r - l + 1) \le 3 \cdot 10^5$ and $(r - l)$ is always odd.

You want to split these numbers into exactly $\frac{r - l + 1}{2}$ pairs in
such a way that for each pair $(i, j)$ the greatest common divisor of $i$ and
$j$ is equal to $1$. Each number should appear in exactly one of the pairs.

Print the resulting pairs or output that no solution exists. If there are multiple solutions, print any of them.

\medskip
\textbf{Input.}

The only line contains two integers $l$ and $r$ ($1 \le l < r \le 10^{18}$, $r - l + 1 \le 3 \cdot 10^5$, $(r - l)$ is odd).

\medskip
\textbf{Output.}

If any solution exists, print ``YES'' in the first line. Each of the next $\frac{r - l + 1}{2}$ lines should contain some pair of integers. GCD of numbers in each pair should be equal to $1$. All $(r - l + 1)$ numbers should be pairwise distinct and should have values from $l$ to $r$ inclusive.

If there are multiple solutions, print any of them.

If there exists no solution, print ``NO''.

\medskip
\textbf{Example.}
\begin{flushleft}\ttfamily
Input:\\
1 8\\[4pt]
Output:\\
YES\\
2 7\\
4 1\\
3 8\\
6 5
\end{flushleft}
\end{tcolorbox}

\textbf{Codex-generated postcondition: weaker than relative primality.}
\gptfivethreecodex{} makes a mistake in its specification (Listing~\ref{lst:case-study-1051b-codex}) that is revealed by the post-soundness hacks given in Table~\ref{tab:case-study-1051b-hacks}. Instead of checking
\texttt{gcd(a,b) == 1}, it checks only that the two numbers are not both even.
That condition is necessary for relative primality, but far from sufficient.
For example, $(3,6)$ satisfies this parity check because one endpoint is odd,
but its gcd is $3$.

\begin{lstlisting}[language=Verus,caption={Snippet from postcondition generated by \gptfivethreecodex() for Codeforces 1051B that fails two post-soundness hack.},label={lst:case-study-1051b-codex}]
exec_spec_unverified! {
    pub open spec fn line_ok(iv: Interval, p: PairLine) -> bool {
        &&& iv.l <= p.a
        &&& p.a <= iv.r
        &&& iv.l <= p.b
        &&& p.b <= iv.r
        &&& p.a != p.b
        &&& (p.a % 2 != 0 || p.b % 2 != 0)
    }

    pub open spec fn post_spec(in1: In1, out: Out) -> bool {
        &&& pre_spec(in1)
        &&& out.is_yes
        &&& {
            let iv = in1.intervals[0];
            &&& forall |i: i64|
                0 <= i < out.lines.len() ==>
                    line_ok(iv, out.lines[i as int])
            &&& (out.lines.len() as i128) == (iv.r - iv.l + 1) / 2
            &&& forall |i: i64, j: i64|
                0 <= i < out.lines.len() && i < j < out.lines.len() ==>
                    disjoint_lines(out.lines[i as int], out.lines[j as int])
        }
    }
}
\end{lstlisting}

On both \texttt{hack\_489486\_\_participant} and
\texttt{hack\_490353\_\_participant}, the invalid output contains the pair
$(3,6)$. \gptfivethreecodex{}'s specification accepts this pair because
\texttt{3 \% 2 != 0}, so the evaluator observes a post-soundness failure:
the generated specification permits an output that Codeforces rejects.

\begin{table*}
\centering
\scriptsize
\setlength{\tabcolsep}{3pt}
\begin{tabular}{p{0.17\textwidth}p{0.20\textwidth}p{0.40\textwidth}p{0.17\textwidth}}
\toprule
Bucket / hack & Original input and output & Verus representation & Commentary \\
\midrule
\begin{minipage}[t]{\linewidth}\raggedright
\texttt{post\_sound}

\medskip
\texttt{hack\_489486}\\
\texttt{\_\_participant}
\end{minipage}
&
\begin{minipage}[t]{\linewidth}\raggedright
\textbf{Input:} \texttt{3 6}

\medskip
\textbf{Output:}
\texttt{YES};
\texttt{4 5};
\texttt{3 6}
\end{minipage}
&
\begin{minipage}[t]{\linewidth}\ttfamily\raggedright
let exec\_in1 = ExecIn1 \{\\
\ \ \ \ intervals: vec![\\
\ \ \ \ \ \ \ \ ExecInterval \{ l: 3, r: 6 \},\\
\ \ \ \ ],\\
\};\\

\medskip
let exec\_out = ExecOut \{\\
\ \ \ \ is\_yes: true,\\
\ \ \ \ lines: vec![\\
\ \ \ \ \ \ \ \ ExecPairLine \{ a: 4, b: 5, count: 2 \},\\
\ \ \ \ \ \ \ \ ExecPairLine \{ a: 3, b: 6, count: 2 \},\\
\ \ \ \ ],\\
\};
\end{minipage}
&
\begin{minipage}[t]{\linewidth}\raggedright
Invalid output. The pair $(3,6)$ has gcd $3$, so the postcondition should reject it. \\
\end{minipage} \\
\midrule
\begin{minipage}[t]{\linewidth}\raggedright
\texttt{post\_sound}

\medskip
\texttt{hack\_490353}\\
\texttt{\_\_participant}
\end{minipage}
&
\begin{minipage}[t]{\linewidth}\raggedright
\textbf{Input:} \texttt{3 6}

\medskip
\textbf{Output:}
\texttt{YES};
\texttt{3 6};
\texttt{4 5}
\end{minipage}
&
\begin{minipage}[t]{\linewidth}\ttfamily\raggedright
let exec\_in1 = ExecIn1 \{\\
\ \ \ \ intervals: vec![\\
\ \ \ \ \ \ \ \ ExecInterval \{ l: 3, r: 6 \},\\
\ \ \ \ ],\\
\};\\

\medskip
let exec\_out = ExecOut \{\\
\ \ \ \ is\_yes: true,\\
\ \ \ \ lines: vec![\\
\ \ \ \ \ \ \ \ ExecPairLine \{ a: 3, b: 6, count: 2 \},\\
\ \ \ \ \ \ \ \ ExecPairLine \{ a: 4, b: 5, count: 2 \},\\
\ \ \ \ ],\\
\};
\end{minipage}
&
\begin{minipage}[t]{\linewidth}\raggedright
Invalid output for the same reason: the pair $(3,6)$ is not relatively prime. \\
\end{minipage} \\
\bottomrule
\end{tabular}
\caption{Raw hack test cases for Codeforces 1051B. Both are post-soundness cases that should be rejected because the output contains the non-coprime pair $(3,6)$.}
\label{tab:case-study-1051b-hacks}
\end{table*}

%% file: paper_sections/appendix/case_study_latex/problem_1027__C.tex
\subsection{Case Study: Problem 1027C}
\label{app:case-study-1027c}

For this case study, some generated specifications capture part of the task, but miss a condition that calls for an optimized output. A correct
output must not only form a rectangle from a list of available sticks; it must also choose a
rectangle minimizing $\frac{P^2}{S}$ among all feasible choices for that list of sticks.

\begin{tcolorbox}[colback=blue!3,colframe=blue!40,fontupper=\scriptsize,title={\textbf{Informal specification (natural language) $s_I$ for Codeforces 1027-C}}]
\textbf{Problem link.}\quad \url{https://codeforces.com/problemset/problem/1027/C}
\hfill{\scriptsize\textit{2s / 256MB}}

\medskip
You have $n$ sticks of the given lengths.
Your task is to choose exactly four of them in such a way that they can form a rectangle.
No sticks can be cut to pieces, each side of the rectangle must be formed by a single stick.
No stick can be chosen multiple times.
It is guaranteed that it is always possible to choose such sticks.

Let $S$ be the area of the rectangle and $P$ be the perimeter of the rectangle.
The chosen rectangle should have the value $\frac{P^2}{S}$ minimal possible.
The value is taken without any rounding.
If there are multiple answers, print any of them.
Each testcase contains several lists of sticks, for each of them you are required to solve the problem separately.

\medskip
\textbf{Input.}\quad
The first line contains a single integer $T$ ($T \ge 1$) — the number of lists of sticks in the testcase.
Then $2T$ lines follow — lines $(2i - 1)$ and $2i$ of them describe the $i$-th list.
The first line of the pair contains a single integer $n$ ($4 \le n \le 10^6$) — the number of sticks in the $i$-th list.
The second line of the pair contains $n$ integers $a_1, a_2, \dots, a_n$ ($1 \le a_j \le 10^4$) — lengths of the sticks in the $i$-th list.
It is guaranteed that for each list there exists a way to choose four sticks so that they form a rectangle.
The total number of sticks in all $T$ lists doesn't exceed $10^6$ in each testcase.

\medskip
\textbf{Output.}\quad
Print $T$ lines. The $i$-th line should contain the answer to the $i$-th list of the input.
That is the lengths of the four sticks you choose from the $i$-th list, so that they form a rectangle and the value $\frac{P^2}{S}$ of this rectangle is minimal possible.
You can print these four lengths in arbitrary order.
If there are multiple answers, print any of them.

\medskip
\textbf{Example.}

\textbf{Input}
\begin{flushleft}\ttfamily
3\\
4\\
7 2 2 7\\
8\\
2 8 1 4 8 2 1 5\\
5\\
5 5 5 5 5
\end{flushleft}
\textbf{Output}
\begin{flushleft}\ttfamily
2 7 7 2\\
2 2 1 1\\
5 5 5 5
\end{flushleft}

\medskip
\textbf{Note.}\quad
There is only one way to choose four sticks in the first list, they form a rectangle with sides $2$ and $7$, its area is $2 \cdot 7 = 14$, perimeter is $2(2 + 7) = 18$. $\frac{18^2}{14} \approx 23.143$.
The second list contains subsets of four sticks that can form rectangles with sides $(1, 2)$, $(2, 8)$ and $(1, 8)$. Their values are $\frac{6^2}{2} = 18$, $\frac{20^2}{16} = 25$ and $\frac{18^2}{8} = 40.5$, respectively. The minimal one of them is the rectangle $(1, 2)$.
You can choose any four of the $5$ given sticks from the third list, they will form a square with side $5$, which is still a rectangle with sides $(5, 5)$.
\end{tcolorbox}

\begin{lstlisting}[language=Verus,caption={Skeleton template for Codeforces 1027C. The LLM fills in \texttt{pre\_spec} and \texttt{post\_spec}; the input/output structures are fixed by the benchmark conversion pipeline.},label={lst:case-study-1027c-template}]
verus! {
    exec_spec_unverified! {
        pub struct In1 {
            pub ns: Seq<i64>,
            pub sticks: Seq<Seq<i64>>,
        }

        pub struct Rectangle {
            pub s1: i64,
            pub s2: i64,
            pub s3: i64,
            pub s4: i64,
        }

        pub struct Out {
            pub rectangles: Seq<Rectangle>,
        }

        pub open spec fn pre_spec(in1: In1) -> bool {
        }

        pub open spec fn post_spec(in1: In1, out: Out) -> bool {
        }
    }
}
\end{lstlisting}

\begin{table*}[t]
\centering
\tiny
\setlength{\tabcolsep}{3pt}
\begin{tabular}{p{0.16\textwidth}p{0.23\textwidth}p{0.39\textwidth}p{0.16\textwidth}}
\toprule
Model / hack & Original input and output & Verus representation & Commentary \\
\midrule
\begin{minipage}[t]{\linewidth}\raggedright
\gptfivethreecodex{}

\medskip
\texttt{post\_sound}

\medskip
\texttt{hack\_477544}\\
\texttt{\_\_participant}
\end{minipage}
&
\begin{minipage}[t]{\linewidth}\raggedright
The full testcase has $T=10$ lists; the checker reports the first failure on
list $8$.

\medskip
\textbf{List 8 input:}\\
\texttt{17}\\
\texttt{3 9 9 1 1 1 1 15}\\
\texttt{15 1 1 1 1 1 1 3 1}

\medskip
\textbf{Participant line 8:}\\
\texttt{9 9 15 15}

\medskip
\textbf{Jury line 8:}\\
\texttt{1 1 1 1}
\end{minipage}
&
\begin{minipage}[t]{\linewidth}\ttfamily\raggedright
let exec\_in1 = ExecIn1 \{\\
\ \ \ \ ns: vec![5, 4, 6, 5, 6, 7, 16, 17, 12, 4],\\
\ \ \ \ sticks: vec![\\
\ \ \ \ \ \ \ \ vec![10, 4, 10, 4, 4],\\
\ \ \ \ \ \ \ \ ...\\
\ \ \ \ \ \ \ \ vec![3, 9, 9, 1, 1, 1, 1, 15, 15,\\
\ \ \ \ \ \ \ \ \ \ \ \ \ 1, 1, 1, 1, 1, 1, 3, 1],\\
\ \ \ \ \ \ \ \ ...\\
\ \ \ \ ],\\
\};\\

\medskip
let exec\_out = ExecOut \{\\
\ \ \ \ rectangles: vec![\\
\ \ \ \ \ \ \ \ ExecRectangle \{ s1: 4, s2: 4, s3: 10, s4: 10 \},\\
\ \ \ \ \ \ \ \ ...\\
\ \ \ \ \ \ \ \ ExecRectangle \{ s1: 9, s2: 9, s3: 15, s4: 15 \},\\
\ \ \ \ \ \ \ \ ...\\
\ \ \ \ ],\\
\};
\end{minipage}
&
\begin{minipage}[t]{\linewidth}\raggedright
Invalid output. The sticks \texttt{9 9 15 15} form a rectangle, but not the
minimum-ratio rectangle for list $8$. The checker reports that the participant's
ratio is greater than the jury's ratio.
\end{minipage} \\
\midrule
\begin{minipage}[t]{\linewidth}\raggedright
\kimiktwosix{}

\medskip
\texttt{post\_sound}

\medskip
\texttt{hack\_477559}\\
\texttt{\_\_participant}
\end{minipage}
&
\begin{minipage}[t]{\linewidth}\raggedright
\textbf{Input:}\\
\texttt{1}\\
\texttt{4}\\
\texttt{1 1 10000 10000}

\medskip
\textbf{Participant output:}\\
\texttt{12 12 1 1}

\medskip
\textbf{Jury output:}\\
\texttt{1 1 10000 10000}
\end{minipage}
&
\begin{minipage}[t]{\linewidth}\ttfamily\raggedright
let exec\_in1 = ExecIn1 \{\\
\ \ \ \ ns: vec![4],\\
\ \ \ \ sticks: vec![\\
\ \ \ \ \ \ \ \ vec![1, 1, 10000, 10000],\\
\ \ \ \ ],\\
\};\\

\medskip
let exec\_out = ExecOut \{\\
\ \ \ \ rectangles: vec![\\
\ \ \ \ \ \ \ \ ExecRectangle \{ s1: 12, s2: 12, s3: 1, s4: 1 \},\\
\ \ \ \ ],\\
\};
\end{minipage}
&
\begin{minipage}[t]{\linewidth}\raggedright
Invalid output. The length \texttt{12} is not present in the input list, but the
generated postcondition only checks that the four output numbers form two equal
pairs.
\end{minipage} \\
\midrule
\begin{minipage}[t]{\linewidth}\raggedright
\kimiktwosix{}

\medskip
\texttt{post\_sound}

\medskip
\texttt{hack\_476338}\\
\texttt{\_\_participant}

\medskip
\texttt{hack\_477553}\\
\texttt{\_\_participant}
\end{minipage}
&
\begin{minipage}[t]{\linewidth}\raggedright
\textbf{Input:}\\
\texttt{1}\\
\texttt{4}\\
\texttt{1 1 10000 10000}

\medskip
\textbf{Participant outputs:}\\
\texttt{0 0 0 0}\\
\texttt{-1 -1 -1 -1}

\medskip
\textbf{Checker:}\\
Integer \texttt{0} or \texttt{-1} violates the range $[1,10000]$.
\end{minipage}
&
\begin{minipage}[t]{\linewidth}\ttfamily\raggedright
let exec\_in1 = ExecIn1 \{\\
\ \ \ \ ns: vec![4],\\
\ \ \ \ sticks: vec![\\
\ \ \ \ \ \ \ \ vec![1, 1, 10000, 10000],\\
\ \ \ \ ],\\
\};\\

\medskip
let exec\_out = ExecOut \{\\
\ \ \ \ rectangles: vec![\\
\ \ \ \ \ \ \ \ ExecRectangle \{ s1: 0, s2: 0, s3: 0, s4: 0 \},\\
\ \ \ \ ],\\
\};\\

\medskip
let exec\_out = ExecOut \{\\
\ \ \ \ rectangles: vec![\\
\ \ \ \ \ \ \ \ ExecRectangle \{ s1: -1, s2: -1, s3: -1, s4: -1 \},\\
\ \ \ \ ],\\
\};
\end{minipage}
&
\begin{minipage}[t]{\linewidth}\raggedright
Invalid outputs. Kimi's postcondition accepts them because the four values are
equal, even though the values are outside the valid stick-length domain.
\end{minipage} \\
\midrule
\begin{minipage}[t]{\linewidth}\raggedright
\kimiktwosix{}

\medskip
\texttt{post\_sound}

\medskip
\texttt{hack\_477544}\\
\texttt{\_\_participant}
\end{minipage}
&
\begin{minipage}[t]{\linewidth}\raggedright
Same original testcase as the \gptfivethreecodex{} row above.

\medskip
\textbf{List 8 input:}\\
\texttt{17}\\
\texttt{3 9 9 1 1 1 1 15}\\
\texttt{15 1 1 1 1 1 1 3 1}

\medskip
\textbf{Participant line 8:}\\
\texttt{9 9 15 15}

\medskip
\textbf{Jury line 8:}\\
\texttt{1 1 1 1}
\end{minipage}
&
\begin{minipage}[t]{\linewidth}\ttfamily\raggedright
let exec\_in1 = ExecIn1 \{\\
\ \ \ \ ns: vec![5, 4, 6, 5, 6, 7, 16, 17, 12, 4],\\
\ \ \ \ sticks: vec![\\
\ \ \ \ \ \ \ \ vec![10, 4, 10, 4, 4],\\
\ \ \ \ \ \ \ \ ...\\
\ \ \ \ \ \ \ \ vec![3, 9, 9, 1, 1, 1, 1, 15, 15,\\
\ \ \ \ \ \ \ \ \ \ \ \ \ 1, 1, 1, 1, 1, 1, 3, 1],\\
\ \ \ \ \ \ \ \ ...\\
\ \ \ \ ],\\
\};\\

\medskip
let exec\_out = ExecOut \{\\
\ \ \ \ rectangles: vec![\\
\ \ \ \ \ \ \ \ ExecRectangle \{ s1: 4, s2: 4, s3: 10, s4: 10 \},\\
\ \ \ \ \ \ \ \ ...\\
\ \ \ \ \ \ \ \ ExecRectangle \{ s1: 9, s2: 9, s3: 15, s4: 15 \},\\
\ \ \ \ \ \ \ \ ...\\
\ \ \ \ ],\\
\};
\end{minipage}
&
\begin{minipage}[t]{\linewidth}\raggedright
Invalid output. The rectangle is made of two equal pairs, so Kimi's
postcondition accepts it, but it is not the minimum-ratio rectangle.
\end{minipage} \\
\bottomrule
\end{tabular}
\caption{Post-soundness hack cases for Codeforces 1027C. The \texttt{...}
markers indicate that some repeated input rows or Verus data-structure entries
are omitted for readability.}
\label{tab:case-study-1027c-post-hacks}
\end{table*}

\begin{table*}[t]
\centering
\scriptsize
\setlength{\tabcolsep}{3pt}
\begin{tabular}{p{0.16\textwidth}p{0.23\textwidth}p{0.39\textwidth}p{0.16\textwidth}}
\toprule
Model / hack & Original input & Verus representation & Commentary \\
\midrule
\begin{minipage}[t]{\linewidth}\raggedright
\kimiktwosix{}

\medskip
\texttt{pre\_sound}

\medskip
\texttt{hack\_477694}
\end{minipage}
&
\begin{minipage}[t]{\linewidth}\raggedright
\textbf{Input:}\\
\texttt{1}\\
\texttt{100}\\
\texttt{1000 1001 1002}\\
\texttt{...}\\
\texttt{1097 1098 1099}

\medskip
\textbf{Validator:}\\
No rectangle can be formed in the first list of sticks.
\end{minipage}
&
\begin{minipage}[t]{\linewidth}\ttfamily\raggedright
let exec\_in1 = ExecIn1 \{\\
\ \ \ \ ns: vec![100],\\
\ \ \ \ sticks: vec![\\
\ \ \ \ \ \ \ \ vec![1000, 1001, 1002, 1003,\\
\ \ \ \ \ \ \ \ \ \ \ \ \ ...,\\
\ \ \ \ \ \ \ \ \ \ \ \ \ 1096, 1097, 1098, 1099],\\
\ \ \ \ ],\\
\};
\end{minipage}
&
\begin{minipage}[t]{\linewidth}\raggedright
Invalid input. The problem statement guarantees that every list admits some
rectangle, but this list has all distinct stick lengths, so no pair of equal
sides exists.
\end{minipage} \\
\bottomrule
\end{tabular}
\caption{Pre-soundness hack case for Codeforces 1027C. The \texttt{...} markers
indicate that consecutive stick lengths are omitted for readability.}
\label{tab:case-study-1027c-pre-hacks}
\end{table*}

\textbf{Codex-generated postcondition: missing optimality condition.}
As shown in Listing~\ref{lst:case-study-1027c-codex}, \gptfivethreecodex{} introduces a helper named \texttt{score\_leq}, which is
close to the comparison needed for minimizing $\frac{P^2}{S}$. However, the
postcondition never uses this helper. It only checks that each output rectangle
can be formed from two feasible pairs of sticks.

This accepts outputs such as \texttt{9 9 15 15} on list $8$ of
\texttt{hack\_477544\_\_participant} (Table~\ref{tab:case-study-1027c-post-hacks}). The rectangle is feasible, but it is not
optimal: \texttt{1 1 1 1} achieves a smaller value of $\frac{P^2}{S}$.

\begin{lstlisting}[language=Verus,caption={Snippet from postcondition generated by \gptfivethreecodex{} for Codeforces 1027C that fails a post-soundness hack.},label={lst:case-study-1027c-codex}]
exec_spec_unverified! {
    pub open spec fn score_leq(a: i64, b: i64,
                               c: i64, d: i64) -> bool {
        (a + b) * (a + b) * c * d <=
            (c + d) * (c + d) * a * b
    }

    pub open spec fn rectangle_valid_for_sticks(
        st: Seq<i64>, r: Rectangle
    ) -> bool {
        exists |x: i64, y: i64|
            1 <= x <= 10_000
            && 1 <= y <= 10_000
            && feasible_pair(st, x, y)
            && rectangle_matches_pair(r, x, y)
    }

    pub open spec fn post_spec(in1: In1, out: Out) -> bool {
        out.rectangles.len() == in1.sticks.len()
        && forall |i: i64|
            0 <= i < in1.sticks.len() ==>
                #[trigger] rectangle_valid_for_sticks(
                    in1.sticks[i as int],
                    out.rectangles[i as int])
    }
}
\end{lstlisting}

\textbf{Kimi-generated postcondition: missing several checks.}
\kimiktwosix{} writes an even weaker postcondition (Listing~\ref{lst:case-study-1027c-kimi}): it only checks that each output
contains two equal pairs. This postcondition misses several parts of the problem statement, as it does not require the lengths to appear in the input list, stay in the valid output range, or minimize the desired metric.

For \texttt{hack\_477559\_\_participant} (Table~\ref{tab:case-study-1027c-post-hacks}), this postcondition accepts
\texttt{12 12 1 1} even though the input contains only
\texttt{1 1 10000 10000}.

\textbf{Kimi-generated precondition: missing check that a rectangle can be formed.} The same specification also accepts invalid inputs
such as \texttt{hack\_477694} (Table~\ref{tab:case-study-1027c-pre-hacks}), where all stick lengths are distinct--this violates the problem's guarantee that it is possible to form at least one rectangle from the input.

\begin{lstlisting}[language=Verus,caption={Pre- and postcondition generated by \kimiktwosix{} for Codeforces 1027C that fails a post-soundness hack and a pre-soundness hack.},label={lst:case-study-1027c-kimi}]
exec_spec_unverified! {
    pub open spec fn pre_spec(in1: In1) -> bool {
        in1.ns.len() >= 1
        && in1.ns.len() == in1.sticks.len()
        && forall |i: i64|
            0 <= i < in1.ns.len() ==>
                in1.ns[i as int] >= 4
                && in1.sticks[i as int].len() as i64
                    == in1.ns[i as int]
                && forall |j: i64|
                    0 <= j < in1.sticks[i as int].len() ==>
                        1 <= in1.sticks[i as int][j as int]
                        && in1.sticks[i as int][j as int] <= 10000
    }

    pub open spec fn is_two_pairs(rect: Rectangle) -> bool {
        (rect.s1 == rect.s2 && rect.s3 == rect.s4)
        || (rect.s1 == rect.s3 && rect.s2 == rect.s4)
        || (rect.s1 == rect.s4 && rect.s2 == rect.s3)
    }

    pub open spec fn post_spec(in1: In1, out: Out) -> bool {
        out.rectangles.len() == in1.ns.len()
        && forall |i: i64|
            0 <= i < in1.ns.len() ==>
                is_two_pairs(out.rectangles[i as int])
    }
}
\end{lstlisting}

%% file: paper_sections/appendix/case_study_latex/problem_1028_C.tex
\subsection{Case Study: Problem 1028C}
\label{app:case-study-1028c}

This case study highlights a pre-soundness failure. The problem statement ensures that
some $n-1$ of the $n$ rectangles in the input have a common point. Both \gptfivethreecodex{}
and \geminithreeonepro{} accept an input that does not satisfy this property.

\begin{tcolorbox}[colback=blue!3,colframe=blue!40,fontupper=\scriptsize,title={\textbf{Informal specification (natural language) $s_I$ for Codeforces 1028-C}}]
\textbf{Problem link.}\quad
\url{https://codeforces.com/problemset/problem/1028/C}

\medskip
You are given $n$ rectangles on a plane with coordinates of their bottom left
and upper right points. Some $(n-1)$ of the given $n$ rectangles have some
common point. A point belongs to a rectangle if this point is strictly inside
the rectangle or belongs to its boundary.

Find any point with integer coordinates that belongs to at least $(n-1)$ given rectangles.

\medskip
\textbf{Input.}

The first line contains a single integer $n$ ($2 \le n \le 132\,674$) --- the number of given rectangles.

Each the next $n$ lines contains four integers $x_1$, $y_1$, $x_2$ and $y_2$ ($-10^9 \le x_1 < x_2 \le 10^9$, $-10^9 \le y_1 < y_2 \le 10^9$) --- the coordinates of the bottom left and upper right corners of a rectangle.

\medskip
\textbf{Output.}

Print two integers $x$ and $y$ --- the coordinates of any point that belongs to at least $(n-1)$ given rectangles.

\medskip
\textbf{Example.}
\begin{flushleft}\ttfamily
Input:\\
3\\
0 0 1 1\\
1 1 2 2\\
3 0 4 1\\
Output:\\
1 1
\end{flushleft}
\end{tcolorbox}

\textbf{Codex- and Gemini-generated precondition: missing common point condition.}
Both generated preconditions (Listing~\ref{lst:case-study-1028c-codex-gemini}) check that the number of rectangles and the coordinates of each rectangle satisfy the bounds given in the problem. However, they do not require that there exists a point contained in at least $n-1$ rectangles. Pre-soundness hack \texttt{hack\_483020} (Table~\ref{tab:case-study-1028c-hacks}) exposes this missing condition.
\todoanmol{PDF note p.~41: Natalie's comment: it is not clear which models generated the precondition shown.}

\begin{lstlisting}[language=Verus,caption={Postcondition for Codeforces 1028C that fails a pre-soundness hack.},label={lst:case-study-1028c-codex-gemini}]
exec_spec_unverified! {
    pub open spec fn rect_valid(r: Rect) -> bool {
        &&& -1_000_000_000 <= r.x1
        &&& r.x1 < r.x2
        &&& r.x2 <= 1_000_000_000
        &&& -1_000_000_000 <= r.y1
        &&& r.y1 < r.y2
        &&& r.y2 <= 1_000_000_000
    }

    pub open spec fn pre_spec(in1: In1) -> bool {
        &&& 2 <= in1.n && in1.n <= 132674
        &&& in1.rectangles.len() as i64 == in1.n
        &&& forall |i: usize|
            0usize <= i < in1.rectangles.len() ==>
                rect_valid(#[trigger] in1.rectangles[i as int])
    }
}
\end{lstlisting}

\begin{table*}[h]
\centering
\scriptsize
\setlength{\tabcolsep}{3pt}
\begin{tabular}{p{0.18\textwidth}p{0.24\textwidth}p{0.36\textwidth}p{0.16\textwidth}}
\toprule
Models / hack & Original input & Verus representation & Commentary \\
\midrule
\begin{minipage}[t]{\linewidth}\raggedright
\gptfivethreecodex{} and \geminithreeonepro{}

\medskip
\texttt{pre\_sound}

\medskip
\texttt{hack\_483020}
\end{minipage}
&
\begin{minipage}[t]{\linewidth}\raggedright
\textbf{Input:}\\
\texttt{3}\\
\texttt{0 0 2 2}\\
\texttt{0 4 2 6}\\
\texttt{0 8 2 10}

\medskip
\textbf{Validator:}\\
No point lies in at least $2$ rectangles.
\end{minipage}
&
\begin{minipage}[t]{\linewidth}\ttfamily\raggedright
let exec\_in1 = ExecIn1 \{\\
\ \ \ \ n: 3,\\
\ \ \ \ rectangles: vec![\\
\ \ \ \ \ \ \ \ ExecRect \{ x1: 0, y1: 0, x2: 2, y2: 2 \},\\
\ \ \ \ \ \ \ \ ExecRect \{ x1: 0, y1: 4, x2: 2, y2: 6 \},\\
\ \ \ \ \ \ \ \ ExecRect \{ x1: 0, y1: 8, x2: 2, y2: 10 \},\\
\ \ \ \ ],\\
\};
\end{minipage}
&
\begin{minipage}[t]{\linewidth}\raggedright
Invalid input. The rectangles are individually well formed, but they are
vertically disjoint, so no two rectangles share a point.
\end{minipage} \\
\bottomrule
\end{tabular}
\caption{Pre-soundness hack for Codeforces 1028C. The input should be rejected
because it violates the benchmark guarantee that some $n-1$ rectangles have a
common point.}
\label{tab:case-study-1028c-hacks}
\end{table*}

%% file: paper_sections/appendix/case_study_latex/problem_2074_D.tex
\caseStudyBarrier
\subsection{Case Study: Problem 2074D}
\label{app:case-study-2074d}

This case study shows a different failure mode that is revealed by post-completeness tests: the generated specification attempts
to compute the exact answer, but the executable specification is itself too
complicated and rejects correct outputs. For this problem, \geminithreeonepro{} writes a
large interval-union specification and fails post-completeness on valid answers.
In contrast, \opusfoursix{} uses a simpler column-wise
characterization of the union of circles, which is the key idea needed for this problem.

\begin{tcolorbox}[colback=blue!3,colframe=blue!40,fontupper=\scriptsize,title={\textbf{Informal specification (natural language) $s_I$ for Codeforces 2074-D}}]
\textbf{Problem link.}\quad
\url{https://codeforces.com/problemset/problem/2074/D}

\medskip
The pink soldiers drew $n$ circles with their center on the $x$-axis of the
plane. Also, they have told that the sum of radii is exactly $m$$^{\text{*}}$.
Please find the number of integer points inside or on the border of at least one
circle. Formally, the problem is defined as follows.
You are given an integer sequence $x_1,x_2,\ldots,x_n$ and a positive integer
sequence $r_1,r_2,\ldots,r_n$, where it is known that $\sum_{i=1}^n r_i = m$.
You must count the number of integer pairs $(x,y)$ that satisfy the following condition.

\medskip
\hspace{1em}There exists an index $i$ such that $(x-x_i)^2 + y^2 \le r_i^2$ ($1 \le i \le n$).

\medskip
$^{\text{*}}$Is this information really useful? Don't ask me; I don't really know.

\medskip
\textbf{Input.}

Each test contains multiple test cases. The first line contains the number of test cases $t$ ($1 \le t \le 10^4$). The description of the test cases follows.

The first line of each test case contains two integers $n$ and $m$ ($1 \le n \le m \le 2\cdot 10^5$).

The second line of each test case contains $x_1,x_2,\ldots,x_n$ --- the centers of the circles ($-10^9 \le x_i \le 10^9$).

The third line of each test case contains $r_1,r_2,\ldots,r_n$ --- the radii of the circles ($1 \le r_i$, $\sum_{i=1}^n r_i = m$).

It is guaranteed that the sum of $m$ over all test cases does not exceed $2\cdot 10^5$.

\medskip
\textbf{Output.}

For each test case, output the number of integer points satisfying the condition on a separate line.

\medskip
\textbf{Example.}
\begin{flushleft}\ttfamily
Input:\\
\mbox{}\\
4\\
2 3\\
0 0\\
1 2\\
2 3\\
0 2\\
1 2\\
3 3\\
0 2 5\\
1 1 1\\
4 8\\
0 5 10 15\\
2 2 2 2\\
Output:\\
\mbox{}\\
13\\
16\\
14\\
52
\end{flushleft}

\medskip
\textbf{Note.}

On the first test case, the circle with $r_1=1$ is completely inside the circle
with $r_2=2$. Therefore, you only have to count the number of integer points
inside the latter. There are $13$ integer points such that $x^2+y^2 \le 2^2$,
so the answer is $13$.

On the second test case, the circle with $r_1=1$ is not completely inside the
circle with $r_2=2$. There are $3$ additional points that are inside the first
circle but not inside the second circle, so the answer is $3+13=16$.
\end{tcolorbox}

\textbf{Gemini-generated postcondition: computing incorrect result.}
\geminithreeonepro{} tries to compute the answer in its specification (Listing~\ref{lst:case-study-2074d-gemini}) by constructing horizontal intervals,
sorting them, merging overlaps, and summing the resulting disjoint intervals.
This is a plausible implementation strategy, but it makes the specification
large and fragile. The valid outputs given by the post-completeness hacks in
Table~\ref{tab:case-study-2074d-hacks} are rejected by this postcondition.
The failures are especially revealing because the table includes very small
inputs where the correct count can be checked directly. For example,
\texttt{hack\_1116159\_\_participant} has only two circles: radius $2$ centered
at $0$ and radius $1$ centered at $1$. The exact union contains $13$ integer
points, yet \geminithreeonepro{}'s executable postcondition rejects \texttt{13}. Similarly,
\texttt{hack\_1118906\_\_participant} contains five identical radius-$2$ circles
centered at $0$, so the union is just one radius-$2$ circle and again contains
$13$ points.

\begin{lstlisting}[language=Verus,caption={Snippet from postcondition generated by \geminithreeonepro{} for Codeforces 2074D that fails post-completeness hacks.},label={lst:case-study-2074d-gemini}]
exec_spec_unverified! {
    pub struct YInterval {
        pub y: i64,
        pub start: i64,
        pub end: i64,
    }

    pub open spec fn compute_union_length(s: Seq<YInterval>) -> i64 {
        if s.len() == 0 {
            0
        } else {
            let sorted = merge_sort(s);
            let disjoint = merge_all_intervals(
                sorted, 0, sorted.len() as usize);
            sum_disjoint_intervals(disjoint, 0, disjoint.len() as usize)
        }
    }

    pub open spec fn count_points(tc: TestCase) -> i64 {
        let intervals = collect_all_y_intervals(tc, 0, tc.n as usize);
        compute_union_length(intervals)
    }

    pub open spec fn post_spec(in1: In1, out: Out) -> bool {
        out.answers.len() == in1.testcases.len()
        && forall |i: usize| 0 <= i < in1.testcases.len() ==>
            out.answers[i as int] == count_points(in1.testcases[i as int])
    }
}
\end{lstlisting}

\textbf{Opus-generated postcondition: a simpler, correct characterization.}
The specification generated by \opusfoursix{} (Listing~\ref{lst:case-study-2074d-opus}) uses the geometric observation that for each integer $x$-coordinate, the
covered integer $y$-values form a symmetric interval
$[-h(x),h(x)]$, where $h(x)$ is the maximum vertical radius among circles
covering that $x$. Counting each $x$ once avoids having to maintain and merge
many horizontal intervals.

This contrast is informative: both specifications attempt to compute the same mathematical object, but higher-level mathematical reasoning allows \opusfoursix{} to express the desired result using a simpler decomposition, whereas the approach taken by \geminithreeonepro{} results in a complex, and ultimately incorrect, computation.

\begin{lstlisting}[language=Verus,caption={Snippet from postcondition generated by \opusfoursix{} for Codeforces 2074D that passes all test cases in our benchmark.},label={lst:case-study-2074d-opus}]
exec_spec_unverified! {
    pub open spec fn max_y_at_x(tc: TestCase, xval: i64) -> i64 {
        max_y_at_x_helper(tc, xval, 0usize, 0i64, 200001usize)
    }

    pub open spec fn contrib_at_dx(tc: TestCase, i: usize, dx: i64) -> i64 {
        let xval = (tc.x[i as int] + dx) as i64;
        if is_first_cover(tc, i, xval) {
            let my = max_y_at_x(tc, xval);
            (2i64 * my + 1i64) as i64
        } else {
            0i64
        }
    }

    pub open spec fn compute_answer(tc: TestCase) -> i64 {
        sum_circles(tc, 0usize, 200001usize)
    }

    pub open spec fn post_spec(in1: In1, out: Out) -> bool {
        pre_spec(in1)
        && out.answers.len() == in1.testcases.len()
        && check_answers(in1, out, 0usize, 10001usize)
    }
}
\end{lstlisting}

\begin{table*}[t]
\centering
\scriptsize
\setlength{\tabcolsep}{3pt}
\begin{tabular}{p{0.18\textwidth}p{0.28\textwidth}p{0.32\textwidth}p{0.16\textwidth}}
\toprule
Model / hack & Original input and output & Verus representation & Commentary \\
\midrule
\begin{minipage}[t]{\linewidth}\raggedright
\geminithreeonepro{}

\medskip
\texttt{post\_complete}

\medskip
\texttt{hack\_1116127}\\
\texttt{\_\_participant}
\end{minipage}
&
\begin{minipage}[t]{\linewidth}\raggedright
This is the sample-style multi-testcase input.

\medskip
\textbf{Input:}\\
\texttt{4}\\
\texttt{2 3; 0 0; 1 2}\\
\texttt{2 3; 0 2; 1 2}\\
\texttt{3 3; 0 2 5; 1 1 1}\\
\texttt{4 8; 0 5 10 15; 2 2 2 2}

\medskip
\textbf{Expected output:}\\
\texttt{13 16 14 52}
\end{minipage}
&
\begin{minipage}[t]{\linewidth}\ttfamily\raggedright
let exec\_out = ExecOut \{\\
\ \ \ \ answers: vec![13, 16, 14, 52],\\
\};\\

\medskip
let exec\_in1 = ExecIn1 \{\\
\ \ \ \ testcases: vec![\\
\ \ \ \ \ \ \ \ ExecTestCase \{ n: 2, m: 3, x: vec![0, 0], r: vec![1, 2] \},\\
\ \ \ \ \ \ \ \ ...\\
\ \ \ \ ],\\
\};
\end{minipage}
&
\begin{minipage}[t]{\linewidth}\raggedright
Valid output rejected by Gemini's postcondition, so this is a
post-completeness failure.
\end{minipage} \\
\midrule
\begin{minipage}[t]{\linewidth}\raggedright
\geminithreeonepro{}

\medskip
\texttt{post\_complete}

\medskip
\texttt{hack\_1116159}\\
\texttt{\_\_participant}
\end{minipage}
&
\begin{minipage}[t]{\linewidth}\raggedright
\textbf{Input:}\\
\texttt{1}\\
\texttt{2 3}\\
\texttt{0 1}\\
\texttt{2 1}

\medskip
\textbf{Expected output:}\\
\texttt{13}
\end{minipage}
&
\begin{minipage}[t]{\linewidth}\ttfamily\raggedright
let exec\_in1 = ExecIn1 \{\\
\ \ \ \ testcases: vec![\\
\ \ \ \ \ \ \ \ ExecTestCase \{\\
\ \ \ \ \ \ \ \ \ \ \ \ n: 2, m: 3,\\
\ \ \ \ \ \ \ \ \ \ \ \ x: vec![0, 1],\\
\ \ \ \ \ \ \ \ \ \ \ \ r: vec![2, 1],\\
\ \ \ \ \ \ \ \ \},\\
\ \ \ \ ],\\
\};\\

\medskip
let exec\_out = ExecOut \{ answers: vec![13] \};
\end{minipage}
&
\begin{minipage}[t]{\linewidth}\raggedright
Small overlapping-circles case. The expected answer is accepted by the contest
checker but rejected by Gemini's executable postcondition.
\end{minipage} \\
\midrule
\begin{minipage}[t]{\linewidth}\raggedright
\geminithreeonepro{}

\medskip
\texttt{post\_complete}

\medskip
\texttt{hack\_1118906}\\
\texttt{\_\_participant}
\end{minipage}
&
\begin{minipage}[t]{\linewidth}\raggedright
\textbf{Input:}\\
\texttt{1}\\
\texttt{5 10}\\
\texttt{0 0 0 0 0}\\
\texttt{2 2 2 2 2}

\medskip
\textbf{Expected output:}\\
\texttt{13}
\end{minipage}
&
\begin{minipage}[t]{\linewidth}\ttfamily\raggedright
let exec\_in1 = ExecIn1 \{\\
\ \ \ \ testcases: vec![\\
\ \ \ \ \ \ \ \ ExecTestCase \{\\
\ \ \ \ \ \ \ \ \ \ \ \ n: 5, m: 10,\\
\ \ \ \ \ \ \ \ \ \ \ \ x: vec![0, 0, 0, 0, 0],\\
\ \ \ \ \ \ \ \ \ \ \ \ r: vec![2, 2, 2, 2, 2],\\
\ \ \ \ \ \ \ \ \},\\
\ \ \ \ ],\\
\};\\

\medskip
let exec\_out = ExecOut \{ answers: vec![13] \};
\end{minipage}
&
\begin{minipage}[t]{\linewidth}\raggedright
All circles coincide, so the union is just one radius-$2$ circle with $13$
integer points. Gemini still rejects the correct answer.
\end{minipage} \\
\bottomrule
\end{tabular}
\caption{Representative post-completeness failures for \geminithreeonepro{} on Codeforces
2074D. In each case, the displayed output is a correct contest answer that the
generated postcondition rejects. Since the task asks for exact counts, each
displayed output is the unique expected answer for its input, not one of several
acceptable witnesses.}
\label{tab:case-study-2074d-hacks}
\end{table*}

%% file: paper_sections/appendix/appendix_prompt.tex
\section{Specification Generation Background Prompt}
\label{app:full-prompt}

Here is the prompt provided to the model to describe the task to it.

\begin{lstlisting}[language={},basicstyle=\ttfamily\scriptsize,breaklines=true,columns=fullflexible,numbers=left,numberstyle=\tiny\color{gray},numbersep=6pt]
You are solving a Verus specification task.

## Workspace

- Edit: `/home/dev/solve.rs`
- Problem statement: `/home/problem_artifacts/problem_statement.md`
- Visible sample tests and snippets: `/home/symbolic_tests`
- Task-specific artifacts: `/home/task_specific_artifacts`
- Examples: `/home/examples`
- Verus docs: `/home/verus_documentation`
- Readable evaluator source: `/home/evaluator_scripts/specgen_evaluator`
- Attempts/logs (agent): `/home/attempts` and `/logs/agent/compressed_verdicts`

## Useful command during generation (sample evaluation)

```bash
verus_gym_specgen_check
```

The default CLI arguments evaluate your submitted `/home/dev/solve.rs` on the
visible sample tests and write feedback under `/home/attempts`.

## Task instructions

## Verus Specification Task
You are working on a benchmark built from real competitive-programming problems
and test data. For each task, you receive a natural-language problem statement
and write formal Verus specifications that capture the valid inputs and correct
outputs for that problem.

The environment has already done the problem-specific setup for you. The
concrete input/output structure will be visible in `/home/dev/solve.rs` and in
the generated testcase snippets, but you do not need to implement that
conversion yourself. In particular, the task bundle already includes:

- Chosen logical types (`In1`, `Out`, ...) for the problem.
- Parsed real `.in` / `.out` files into these types.
- Pre-generated **Verus-ready definition snippets** (files like `out.input_defn`
  and, for postcondition testcases, `out.gt_output_defn`) for each symbolic
  testcase.

Your job is:

> Given the problem statement and pre-generated Verus snippets, write
> **logical specifications** `pre_spec` and `post_spec` so that they correctly
> classify real inputs/outputs, i.e. they accept valid inputs/outputs and reject invalid ones.

The proof helpers in `/home/dev/solve.rs` are only there to help Verus establish
the fixed testcase assertions against your specs. You are not proving an
algorithm or implementing a solver for the problem.

You do **not** change parsing, printing, or snippet generation. You only edit the
**specs and proofs** in `/home/dev/solve.rs`.

To check your current implementation against the visible sample tests, run:

```bash
verus_gym_specgen_check
```

In this environment, the default CLI arguments evaluate the current
`/home/dev/solve.rs` and write feedback/artifacts under `/home/attempts`.

---

## 1. Files and environment overview

In this environment you will primarily work with `/home/dev/solve.rs` and the
problem artifacts under `/home/problem_artifacts`.

The current problem statement is at
`/home/problem_artifacts/problem_statement.md`. The visible sample testcases are
under `/home/symbolic_tests/<bucket>/<testcase_id>/`, where `<bucket>` is one of
`pre_sound`, `pre_complete`, `post_sound`, or `post_complete`. Some bucket
directories may be empty in the visible sample set. Non-empty testcase
directories contain:

- `pre_*` testcases: `test.in` and `out.input_defn`.
- `post_*` testcases: `test.in`, `test.out`, `out.input_defn`, and
  `out.gt_output_defn`.

The logical input type is `In1`, and the logical output type is `Out`. Their
executable counterparts are `ExecIn1` and `ExecOut`. You can inspect all of
these definitions in `/home/dev/solve.rs`.

Below are more details:

- `/home/dev/solve.rs`
  - This is the Verus file you edit for this benchmark.
  - It already contains:
    - Logical/spec types (`In1`, `Out`, ...) defined via `exec_spec_unverified! { ... }`.
      - This also generates executable counterparts (`ExecIn1`, `ExecOut`, ...) and
        runtime wrappers `exec_pre_spec` / `exec_post_spec` for exec fallback.
    - Stub specs:
      - `pub open spec fn pre_spec(in1: In1) -> bool { ... }`
        - Should return `true` exactly for inputs satisfying the problem
          statement's input constraints.
      - `pub open spec fn post_spec(in1: In1, out: Out) -> bool { ... }`
        - Should return `true` exactly for outputs that are correct for the
          given input according to the problem statement.
    - Proof helpers:
      - `proof fn pre_spec_soundness_proof(in1: In1) { ... }`
      - `proof fn pre_spec_completeness_proof(in1: In1) { ... }`
      - `proof fn post_spec_soundness_proof(in1: In1, out: Out) { ... }`
      - `proof fn post_spec_completeness_proof(in1: In1, out: Out) { ... }`
    - Four **check wrappers**:
      - `fn check_pre_spec_completeness() { ... }`
      - `fn check_pre_spec_soundness() { ... }`
      - `fn check_post_spec_completeness() { ... }`
      - `fn check_post_spec_soundness() { ... }`
        Each of these contains **paste markers** such as:
        - `// __PASTE_out.input_defn__`
        - `// __PASTE_out.gt_output_defn__`
        and calls to the corresponding proof helpers.
        The **assertion lines are fixed** by the provided testcase snippets and are not
        model-editable.
    - Runtime exec fallback entrypoints (fixed):
      - `pub fn main_exec_pre_spec_check() { ... }`
      - `pub fn main_exec_post_spec_check() { ... }`
      - `fn main() { unimplemented!(...) }` (placeholder; evaluator replaces it)

- `/home/examples/solve.rs`
  - A **worked code example** of a Verus specification file for another problem.
  - Use it as a reference for Verus structure, proof-helper style, and common
    spec patterns. The example specs may be incomplete or imperfect; for now,
    use this file mainly to understand the expected spec/proof format.
- `/home/examples/problem_statement_for_example.md`
  - The natural-language problem statement corresponding to
    `/home/examples/solve.rs`.
  - Use the pair as an example of how a statement can map to Verus spec format.
    Do not treat it as semantic guidance for your current task.

- `/home/problem_artifacts/problem_statement.md`
  - The full **natural-language problem statement** for the programming problem you are working on.
  - You must base both `pre_spec` and `post_spec` on this statement.

- `/home/symbolic_tests/`
  - Contains **sample symbolic test buckets**. Some buckets may be empty in the
    visible sample set.
    - `pre_sound/` -- inputs that must violate the precondition (invalid inputs).
    - `pre_complete/` -- inputs that must satisfy the precondition (valid inputs).
    - `post_sound/` -- (input, output) pairs where the output is **incorrect**.
    - `post_complete/` -- (input, output) pairs where the output is **correct**.
  - For each testcase id `T`, you will see:
    - `pre_*` testcases: `test.in` for precondition examples, plus
      `out.input_defn` to see the Verus encoding of input.
      Do not expect `test.out` in `pre_*` directories.
    - `post_*` testcases: `test.in`, `test.out`, `out.input_defn`,
      and `out.gt_output_defn`.
      These are **pre-generated Verus snippets** that encode the same data as
      the testcase files, but in Verus syntax (e.g., `let exec_in1: ExecIn1 = ...;`).
  - **Important:** these are only the **visible** sample tests. Local/debug
    feedback may be based on this visible subset, while final scoring uses the
    full evaluator suite, including hidden tests derived from the same fixed
    conversion setup.

- `/home/evaluator_scripts/specgen_evaluator/`
  - Readable Python evaluator source copied into the container for reference.
  - The active evaluation flow is in
    `/home/evaluator_scripts/specgen_evaluator/runner/evaluate_specgen.py`,
    with helpers under `runner/`, `phases/`, and `models/`.
  - The command you run is still `verus_gym_specgen_check`; it is installed
    from the trusted runtime bundle under `/opt/verus_gym_specgen_check_runtime`.
  - Editing files under `/home/evaluator_scripts` does not affect scoring.

- `/home/verus_documentation`
  - Verus guide and API docs. The file
    `/home/verus_documentation/guide/src/SUMMARY.md`
    lists all chapters and paths.
  - For `exec_spec` / `exec_spec_unverified`, you may find it helpful to read:
    `/home/verus_documentation/guide/src/exec_spec.md`
- `/home/verus-x86-linux`
  - The code for Verus.
  - You may also find it helpful to inspect the `exec_spec` implementation in the Verus source tree:
    - `/home/verus-x86-linux/vstd/contrib/exec_spec/`
    - `/home/verus-x86-linux/builtin_macros/src/contrib/exec_spec.rs`

---

## 2. What is fixed vs what you may edit

In `/home/dev/solve.rs`, **you may edit**:

- The bodies of:
  - `spec fn pre_spec(...) -> bool`
  - `spec fn post_spec(...) -> bool`
- The bodies of the four proof helpers:
  - `proof fn pre_spec_soundness_proof(...)`
  - `proof fn pre_spec_completeness_proof(...)`
  - `proof fn post_spec_soundness_proof(...)`
  - `proof fn post_spec_completeness_proof(...)`
- Any **additional spec or proof helpers** you introduce (e.g., helper predicates or lemmas),
  as long as they stay within the `verus! { ... }` block and do not break existing signatures.

You **must not**:

- Change the signatures or names of:
  - `pre_spec`, `post_spec`, the four `*_proof` functions, or any `check_*` wrapper.
- Edit the bodies or structure of the four `check_*` functions:
  - Do **not** remove or move:
    - The `// __PASTE_out.*__` marker comments.
    - The `proof { ... }` blocks that call your `*_proof` functions.
    - The `assert(...)` lines; they have been wired from the testcase snippets.
- Edit the runtime exec fallback functions or the placeholder `main()`:
  - `main_exec_pre_spec_check`, `main_exec_post_spec_check`, or `main`.
- Change type or struct definitions such as `In1`, `Out`, or any additional input types.
  - These are synchronized with the fixed conversion layer (`model_mod.rs`, `main.rs`),
    which you **cannot** see or edit in this task.

If you break these constraints (e.g., by deleting `check_post_spec_soundness` or
removing `// __PASTE_out.input_defn__`), the evaluator will reject your submission
with a **syntax / shape error** before running any proofs.

## 3. What `pre_spec` and `post_spec` should mean

Conceptually, we think in terms of sets:

- Let `I` be all possible inputs (each encoded as a `.in` file).
- Let `I_correct subset I` be those inputs that satisfy **all** constraints in the problem statement.
- Let `I_wrong = I \ I_correct` be structurally well-formed but logically invalid inputs.

For each valid input `i in I_correct`:

- Let `O_i` be all possible outputs (`.out` files).
- Let `O_i_correct subset O_i` be outputs that satisfy the problem's requirements on `i`.
- Let `O_i_wrong = O_i \ O_i_correct` be incorrect outputs for that input.

Your specs should capture:

- `pre_spec(in1, ...)` is **true** iff the logical input corresponds to a valid `.in` file:
  - Completeness: `pre_spec(i)` holds for every `i in I_correct`.
  - Soundness:    `pre_spec(i)` does **not** hold for any `i in I_wrong`.

- `post_spec(in1, ..., out)` is **true** iff `out` is a correct solution for `in1`:
  - Completeness: for all `i in I_correct` and all `o in O_i_correct`,
    `post_spec(i, o)` holds.
  - Soundness:    for all `i in I_correct` and all `o in O_i_wrong`,
    `post_spec(i, o)` does **not** hold.

In practice:

- `pre_spec` should encode **all structural and range constraints** on the input
  (e.g., bounds on `n`, array lengths, value ranges) that appear in the problem statement.
- `post_spec` should encode the **mathematical correctness** of the output:
  - Not just format (e.g., "`out.len() == n`"), but the intended semantics
    (e.g., "this is a valid matching with minimal cost", "this sequence satisfies the constraints").

Weak specs such as "`post_spec` always returns true" or specs that only check trivial
format properties will almost certainly fail the soundness/completeness tests.

---

## 4. How `.in` / `.out` become symbolic testcases

You **do not** implement the conversion, but it helps to understand
what is happening under the hood.

For each testcase, we have already extracted:

- **Input snippets**:

  ```rust
  // out.input_defn
  let exec_in1: ExecIn1 = ...;

  // out.assert_pre_spec
  assert(pre_spec(exec_in1.deep_view()));
  ```

- **Output snippets** (for post categories):

  ```rust
  // out.gt_output_defn
  let exec_out: ExecOut = ...;

  // out.assert_post_spec_gt
  assert(post_spec(exec_in1.deep_view(), exec_out.deep_view()));
  ```
These snippets are generated from real `.in` / `.out` files by the fixed
conversion layer. For an input testcase, `out.input_defn` constructs an
`ExecIn1` value corresponding to `test.in`; calling `exec_in1.deep_view()`
produces the logical `In1` value passed to `pre_spec` and `post_spec`. For an
output testcase, `out.gt_output_defn` constructs an `ExecOut` value
corresponding to the candidate output in `test.out`; calling
`exec_out.deep_view()` produces the logical `Out` value passed to `post_spec`.

The evaluator plugs these snippets into your `solve.rs`
inside the `check_*` wrappers.

---

## 4.1 exec_spec_unverified runtime fallback (SMT -> exec)

The evaluator primarily tries to establish correctness **symbolically** via SMT.
For each bucket, it checks the expected polarity:

- Completeness buckets should prove `pre_spec(...)` or `post_spec(...)`.
- Soundness buckets should prove `!pre_spec(...)` or `!post_spec(...)`.

When SMT cannot decide the expected assertion, the evaluator may fall back to
**runtime execution** using the exec versions generated by
`exec_spec_unverified!`:
- `exec_pre_spec(&ExecIn1) -> bool`
- `exec_post_spec(&ExecIn1, &ExecOut) -> bool`

Practical implication: write specs that are both SMT-friendly *and* exec-friendly.
Also note: inside `exec_spec_unverified!`, quantifier variables should use concrete Rust
types (e.g., `i64`, `usize`), not `int`/`nat`.

---

## 5. How the evaluator uses your code

The evaluator runs the following steps for each testcase:

1. **Starting point**:

   - It uses the `/home/dev/solve.rs` file you edited, with the provided fixed
     types and printers.
   - It also uses four **derived files** that extract only:
     - Your `pre_spec` / `post_spec`,
     - Your four `*_proof` helpers,
     - The relevant `check_*` wrapper,
     - And any helper functions you introduced.

2. **Snippet injection**:

   - For each testcase and category, it reads:
     - `out.input_defn` and possibly `out.gt_output_defn`.
   - It then substitutes these into the appropriate `check_*` function bodies:

     ```rust
     // __PASTE_out.input_defn__
     // __PASTE_out.gt_output_defn__
     ```

   - The assertions inside the `check_*` functions are already wired to use
     the correct sign:
     - For completeness tests:
       - `assert(pre_spec(exec_in1.deep_view()));`
       - `assert(post_spec(exec_in1.deep_view(), exec_out.deep_view()));`
     - For soundness tests:
       - `assert(!pre_spec(exec_in1.deep_view()));`
       - `assert(!post_spec(exec_in1.deep_view(), exec_out.deep_view()));`
     These were filled in automatically from the snippets.

3. **Verus execution**:

   - For each testcase, the evaluator runs Verus on the derived check wrapper
     for that testcase's bucket:
     - `pre_complete` uses `check_pre_spec_completeness`.
     - `pre_sound` uses `check_pre_spec_soundness`.
     - `post_complete` uses `check_post_spec_completeness`.
     - `post_sound` uses `check_post_spec_soundness`.
   - It records:
     - Whether verification of the corresponding `assert(...)` succeeded.
     - Any Verus errors, plus paths to the `.rs`, `.stdout`, and `.stderr` logs
       under `/home/attempts/<run_id>/snippets/...`.

4. **Pass/fail logic**:

   - For **pre-completeness** tests (`pre_complete`):
     - The inputs are known valid.
     - The test passes if Verus can prove `assert(pre_spec(...))`.
   - For **pre-soundness** tests (`pre_sound`):
     - The inputs are invalid.
     - The test passes if Verus can prove `assert(!pre_spec(...))`.
   - For **post-completeness** tests (`post_complete`):
     - `(in1, out)` pairs are known correct.
     - The test passes if Verus can prove `assert(post_spec(exec_in1.deep_view(), exec_out.deep_view()))`.
   - For **post-soundness** tests (`post_sound`):
     - `(in1, out)` are known to be logically incorrect outputs.
     - The test passes if Verus can prove `assert(!post_spec(exec_in1.deep_view(), exec_out.deep_view()))`.

5. **Visible vs hidden tests**:

   - Running `verus_gym_specgen_check` in this environment may give local/debug
     feedback on only the visible sample tests.
   - Final scoring uses the full evaluator suite, including hidden tests not
     present under `/home/symbolic_tests`.
   - Use visible failures as diagnostics, but write specs from the problem
     statement rather than tuning only to the visible sample.

---

## 6. What to put inside the proof blocks

Each `check_*` wrapper looks roughly like:

```rust
fn check_pre_spec_completeness() {
    // __PASTE_out.input_defn__

    proof {
        pre_spec_completeness_proof(exec_in1.deep_view());
    }

    assert(pre_spec(exec_in1.deep_view()));
}
```

The proof functions are called inside `proof { ... }` blocks. Their role is to
help Verus discharge the verification conditions created by the `assert(...)`
lines.

Guidelines:

- You **may**:
  - Introduce additional lemmas/proof helpers and call them from the four
    main proof functions.
  - Use Verus ghost/state features (e.g., `assert_by`, `reveal`, or helper specs)
    to structure your reasoning.
- You **do not** have to fully prove deep properties if they are not needed:
  - It is acceptable to leave a proof body mostly empty if Verus can already
    verify the assertions from the specs alone.
  - However, if you strengthen the specs or introduce more complex conditions,
    you may need to fill in more detailed proofs.

The proof helpers are generic over `in1` or `(in1, out)`, not tied to a single
testcase. Use them for general lemmas and proof steps that help Verus prove the
concrete assertions after testcase snippets are injected.

If verification fails, the logs in
`/home/attempts/<run_id>/snippets/<category>/<test_id>/`
will contain the `.rs`, `.stdout`, and `.stderr` files for each check; the
error messages will point to these paths.

---

## 7. Practical strategy

Here is a suggested workflow:

1. **Understand the problem:**
   - Read `/home/problem_artifacts/problem_statement.md` carefully.
   - Inputs are represented by a single logical input struct `In1`.
   - Outputs are represented by `Out`.
   - Check `/home/examples/solve.rs` for a worked example of a complete Verus specification file.

2. **Inspect the existing types and snippets:**
   - Look at `/home/dev/solve.rs` and confirm the definitions of `In1` and `Out`.
   - Look at a few `out.input_defn` / `out.gt_output_defn` files under `/home/symbolic_tests/*/*/`
     to see concrete examples of how `ExecIn1` / `ExecOut` are constructed.

3. **Sketch `pre_spec`:**
   - Encode all input constraints from the problem statement.

4. **Sketch `post_spec`:**
   - Express the output correctness condition from the problem statement.

5. **Add or refine helpers and proofs:**
   - Introduce helper spec functions to keep `pre_spec`/`post_spec` readable.
   - Fill in the four `*_proof` functions enough to help Verus verify the
     `check_*` assertions.

6. **Use the feedback loop:**
   - After running `verus_gym_specgen_check`, read
     `/home/attempts/<run_id>/natural_language_feedback.txt`.
   - For failing tests, inspect:
     - `/home/attempts/<run_id>/snippets/<category>/<test_id>/*.rs`
     - Corresponding `.stdout` / `.stderr` paths mentioned in the error messages.
   - Adjust specs and proofs to fix the misclassified cases or verification failures.

---

## 8. Summary of your task

- **You edit only**:
  - The bodies of `pre_spec`, `post_spec`.
  - The bodies of the four `*_proof` helpers.
  - Any additional spec/proof helpers you introduce.

- **You do not edit**:
  - Types `In1`, `Out`, etc. (already fixed by the conversion layer).
  - Any fixed parsing, printing, or conversion code.
  - `check_*` wrappers' structure or paste markers.
  - Any conversion-related Rust files (`model_mod.rs`, `main.rs`).

- **Goal**:
  - Make `pre_spec` and `post_spec` match the problem's true notion of
    valid inputs and correct outputs as closely as possible.
  - Ensure that, when evaluated on the symbolic tests derived from real
    `.in` / `.out` files, your specs are both **sound** and **complete**
    to the extent that Verus can prove.

NOTE:
- Verus syntax documentation, tips, and instructions can be found in `/home/verus_documentation`. `/home/verus_documentation/guide/src/SUMMARY.md` contains a list of all the chapters and their paths in the filesystem in the guide.
- Some examples of Verus Code in HumanEval problems can be found in `/home/verus_documentation/humaneval_verus_solved_examples`
- Some other examples can be found in: `/home/verus_documentation/some_more_examples`
- Readable evaluator source under `/home/evaluator_scripts/specgen_evaluator/`
  as a guide while you work. The runnable checker command is
  `verus_gym_specgen_check`.
- Feel free to search anywhere in /home/verus_documentation to find helpful information, documentation and take inspiration from the examples!

Now go ahead and start implementing in /home/dev/solve.rs!
Also, you have a limited budget, so please submit whenever ready but do not wait for too long.
\end{lstlisting}